\documentclass[twocolumn,prb,aps,superscriptaddress,longbibliography]{revtex4-1}
\usepackage[T1]{fontenc}
\usepackage[latin9]{inputenc}

\usepackage{amsmath}
\usepackage{amssymb}
\usepackage{bbm}
\usepackage{braket}
\usepackage{dsfont}
\allowdisplaybreaks

\usepackage{graphicx}
\usepackage[colorlinks=true,citecolor=blue,linkcolor=blue]{hyperref}

\newcommand{\equref}[1]{Eq.~(\ref{#1})}
\newcommand{\equsref}[2]{Eqs.~(\ref{#1}) and (\ref{#2})}
\newcommand{\secref}[1]{Sec.~\ref{#1}}
\newcommand{\figref}[1]{Fig.~\ref{#1}}	

\newcommand{\refcite}[1]{Ref.~\onlinecite{#1}}
\newcommand{\refscite}[1]{Refs.~\onlinecite{#1}}

\newcommand{\appref}[1]{Appendix~\ref{#1}}

\newcommand{\diff}{\mathrm{d}}

\newcommand{\pdagger}{{\phantom{\dagger}}}

\newcommand{\K}{{\text{K}}}
\newcommand{\R}{{\text{R}}}

\newcommand{\sign}{\,\text{sign}}
\renewcommand{\approx}{\simeq}
\renewcommand{\Re}{\text{Re}}

\renewcommand{\vec}[1]{\boldsymbol{#1}}

\newcommand{\ie}{i.e.~}
\newcommand{\cf}{cf.~}

\begin{document}
\title{Mechanism, time-reversal symmetry and topology of superconductivity in noncentrosymmetric systems}

\author{M.\,S.\ Scheurer}
\affiliation{Institut f\"ur Theorie der Kondensierten
Materie, Karlsruher Institut f\"ur Technologie, D-76131 Karlsruhe, Germany}

\date{\today}

\begin{abstract}
 We analyze the possible interaction-induced superconducting instabilities in noncentrosymmetric systems based on symmetries of the normal state. 
 It is proven that pure electron-phonon coupling will always lead to a fully gapped superconductor that does not break time-reversal symmetry and is topologically trivial. 
 We show that topologically nontrivial behavior can be induced by magnetic doping without gapping out the resulting Kramers pair of Majorana edge modes.
 In case of superconductivity arising from the particle-hole fluctuations associated with a competing instability, the properties of the condensate crucially depend on the time-reversal behavior of the order parameter of the competing instability. When the order parameter preserves time-reversal symmetry, we obtain exactly the same properties as in case of phonons. If it is odd under time-reversal, the Cooper channel of the interaction will be fully repulsive leading to sign changes of the gap and making spontaneous time-reversal symmetry breaking possible. To discuss topological properties, we focus on fully gapped time-reversal symmetric superconductors and derive constraints on possible pairing states that yield necessary conditions for the emergence of topologically nontrivial superconductivity. 
 These conditions might serve as a tool in the search for topological superconductors. We also discuss implications for oxides heterostructures and single-layer FeSe.
\end{abstract}
\maketitle

\section{Introduction}
In the past few years, topological phases of matter have attracted considerable interest in condensed matter physics\cite{Bernevig}. A central consequence of topologically nontrivial bulk structures, which can be classified by topological invariants in momentum space, is the emergence of zero-energy modes localized at the edge of the system. 
Several different materials have been experimentally identified as topological insulators (see, e.g., \refscite{Koenig,Hsieh2008}), whereas unambiguous evidence of their superconducting analogues, topological superconductors, is still lacking despite intense research activities\cite{MajoranaReview,FuReview}. 
In case of topological superconductors, the edge modes are Majorana bound states (MBS) that are highly sought-after both because of their exotic non-Abelian statistics and potential application in topological quantum computation\cite{MajoranaReview}.

Concerning the realization of these phases, one has to distinguish between ``intrinsic'' and ``extrinsic'' topological superconductors. In ``extrinsic'' systems, superconductivity is induced in a spin-orbit coupled normal conducting material via injection of Cooper pairs from a trivial superconductor and, potentially, additional external fields are applied (see, e.g., \refcite{MajoranaKouwenhoven}). In ``intrinsic'' topological superconductors, both superconductivity as well as the nontrivial topology arise spontaneously due to the internal interactions of the system.
In particular in the search for the right material for the latter type of topological superconductors, guiding principles are required that go beyond model studies and depend only on very few and easily accessible details of the system, such as symmetries or Fermi surface topologies.

A related, but more fundamental, question is whether there is a direct relation between the mechanism that leads to the superconducting instability and the topology of the resulting phase. E.g., in \refcite{ScheurerSchmalian}, it has been shown for a specific model of the two-dimensional (2D) electron fluid in oxide heterostructures that there is a one-to-one correspondence: For conventional, i.e., electron-phonon induced, superconductivity, the condensate is trivial, whereas in case of an unconventional mechanism, i.e., superconductivity resulting from electronic particle-hole fluctuations, topologically nontrivial behavior is found. This can be used to identify the microscopic origin of the pairing state by determining its topological properties.

One promising class of materials for observing intrinsic topological behavior, are superconductors where inversion symmetry is already broken in the normal state\cite{BauerSigristNCSSC}.
This can be the case in the bulk of a three-dimensional (3D) material if the crystal structure lacks a center of inversion as, e.g., for the heavy-fermion superconductor\cite{NCSSCBulk} CePt$_3$Si, or for 2D superconductivity at interfaces and surfaces. Examples of the latter are given by oxide heterostructures\cite{Reyren,110SC} and single-layer\cite{FeSeObs} FeSe on SrTiO$_3$.
The main difference as compared to centrosymmetric systems is that the broken inversion symmetry together with atomic spin-orbit coupling remove the spin degeneracy of the Fermi surfaces. 
The energetic splitting of the Fermi surfaces defines an additional energy scale $E_{\text{so}}$, which essentially changes the theoretical description of superconductivity and has direct consequences for the possible pairing states\cite{CrunoePhaseFactor,DesignPrinciples,*DesignPrinciplesUnpub,Samokhin}.

In this paper, we address both issues of intrinsic topological superconductors outlined above: We derive simple guiding principles for the search for interaction-induced topological superconductivity and relate the mechanism driving the superconducting instability to the topology of the corresponding condensate. Our analysis indicates that magnetic fluctuations, either from the proximity to a magnetic instability or due to magnetic impurities, are essential for the formation of a time-reversal symmetric topological superconductor. To arrive at these conclusions, we focus on strongly noncentrosymmetric systems in the sense that the spin-orbit splitting $E_{\text{so}}$ exceeds the transition temperature $T_c$ of superconductivity, $E_{\text{so}} \gtrsim  T_c$. Our results are based on exact relations following from symmetries of the system, most notable the time-reversal symmetry (TRS) of the high-temperature phase, and, thus, do not depend on further microscopic details. To describe unconventional pairing, we 
apply an effective low-energy approach\cite{JoergChubukovRev} where processes at high energies are assumed to lead to fluctuations in the particle-hole channel that eventually drive the superconducting instability. 

More specifically, it is shown that, irrespective of whether superconductivity arises from phonons or fluctuations of a time-reversal symmetric order parameter in the particle-hole channel, the resulting condensate will be fully gapped, preserve all point symmetries of the high-temperature phase as well as TRS. 
The invariance under the reversal of the time direction both crucially affects the electromagnetic response of the system and determines the topological classification of the superconductor as well as the structure of the MBS emerging at edges of the system in case of a topologically nontrivial phase\cite{Bernevig}. 
To deduce the associated topological invariant (class DIII\cite{ALclasses}), we apply the topological Hamiltonian approach\cite{TopHam1,TopHam2} using the full Green's function obtained from Eliashberg theory\cite{Eliashberg}. 
This captures interaction effects beyond\cite{PoleExpansion} the mean-field level as the full frequency dependence of the self-energy is taken into account.

We find for phonons or fluctuations of a particle-hole order parameter which is even under time-reversal, that the invariant is generically trivial. 
Naturally, significant residual Coulomb repulsion can lead to a topological phase. We show that also magnetic impurities can drive the system into a topologically nontrivial superconducting state that still preserves TRS. Although the disorder required to stabilize the topological structure of the bulk system locally breaks TRS, the MBS at the edge of the topological domain are shown to be protected if there is a residual reflection symmetry at the boundary.

In case of fluctuations of an order parameter that is odd under time-reversal, the situation is completely different: The interaction is fully repulsive within and between all Fermi surfaces such that the resulting superconducting order parameter will have sign changes, can break any point symmetry as well as the TRS of the high-temperature phase.
We consider the limit where the inversion-symmetry breaking terms induce a momentum-space splitting of the Fermi surfaces that is smaller than the scale on which the spin-orbit texture varies. This roughly corresponds to $E_{\text{so}} \ll \Lambda_t$ where $\Lambda_t$ denotes the bandwidth of the system.
We derive an asymptotic symmetry and show that all possible superconducting order parameters can be grouped into two classes: The relative sign of the order parameter can only be either positive or negative at \textit{all} ``Rashba pairs'' of Fermi surfaces. Here ``Rashba pair'' denotes a pair of Fermi surfaces that merge into one doubly degenerate Fermi surface upon hypothetically switching off the inversion-symmetry breaking terms in the Hamiltonian.
To discuss the implications on the topological properties, we focus on fully gapped (sign changes only between the different Fermi surfaces) and time-reversal symmetric superconductors. The latter assumption is not very restrictive as many noncentrosymmetric point groups do not allow for spontaneous breaking of TRS by a single superconducting phase transition\cite{DesignPrinciples, *DesignPrinciplesUnpub}.
Most importantly, we find that, for one-dimensional (1D) and 2D systems, the total number of time-reversal invariant momenta (TRIM) enclosed by Rashba pairs of Fermi surfaces must be necessarily odd for the superconductor to be topologically nontrivial.
Furthermore, in case of a single Rashba pair (two singly-degenerate Fermi surfaces) enclosing an odd number of TRIM, the resulting superconductor, if fully gapped, will be automatically topological irrespective of the dimensionality of the system (i.e., for 1D, 2D and 3D). 
This confirms and generalizes the correspondence between mechanism and topology found in \refcite{ScheurerSchmalian}. 

Our results imply that one should look for superconducting systems that are close to a particle-hole instability with an order parameter that is odd under time-reversal (e.g., a spin-density wave instability) for realizing a noncentrosymmetric topological superconductor. This leads to strong magnetic fluctuations that can drive a superconducting instability. Even if the superconducting state is due to electron-phonon coupling, the proximity to a magnetic phase might lead to the spontaneous formation\cite{DFToxygen1} of local magnetic moments due to initially nonmagnetic impurities which can induce a transition to a topological phase. Alternatively, intentional magnetic doping can be used to render an electron-phonon superconductor topological. 
As the Fermi surface structure is directly accessible experimentally, e.g., via photoemission experiments, the necessary condition that the total number of TRIM enclosed by Rashba pairs of Fermi surfaces must be odd for having a 2D topological superconductor can be readily applied for ruling out certain candidate systems.  

We believe that this work will serve as a guiding tool in the search for topologically nontrivial superconducting states and, in addition, help determining the pairing mechanism of noncentrosymmetric superconductors.

The remainder of the paper is organized as follows. In \secref{ElectronPhononCoupling}, we introduce the notation used in this work and proof the absence of topological superconductivity in a clean electron-phonon superconductor. To describe unconventional pairing, the analysis will be extended to general bosonic fluctuations in \secref{UnconventionalPairing}.
In \secref{TopSupPhon}, we discuss the modifications when disorder is taken into account and show how magnetic impurities can render an electron-phonon superconductor topological.
Finally, \secref{ApplicationToMaterials} is devoted to illustrating the consequences of our results for two specific materials.

\section{Electron-phonon coupling}
\label{ElectronPhononCoupling}
In this section, we will discuss conventional, i.e., electron-phonon induced, superconductivity in noncentrosymmetric systems. Assuming that the normal phase is time-reversal symmetric, we will proof on a very general level that the resulting superconducting state will be necessarily topologically trivial in the absence of disorder and additional residual electronic interactions. The inclusion of the latter two effects, which make topological superconductivity possible, is postponed to \secref{TopSupPhon}. 

Throughout this work, we consider fermions described by the general noninteracting Hamiltonian
\begin{equation}
 \hat{H}_0 = \sum_{\vec{k}} \hat{c}_{\vec{k}\alpha}^\dagger \left(h_{\vec{k}}\right)_{\alpha\beta} \hat{c}^\pdagger_{\vec{k}\beta}, \label{NonInteractingHam}
\end{equation} 
where the indices $\alpha, \beta = 1,2, \dots 2N$ represent all relevant microscopic degrees of freedom, e.g., spin, orbitals and subbands.
Here and in the following we use hats to denote operators acting in the many-body Fock space. The only symmetry we assume in this section is TRS. Time-reversal is represented by the antiunitary operators $\hat{\Theta}$ and $\Theta$ in Fock and single-particle space, respectively, \ie
\begin{equation}
\hat{\Theta} \hat{c}^\dagger_{\vec{k}\alpha}  \hat{\Theta}^\dagger =  \hat{c}^\dagger_{-\vec{k}\beta} T_{\beta\alpha}, \qquad T^\dagger T = \mathbbm{1}, \label{TRSOfOperators}
\end{equation}
such that $\hat{\Theta}\hat{H}_0\hat{\Theta}^\dagger = \hat{H}_0$ is equivalent to
\begin{equation}
 \Theta h_{-\vec{k}} \Theta^\dagger = h_{\vec{k}}. \label{HTRS}
\end{equation} 
Here, $\Theta = T \mathcal{K}$ with $\mathcal{K}$ denoting complex conjugation. Since we will focus on spin-1/2 fermions, it holds $\Theta^2 = -\mathbbm{1}$ and, hence, $T^T = -T$.

The electron-phonon coupling giving rise to superconductivity is taken to be of the general form
\begin{equation}
 \hat{H}_{\text{el-ph}} = \sum_{\vec{k},\vec{k}',l} \hat{c}^\dagger_{\vec{k}\alpha} g^{(l)}_{\alpha\beta}(\vec{k},\vec{k}') \hat{c}^\pdagger_{\vec{k}'\beta} \left(\hat{b}^\dagger_{\vec{k}'-\vec{k}l} + \hat{b}^\pdagger_{\vec{k}-\vec{k}'l} \right). \label{ElPhCoupl}
\end{equation} 
As mentioned above, further interaction channels are assumed to be irrelevant in this section. In \equref{ElPhCoupl}, $\hat{b}^\dagger_{\vec{q}l}$ and $\hat{b}_{\vec{q}l}$ are the creation and annihilation operators of phonons of branch $l$. The associated coupling matrix $g^{(l)}$ can, by virtue of spin-orbit interaction, couple states of different spin and might have nontrivial structure, e.g., in orbital space. It will not be explicitly specified in this work -- only the constraints resulting from Hermiticity and TRS will be taken into account. The former implies
\begin{equation}
 g^{(l)}_{\alpha\beta}(\vec{k},\vec{k}') = \left(g^{(l)}_{\beta\alpha}(\vec{k}',\vec{k})\right)^*. \label{Hermiticity}
\end{equation} 
To analyze the consequences of the latter, first note that 
\begin{equation}
 \hat{\Theta} \hat{b}^{(\dagger)}_{\vec{q}l}  \hat{\Theta}^\dagger = \hat{b}^{(\dagger)}_{-\vec{q}l}
\end{equation} 
since the deformation of the lattice $\hat{Q}_{\vec{q}} \sim \hat{b}^\dagger_{-\vec{q}} + \hat{b}_{\vec{q}}$ and the conjugate momentum $\hat{P}_{\vec{q}} \sim i(\hat{b}^\dagger_{\vec{q}} - \hat{b}_{-\vec{q}})$ must be even, $\hat{\Theta}\hat{Q}_{\vec{q}}\hat{\Theta}^\dagger = \hat{Q}_{-\vec{q}}$, and odd, $\hat{\Theta}\hat{P}_{\vec{q}}\hat{\Theta}^\dagger = -\hat{P}_{-\vec{q}}$, under time-reversal, respectively.
Using this in \equref{ElPhCoupl}, one immediately finds that TRS demands
\begin{equation}
 g^{(l)}(\vec{k},\vec{k}') = T \left(g^{(l)}(-\vec{k},-\vec{k}')\right)^* T^\dagger. \label{TRSOfg}
\end{equation} 
Finally, the Hamiltonian of the phonons reads
\begin{equation}
 \hat{H}_{\text{ph}} = \sum_{\vec{q},l} \hat{b}^{\dagger}_{\vec{q}l}\hat{b}^{\phantom{\dagger}}_{\vec{q}l} \omega_{\vec{q}l}, \label{PhononicHamiltonian}
\end{equation} 
where the phonon dispersion $\omega_{\vec{q}l}$ satisfies $\omega_{\vec{q}l}>0$ and $\omega_{\vec{q}l} = \omega_{-\vec{q}l}$ due to stability of the crystal and TRS, respectively.

\subsection{Effective electron-electron interaction}
\label{EffectiveElElInteraction}
Restating the system in the action description and integrating out the phonon degrees of freedom yields the effective electron-electron interaction
\begin{align}
  \begin{split}
S_{\text{int}}^{\text{eff}} =& -\sum_l\int_{k_1,k_2,q} \frac{\omega_{\vec{q}l}}{\Omega_n^2+\omega_{\vec{q}l}^2} g^{(l)}_{\alpha\beta}(\vec{k}_1+\vec{q},\vec{k}_1)  \\  & \times  g^{(l)}_{\alpha'\beta'}(\vec{k}_2-\vec{q},\vec{k}_2) \,\,  \bar{c}_{k_1+q\alpha}\bar{c}_{k_2-q\alpha'} c_{k_2\beta'}c_{k_1\beta}. \label{MicroscopicInteraction}  \end{split}
\end{align} 
Here $\bar{c}_{\alpha}$ and $c_{\alpha}$ are the Grassmann analogues of the fermionic creation and annihilation operators $\hat{c}_{\alpha}^\dagger$ and $\hat{c}_{\alpha}$. We use $k\equiv (i\omega_n,\vec{k})$, $q\equiv (i\Omega_n,\vec{q})$ with $\int_k$ comprising both momentum and Matsubara summation,
\begin{equation}
 \int_k \dots = \frac{1}{\beta}\sum_{\omega_n} \sum_{\vec{k}} \dots ,
\end{equation} 
where $\beta$ denotes the inverse temperature. 

\begin{figure}[tb]
\begin{center}
\includegraphics[width=0.55\linewidth]{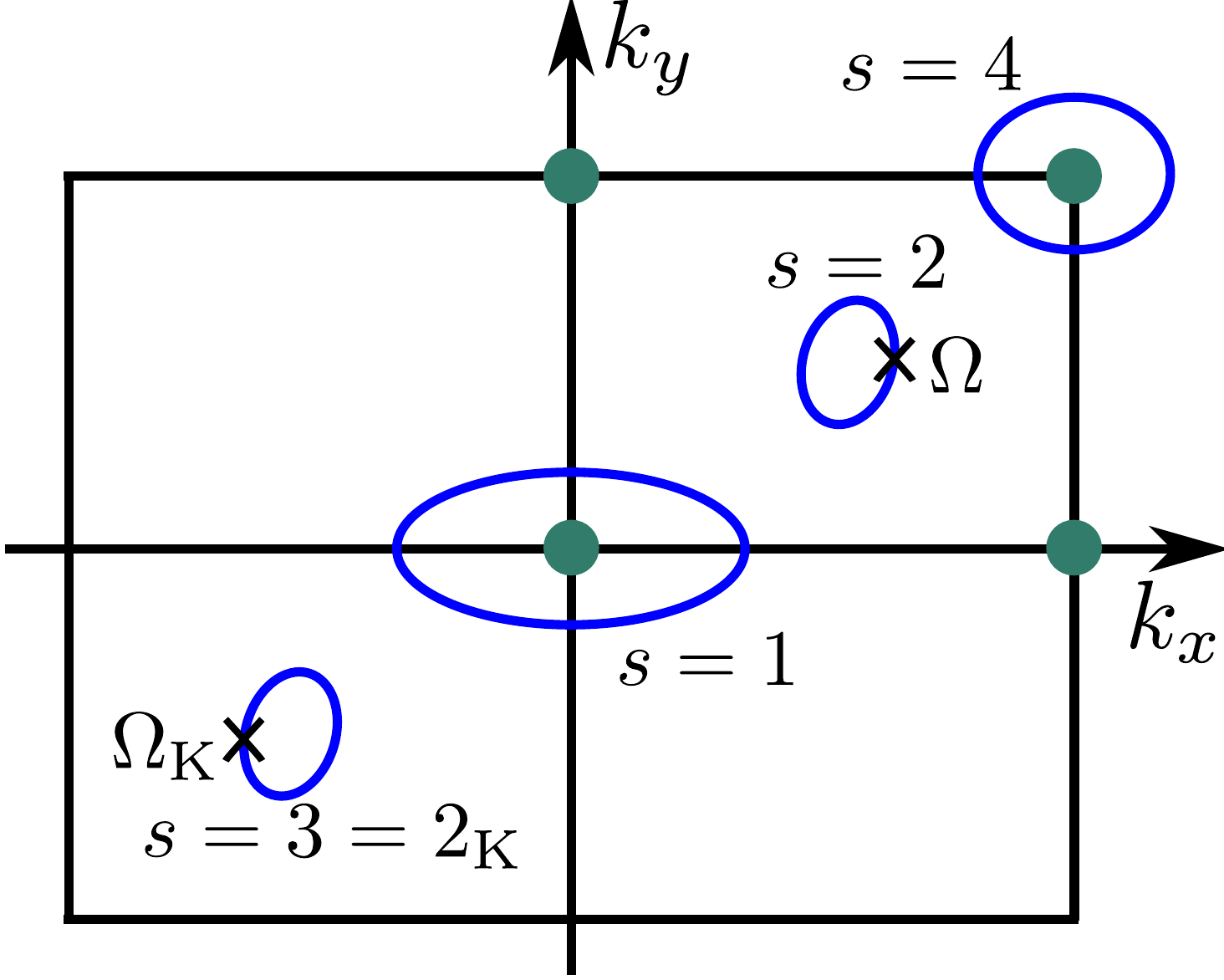}
\caption{(Color online) Illustration of the parameterization of the Fermi surfaces in case of a 2D system. All Fermi surfaces are chosen to be connected such that Kramers partner $\{(s,\Omega),(s_\K,\Omega_\K)\}$ can belong to different Fermi surfaces. All (distinct) TRIM, defined by $\vec{k} = -\vec{k}$, are indicated as green dots.}
\label{FSParameterization}
\end{center}
\end{figure}

For describing superconducting instabilities, it is very convenient to work in the eigenbasis of the noninteracting part $h_{\vec{k}}$ of the high-temperature Hamiltonian. We thus write
\begin{equation}
 c_{k\alpha} = \left(\psi_{\vec{k}s}\right)_{\alpha} f_{ks}, \qquad \bar{c}_{k\alpha} =  \bar{f}_{ks}\left(\psi^*_{\vec{k}s}\right)_{\alpha} \label{TrafoInEigenbasis}
\end{equation} 
where $\psi_{\vec{k}s}$ denote the eigenstates of $h_{\vec{k}}$, \ie $h_{\vec{k}}\psi_{\vec{k}s} = \epsilon_{\vec{k}s}\psi_{\vec{k}s}$. 
If the summation over $s$ includes all $2N$ values, \equref{TrafoInEigenbasis} will just constitute a unitary transformation and thus be exact. 
In the following, we will only take into account the bands that lead to Fermi surfaces and focus on the degrees of freedom in the energetic vicinity of the chemical potential ($-\Lambda < \epsilon_{\vec{k}s} < \Lambda$). Therefore, \equref{TrafoInEigenbasis} has to be understood as a low-energy approximation. We will label the states in such a way that, for each $s$, the Fermi momenta $\{\vec{k} | \epsilon_{\vec{k}s} = 0\}$ form a connected set (see \figref{FSParameterization}) which we will refer to as Fermi surface $s$ in the remainder of the paper. 

Inserting the transformation (\ref{TrafoInEigenbasis}) into \equref{MicroscopicInteraction} yields
\begin{equation}
 S_{\text{int}}^{\text{eff}} = \int_{k_1,k_2,q} V^{s_1s_2}_{s_3s_4}(k_1,k_2,q) \, \bar{f}_{k_1+qs_1}\bar{f}_{k_2-qs_2} f_{k_2s_3}f_{k_1s_4} \label{InteractionInEigenbasis}
\end{equation}
with coupling tensor
\begin{align}
 & V^{s_1s_2}_{s_3s_4}(k_1,k_2,q) \\ &= - \sum_l \frac{\omega_{\vec{q}l}}{\Omega_n^2+\omega_{\vec{q}l}^2} G^{(l)}_{s_1s_4}({\vec{k}_1+\vec{q},\vec{k}_1})G^{(l)}_{s_2s_3}({\vec{k}_2-\vec{q},\vec{k}_2}) \nonumber
\end{align} 
where we have introduced
\begin{equation}
 G_{ss'}^{(l)}(\vec{k},\vec{k}') = \psi_{\vec{k}s}^\dagger \, g^{(l)}(\vec{k},\vec{k}') \psi^\pdagger_{\vec{k}'s'}. \label{DefinitionOfG}
\end{equation} 

The Hermiticity constraint (\ref{Hermiticity}) now becomes
\begin{equation}
 G_{ss'}^{(l)}(\vec{k},\vec{k}') = \left[G_{s's}^{(l)}(\vec{k}',\vec{k}) \right]^*. \label{HermiticityGMatrix}
\end{equation} 
In this paper, we will focus on systems with singly-degenerate Fermi surfaces which requires a center of inversion to be absent. The combination of broken inversion symmetry, e.g., at an interface or in the bulk of a noncentrosymmetric crystal, and atomic spin-orbit coupling will generally lift the degeneracy of the Fermi surfaces. Together with TRS of the normal state Hamiltonian, \equref{HTRS}, the absence of degeneracy of the Fermi surfaces implies
\begin{equation}
 \psi_{\vec{k}s} = e^{i\varphi_{\vec{k}}^s} \Theta \psi_{-\vec{k}s_\K} \label{TRSOfESs}
\end{equation} 
with ($\vec{k}$-dependent) phases $\varphi_{\vec{k}}^s \in \mathbbm{R}$ that are determined by how the phases of the eigenstates $\psi_{\vec{k}s}$ are chosen (gauge symmetry). \equref{TRSOfESs}, which will be used repeatedly throughout the paper, shows that TRS is more restrictive in noncentrosymmetric systems as compared to the situation with inversion symmetry since the eigenstate at $\vec{k}$ with energy close to the Fermi level fully determines the structure of the wavefunction at low energies at momentum $-\vec{k}$. Here and in the following $s_\K$ denotes the Fermi surface consisting of the Kramers partners of the momenta of $s$. Depending on the topology of the Fermi surfaces with respect to the TRIM, both $s_\K = s$ and $s_\K \neq s$ are possible. E.g., for the Fermi surfaces shown in \figref{FSParameterization}, $s=1,4$ and $s=2=3_\K$ are examples of the former and latter case, respectively. 

In this work we will not have to specify the phases $\varphi_{\vec{k}}^s$ explicitly, it will only be taken into account that
\begin{equation}
 e^{i\varphi_{-\vec{k}}^{s_\K}} = - e^{i\varphi_{\vec{k}}^s} \label{ParityOfThePhases}
\end{equation} 
as a consequence of $\Theta^2 = -\mathbbm{1}$.
Using \equsref{TRSOfg}{TRSOfESs}, \ie the consequences of TRS for the electron-phonon coupling and the wavefunctions of the normal state Hamiltonian, it is straightforward to show that
\begin{equation}
 G_{ss'}^{(l)}(\vec{k},\vec{k}') = e^{i(\varphi_{\vec{k}'}^{s'}-\varphi_{\vec{k}}^s)} \left(G_{s^{\phantom{.}}_\K s'_\K}^{(l)}(-\vec{k},-\vec{k}')\right)^*. \label{TRSOfG}
\end{equation} 
This is a central relation for our analysis as it can be used to rewrite the Cooper channel of the interaction (\ref{InteractionInEigenbasis}), i.e., the scattering process of a Kramers pair $\{s,k;s_\K,-k\}$ of quasiparticles into another Kramers pair $\{s',k';s'_\K,-k'\}$ as depicted in \figref{DiagramsOfEliashberg}(a). \equref{TRSOfG} readily yields for this type of scattering event
\begin{equation}
 V^{s's'_\K}_{s_\K s} (k,-k,k'-k) = e^{i(\varphi_{\vec{k}}^s-\varphi_{\vec{k}'}^{s'})} \mathcal{V}_{s's}(k';k) \label{CooperChannelInteraction}
\end{equation} 
where
\begin{align}
 \begin{split}
& \mathcal{V}_{s's}(k';k) \\ &= - \sum_l \frac{\omega_{\vec{k}'-\vec{k}l}}{(\omega_{n'}-\omega_n)^2+\omega_{\vec{k}'-\vec{k}l}^2} \left|G^{(l)}_{s's}({\vec{k}',\vec{k}})\right|^2 < 0. \label{InteractionMatrixElement} \end{split}
\end{align} 
The very same matrix elements also govern the forward scattering processes shown in \figref{DiagramsOfEliashberg}(b) with amplitude $\mathcal{F}_{s's}(k';k) := V^{s's}_{s's}(k,k',k'-k)$: Using the Hermiticity (\ref{Hermiticity}) of the electron-phonon interaction one finds $\mathcal{F}=\mathcal{V}$.

Consequently, the combination of TRS and the fact that the Fermi surfaces are singly degenerate highly constraints the Cooper channel of the interaction. As stated in \equref{CooperChannelInteraction}, it can be written as the product of the time-reversal phases defined in \equref{TRSOfESs} and the forward scattering matrix $\mathcal{V}_{s's}(k';k)$ which only has negative entries.  
We emphasize that this is a very general result since no additional model specific assumptions other than TRS and singly-degenerate Fermi surfaces (such as number/character of relevant orbitals or dimensionality of the system) have been taken into account.  
In the next subsection, we will analyze the consequences for the resulting possible superconducting instabilities using Eliashberg theory\cite{Eliashberg}. 

Before proceeding, a few remarks are in order:
Naturally, the Cooper scattering amplitude (\ref{CooperChannelInteraction}) is a complex number that depends on the phases of the eigenstates $\psi_{\vec{k}s}$ whereas the forward scattering amplitude $\mathcal{V}$ is independent of the phases as it always involves a wavefunction and its complex conjugate in pairs [\cf\figref{DiagramsOfEliashberg}(b)]. Despite the gauge dependence of the Cooper scattering amplitude, the time-reversal and topological properties of the resulting superconducting state are, of course, independent of the time-reversal phases $\varphi_{\vec{k}}^s$ as we will see explicitly below.

We finally list three properties
\begin{subequations}
\begin{align}
 \mathcal{V}_{s's}(k';k) &= \mathcal{V}_{ss'}(k;k'), \label{VPropI} \\
 \mathcal{V}_{s's}(k';k) &= \mathcal{V}_{s'_\K s^{\phantom{.}}_\K}(-k';-k), \label{VPropII} \\
 \mathcal{V}_{s's}\left(i\omega_{n'},\vec{k}';i\omega_n,\vec{k}\right) &= \mathcal{V}_{s's}\left(-i\omega_{n'},\vec{k}';-i\omega_n,\vec{k}\right), \label{VPropIII}
\end{align}\label{VProp}\end{subequations} 
which are readily read off from \equref{InteractionMatrixElement} and will be taken into account in the following.

\subsection{Eliashberg theory}
\label{EliashbergTheory}
The aim of Eliashberg theory\cite{Eliashberg,EliashbergRev} consists of calculating the Nambu Green's function in the superconducting phase. For our purposes, it will be convenient to perform the calculation in the eigenbasis of the normal state Hamiltonian $h_{\vec{k}}$. We thus introduce the Nambu Green's function as
\begin{equation}
 \mathcal{G}_{ss'}(i\omega_n,\vec{k}) := -\frac{1}{\beta}\begin{pmatrix} \braket{f_{ks}\bar{f}_{ks'}}  & \braket{f_{ks}f_{-ks'_\K}} \\ \braket{\bar{f}_{-ks^{\phantom{.}}_\K}\bar{f}_{ks'}} & \braket{\bar{f}_{-ks^{\phantom{.}}_\K}f_{-ks'_\K}}  \end{pmatrix}. \label{PathIntegralDef}
\end{equation} 
According to this ansatz, all Cooper pairs carry zero total momentum excluding the formation of translation-symmetry breaking superconductivity, e.g., Fulde-Ferrell-Larkin-Ovchinnikov\cite{FF,LO} states. 

In addition, we assume that, for determining the superconducting properties, the Green's function can be approximated to be diagonal in Fermi-surface space,
\begin{equation}
  \mathcal{G}_{ss'}(i\omega_n,\vec{k}) = \delta_{s,s'} \mathcal{G}_{s}(i\omega_n,\vec{k}), \label{ParameterizationOfG}
\end{equation} 
which will be referred to as \textit{weak-pairing approximation} in the following. 

When the energetic separation $E_{\text{so}}$ of the Fermi surfaces is larger than the energy range $2\Lambda$ of the low-energy theory, \equref{ParameterizationOfG} is enforced by momentum conservation such that the weak-pairing approximation becomes exact. 
However, even for $E_{\text{so}} < 2\Lambda$, \equref{ParameterizationOfG} can be used as matrix elements of the Green's function (\ref{PathIntegralDef}) with $s\neq s'$ couple single-particle states with energies differing by $E_{\text{so}}$.
In the calculation of the leading superconducting instability this will cut off the Cooper logarithms associated with these processes unless these integrals are first cut off by temperature. In other words, the weak-pairing approximation is expected to be applicable for determining the superconducting properties as long as $E_{\text{so}} \gtrsim  T_c$. This criterion agrees with the explicit check of the validity of the weak-pairing approximation in \refcite{DesignPrinciples,*DesignPrinciplesUnpub}.

Physically, the weak-pairing approximation means that the Cooper pairs are made from the same quantum numbers as the normal state. We emphasize that this is only a statement about the propagator $\mathcal{G}$ and does not restrict the interaction to be diagonal in Fermi-surface space. On the contrary, interband interactions are even essential to have a unique superconducting order parameter as, otherwise, the free energy would be independent of the relative phase of the order parameter at different Fermi surfaces.

Before proceeding with the calculation of the Green's function, let us discuss its antiunitary symmetries (see \appref{TRSOfGF} for more details on the derivation of the following statements).
To begin with charge-conjugation symmetry, it holds
\begin{equation}
 \Xi \mathcal{G}_{s_\K}(i\omega_n,-\vec{k}) \Xi^{-1} = -\mathcal{G}_{s}(i\omega_n,\vec{k}), \quad \Xi = \tau_1 \mathcal{K}, \label{ChargeConjugation}
\end{equation} 
which is just a consequence of the inherent redundancy of the Nambu Green's function in \equref{PathIntegralDef}.

Secondly, the TRS constraint, which is, in the microscopic basis, described by the operator $\Theta$, reads 
\begin{equation}
 \widetilde{\Theta}_{\vec{k}s} \mathcal{G}_{s_\K}(-k) \widetilde{\Theta}_{\vec{k}s}^{-1} = \mathcal{G}_s(k), \quad \widetilde{\Theta}_{\vec{k}s} = \tau_3 e^{-i\varphi_{\vec{k}}^s\tau_3} \mathcal{K}, \label{TRSOfGreesFunc}
\end{equation}
when transformed into the eigenbasis according to \equref{TrafoInEigenbasis}. 
The phases $\varphi_{\vec{k}}^s$ enter because of the relation (\ref{TRSOfESs}) between the wavefunctions of Kramers partners.
Note that the expression for the time-reversal operator stated above yields $\widetilde{\Theta}^2_{\vec{k}s} = \mathbbm{1}$ which, at first sight, seems to disagree with $\mathcal{G}_s$ being a Green's function of spin-1/2 fermions. This can be reconciled by noting that the full time-reversal operator $\widetilde{\Theta}_{\vec{k}s}\mathcal{I}$ that also includes the inversion $\mathcal{I}$ of momentum 
indeed satisfies $(\widetilde{\Theta}_{\vec{k}s}\mathcal{I})^2 = \widetilde{\Theta}_{\vec{k}s}\widetilde{\Theta}_{-\vec{k}s_\K} = -\mathbbm{1}$ as a consequence of \equref{ParityOfThePhases}. Note that this subtlety usually does not play any role as the time-reversal operator in momentum space in many cases (e.g.~in the microscopic basis as in \equref{TRSOfOperators}) does not depend on momentum.
It indicates that the property (\ref{ParityOfThePhases}) of the phases $\varphi_{\vec{k}}^s$ carries the information that the nondegenerate bands of the system microscopically arise from spin-$1/2$ fermions. 

In order to compare our Green's function approach with the mean-field picture, which will be particularly useful when discussing the topological properties below, let us consider the generic superconducting mean-field Hamiltonian
\begin{equation}
 \hat{H}_{\text{MF}} = \sum_{\vec{k}} \hat{c}^\dagger_{\vec{k}} h_{\vec{k}} \hat{c}^\pdagger_{\vec{k}} + \frac{1}{2} \sum_{\vec{k}} \left(\hat{c}^\dagger_{\vec{k}} \Delta_{\vec{k}} \left(c_{-\vec{k}}^\dagger\right)^T + \text{H.c.}  \right). \label{MeanFieldHamiltonian}
\end{equation}  
Performing the transformation into the band basis analogously to \equref{TrafoInEigenbasis}, we get, within the weak-pairing approximation,
\begin{equation}
 \hat{H}_{\text{MF}} = \frac{1}{2} \sum_{\vec{k}} \hat{\Psi}^\dagger_{\vec{k}s} h^{\text{BdG}}_{\vec{k}s} \hat{\Psi}_{\vec{k}s}^\pdagger
\end{equation} 
with Nambu spinor $\hat{\Psi}^\dagger_{\vec{k}s} = \begin{pmatrix} \hat{f}^\dagger_{\vec{k}s} & \hat{f}^\pdagger_{-\vec{k}s_\K}  \end{pmatrix}$ and Bogoliubov-de Gennes (BdG) Hamiltonian
\begin{equation}
       h^{\text{BdG}}_{\vec{k}s}= \begin{pmatrix} \epsilon_{\vec{k}s}  &  \widetilde{\Delta}_{s}(\vec{k})e^{-i\varphi_{\vec{k}}^s} \\  \widetilde{\Delta}_{s}^*(\vec{k})e^{i\varphi_{\vec{k}}^s} & -\epsilon_{\vec{k}s}  \end{pmatrix}. \label{BdGHamiltonian}
\end{equation} 
Here we have introduced the Fermi-surface-diagonal matrix elements $\widetilde{\Delta}_{s}(\vec{k}) = \braket{\psi_{\vec{k}s}|\Delta_{\vec{k}} T^\dagger|\psi_{\vec{k}s}}$ of the order parameter.
 
Demanding that \equref{MeanFieldHamiltonian} be time-reversal symmetric, $\hat{\Theta}\hat{H}_{\text{MF}}\hat{\Theta}^{-1} = \hat{H}_{\text{MF}}$ with $\hat{\Theta}$ as defined in \equref{TRSOfOperators}, one finds that TRS is equivalent to $\widetilde{\Delta}_{s}(\vec{k}) \in \mathbbm{R}$. 
Comparison with \equref{BdGHamiltonian} shows that TRS on the level of the BdG Hamiltonian reads $\widetilde{\Theta}_{\vec{k}s} h^{\text{BdG}}_{-\vec{k}s_\K} \widetilde{\Theta}_{\vec{k}s}^{-1} = h^{\text{BdG}}_{\vec{k}s}$ which is just a special case of \equref{TRSOfGreesFunc} restricted to the mean-field level where 
\begin{equation}
 \mathcal{G}_s(i\omega_n,\vec{k}) = \left(i\omega_n - h^{\text{BdG}}_{\vec{k}s}\right)^{-1}. \label{MeanFieldGreensFunction}
\end{equation} 
The relation (\ref{MeanFieldGreensFunction}) between the mean-field Green's function $\mathcal{G}$ and the weak-pairing representation of the general multiband mean-field Hamiltonian (\ref{MeanFieldHamiltonian}) will be relevant in \secref{TopologicalProperties} when discussing topological properties of the superconducting phase beyond mean-field.

\begin{figure}[tb]
\begin{center}
\includegraphics[width=\linewidth]{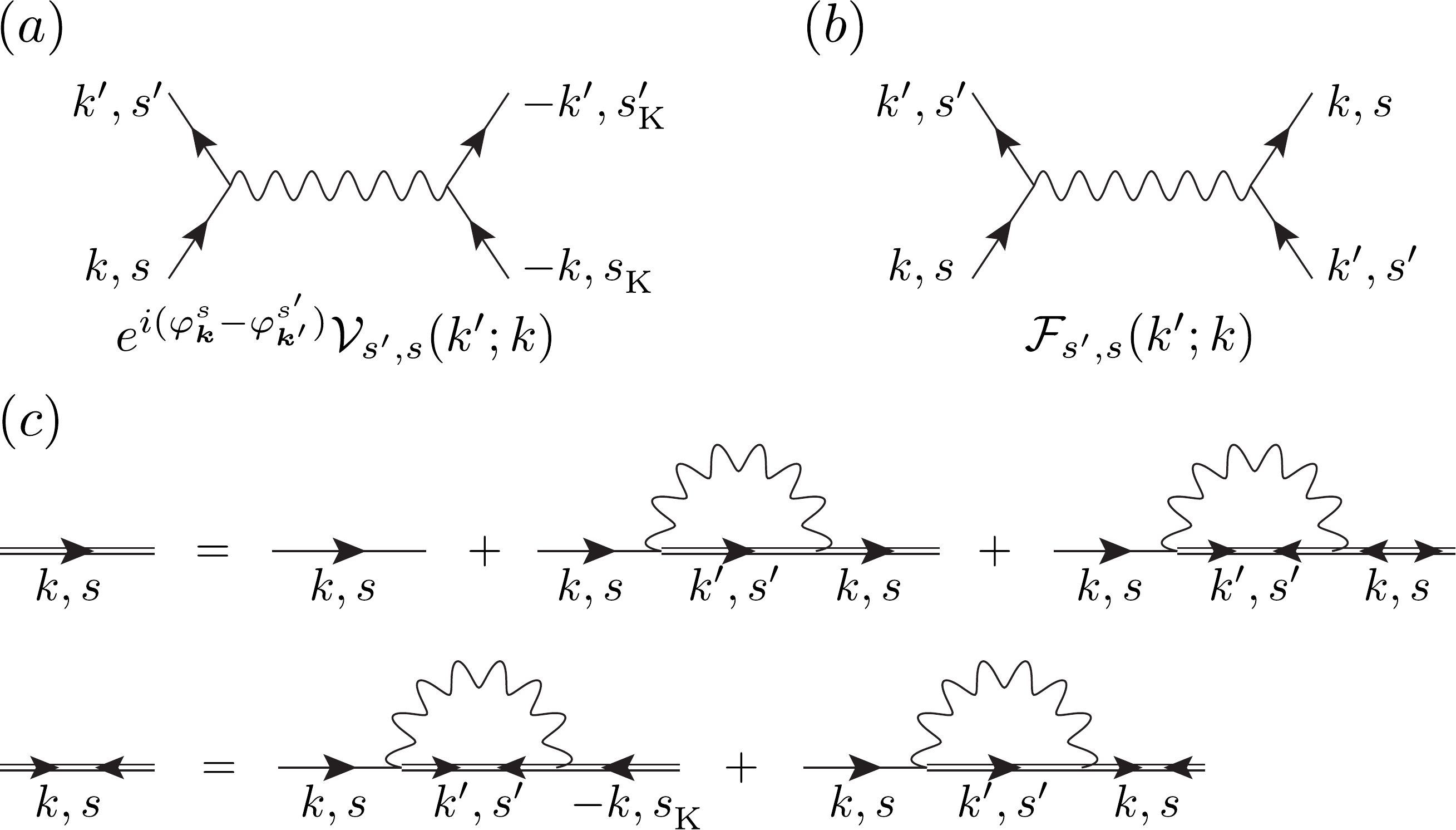}
\caption{To calculate the superconducting properties we only need two specific channels of the effective electron-electron interaction: The scattering of Cooper pairs as shown in (a) and forward scattering (b). Applying the weak-pairing approximation (\ref{ParameterizationOfG}) to  Eliashberg theory, the normal and anomalous components of the Nambu Green's function follow from the self-consistency relations represented diagrammatically in (c).}
\label{DiagramsOfEliashberg}
\end{center}
\end{figure}

Let us next calculate the Green's function $\mathcal{G}$ within Eliashberg approximation, i.e., by solving the Dyson equations for the electronic and anomalous Green's function represented diagrammatically in \figref{DiagramsOfEliashberg}(c). The validity of this approach goes beyond weak coupling. It is controlled in the limit $m/M \ll 1$ with $m$ ($M$) denoting the mass of the electrons (ions) where vertex corrections can be neglected according to Migdal's theorem\cite{Migdal}.
As anticipated by our discussion above, we now see directly that only the forward, \figref{DiagramsOfEliashberg}(a), and the Cooper, \figref{DiagramsOfEliashberg}(b), scattering amplitudes of the phonon mediated interaction enter.

We parameterize the Green's function according to 
\begin{align}
\mathcal{G}^{-1}_s(k) =  i\omega_n Z_s(k) \tau_0 - \widetilde{\epsilon}_{s}(k)\tau_3 - \begin{pmatrix} 0 & \Phi_s(k) \\ \overline{\Phi}_s(k) & 0 \end{pmatrix} \label{GreensFunctionParameterize}
\end{align}
with quasiparticle weight $Z_s(k)$, $\widetilde{\epsilon}_{s}(k) = \epsilon_{\vec{k}s} + \delta\epsilon_{s}(k)$, where $\delta\epsilon_{s}(k)$ is the band renormalization, and anomalous self-energies $\Phi_s(k)$ and $\overline{\Phi}_s(i\omega_n,\vec{k}) = \Phi_s^*(-i\omega_n,\vec{k})$. 
These quantities, which uniquely determine $\mathcal{G}$, follow from the self-consistency equations
\begin{subequations}\begin{align}
Z_s(k) &= 1 + \frac{2}{i \omega_n} \sum_{s'}\int_{k'}  \mathcal{V}_{ss'}(k;k') \frac{i\omega_{n'}Z_{s'}(k')}{\mathcal{D}_{s'}(k')}, \label{GeneralEliashbergZ} \\
\widetilde{\Phi}_s(k) &= 2 \sum_{s'}\int_{k'} \mathcal{V}_{ss'}(k;k') \frac{\widetilde{\Phi}_{s'}(k')}{\mathcal{D}_{s'}(k')}, \\
\delta\epsilon_{s}(k) &= - 2 \sum_{s'}\int_{k'}  \mathcal{V}_{ss'}(k;k') \frac{\widetilde{\epsilon}_{s'}(k')}{\mathcal{D}_{s'}(k')}, \label{GeneralEliashbergeps} 
\end{align}\label{GeneralEliashbergEquations}\end{subequations}
where we have introduced
\begin{equation}
 \mathcal{D}_s(k) = \left[i\omega_{n} Z_{s}(k)\right]^2 - \left[\widetilde{\epsilon}^{\,2}_{s}(k) + \overline{\Phi}_{s}(k)\Phi_{s}(k)\right]
\end{equation} 
for notational convenience.
Here \equsref{VPropI}{VPropII} have been taken into account to write the expressions in more compact form.
The additional factor of $2$ on the right-hand sides of \equref{GeneralEliashbergEquations} (as compared to the more frequently encountered form of the Eliashberg equations for spinfull fermions) 
arises since, in the band basis, the theory looks as if we were considering spinless particles making more contractions of the interaction vertex possible. The time-reversal phases $\varphi_{\vec{k}}^s$ of the Cooper amplitude in \equref{CooperChannelInteraction} that have been absorbed by defining $\widetilde{\Phi}_{s}(k) := \Phi_s(k) e^{i\varphi_{\vec{k}}^s}$ are reminiscent of the fact that we are considering not truly spinless particles, but spin-$1/2$ particles with singly-degenerate bands.

In this work, we will focus on the vicinity of the critical temperature of the superconducting transition and, hence, linearize the Eliashberg equations (\ref{GeneralEliashbergEquations}) in $\Phi$. To proceed further, let us rewrite the momentum summation as an angular integration over the Fermi surfaces and an energy integration (momentum perpendicular to the Fermi surface) subject to an energetic cutoff $\Lambda$, which is a characteristic energy scale of the phonons (e.g., the Debye energy). More explicitly, we replace
\begin{equation}
 \sum_{s}\int_{k} \dots \rightarrow \beta^{-1} \sum_{\omega_n} \sum_{s} \int_{-\Lambda}^{\Lambda} \diff \epsilon \int_s\diff\Omega \, \rho_s(\Omega) \dots,
\end{equation} 
where $\rho_s(\Omega) >0$ denotes the angle-resolved density of states that is taken to be independent of $\epsilon$. The dimensionality of $\int_s\diff\Omega$ is set by the dimensionality of the Fermi surface $s$. For the general purposes of this paper, we do not have to specify any parameterization, we will, as illustrated in \figref{FSParameterization}, only apply the convention that the Kramers partner of the state $(s,\Omega)$ is given by $(s_\K,\Omega_\K)$.

In addition, we take the interaction $\mathcal{V}$, $\delta\epsilon$ and $\widetilde{\Phi}$ as well as the quasiparticle residue $Z$ to be only weakly dependent on the momentum perpendicular to the Fermi surface (valid for $m/M \ll 1$) and set
\begin{align}
\begin{split}
 \mathcal{V}_{ss'}(i\omega_n,\vec{k};i\omega_{n'},\vec{k}') &\approx \mathcal{V}_{ss'}(i\omega_n,\Omega;i\omega_{n'},\Omega'), \\ \widetilde{\Phi}_s(i\omega_n,\vec{k}) &\approx \widetilde{\Phi}_s(i\omega_{n},\Omega) \label{SeparationApprox}\end{split}
\end{align}  
and similarly for $\delta\epsilon$ and $\widetilde{\Phi}$.
With these approximations, the Eliashberg equations (\ref{GeneralEliashbergEquations}) become (for $\Lambda\rightarrow \infty$)
\begin{subequations}
\begin{align}
\begin{split}
Z_s(i\omega_{n},\Omega) &= 1 - \frac{2}{2n+1} \sum_{n'}\sum_{s'}\int_{s'}\diff\Omega' \rho_{s'}(\Omega') \\ & \times \mathcal{V}_{ss'}(i\omega_n,\Omega;i\omega_{n'},\Omega') \sign(2n'+1),  \label{EliashbergQPResidue}\end{split} \\ 
\begin{split}
 \delta_s(i\omega_{n},\Omega) &= \sum_{n'}\sum_{s'}\int_{s'}\diff\Omega' v_{ss'}(i\omega_n,\Omega;i\omega_{n'},\Omega') \\ & \times \delta_{s'}(i\omega_{n'},\Omega'), \label{EliashbergGap}\end{split} \\
\delta\epsilon_s(k)& = 0,
\end{align}\label{LinearizedEliashbergEquations}\end{subequations}
i.e., there is no Fermi velocity correction.
In \equref{LinearizedEliashbergEquations}, the normalized anomalous self-energy 
\begin{equation}
 \delta_s(i\omega_{n},\Omega) :=  \Theta_{s,n}(\Omega) \, \widetilde{\Phi}_s(i\omega_{n},\Omega) \label{DefinitionOfDelta}
\end{equation} 
with\footnote{In the expression for $\Theta_{s,n}(\Omega)$, we have, for simplicity, already taken into account that $Z$ is real. In general, $|Z_s(i\omega_{n},\Omega)|$ has to be replaced by $Z_s(i\omega_{n},\Omega)\sign(\Re(Z_s(i\omega_{n},\Omega)))$ such that $\delta$ and $v$ will only be generically real if $Z \in \mathbbm{R}$.}
\begin{equation}
 \Theta_{s,n}(\Omega) = \frac{\sqrt{\rho_s(\Omega)} }{\sqrt{|2n+1|}\sqrt{|Z_s(i\omega_{n},\Omega)|}}
\end{equation} 
has been introduced in order to render the kernel
\begin{align}
 \begin{split}
&v_{ss'}(i\omega_{n},\Omega;i\omega_{n'},\Omega') \\ & \quad := -2 \Theta_{s,n}(\Omega)\mathcal{V}_{ss'}(i\omega_{n},\Omega;i\omega_{n'},\Omega') \Theta_{s',n'}(\Omega') \label{DefinitionOfv} \end{split}
\end{align} 
of the gap equation (\ref{EliashbergGap}) symmetric. 

Note that, as a consequence of linearizing in $\widetilde{\Phi}$, \equref{EliashbergQPResidue} explicitly determines $Z_s(i\omega_{n},\Omega)$, i.e., it follows without solving a self-consistency equation. We directly see that $Z_s(i\omega_n,\Omega) \in \mathbbm{R}$. In addition, it holds $Z_s(k)=Z_{s_\K}(-k)$ which is a consequence of its definition but can, alternatively, be explicitly seen in \equref{EliashbergQPResidue} using \equref{VPropII}. Together with $Z_s(i\omega_n,\Omega) = Z_s^*(-i\omega_n,\Omega)$ following from \equref{VPropIII}, we can summarize
\begin{equation}
 Z_s(i\omega_n,\Omega) = Z_s(-i\omega_n,\Omega) = Z_{s_\K}(i\omega_n,\Omega_\K) \in \mathbbm{R}. \label{PropertiesOfZ}
\end{equation} 
The properties of the superconducting order parameter follow from the second Eliashberg equation (\ref{EliashbergGap}). 
As opposed to mean-field theory, the temperature dependence of \equref{EliashbergGap} is more complicated and hidden in the kernel $v$ defined in \equref{DefinitionOfv}. However, it can be shown (see \appref{HighTemperature}) that the leading superconducting instability is, as in the mean-field case, determined by the largest eigenvalue of the (symmetric and real) matrix $v$ while the order parameter $\delta_s(i\omega_{n},\Omega)$ belongs to the corresponding eigenspace.

As $v$ only has positive components, we conclude from the Perron-Frobenius theorem\cite{PerronFrobTh}, that the largest eigenvalue of $v$ is nondegenerate with associated eigenvector that can be chosen to have purely positive components as well. 
Therefore, the leading instability is characterized by $\delta_s(i\omega_{n},\Omega) > 0$ and, hence, $\widetilde{\Phi}_s(i\omega_{n},\Omega) > 0$, i.e., the superconductor is fully gapped with the sign of the gap being the same on all Fermi surfaces. Also as a function of Matsubara frequency, the anomalous self-energy does not change sign. Due to the absence of any sign change, the superconducting state cannot break any point-group symmetry and must, therefore, transform under the trivial representation of the point group (see \appref{Representation} for a proof of this conclusion). 

Since $v_{ss'}(i\omega_{n},\Omega;i\omega_{n'},\Omega')$ is invariant under a simultaneous sign change of $\omega_n$ and $\omega_{n'}$, which follows from and \equsref{VPropIII}{PropertiesOfZ}, we know that $\delta_s(-i\omega_{n},\Omega)$ is also a solution of \equref{EliashbergGap}. Due to the absence of degeneracies, we conclude that $\delta_s(i\omega_{n},\Omega) = \pm \delta_s(-i\omega_{n},\Omega)$, i.e., we obtain either an even- or an odd-frequency pairing state. As $\delta_s(i\omega_{n},\Omega) > 0$ odd-frequency pairing can be excluded. In combination with $\widetilde{\Phi}_{s_\K}(-k) = \widetilde{\Phi}_s(k)$ following from Fermi-Dirac statistics and the property (\ref{ParityOfThePhases}) of the time-reversal phases, one has
\begin{equation}
 \widetilde{\Phi}_s(i\omega_{n},\Omega) = \widetilde{\Phi}_s(-i\omega_{n},\Omega) = \widetilde{\Phi}_{s_\K}(i\omega_{n},\Omega_\K) > 0. \label{PropertiesOfPhiTilde}
\end{equation} 
\equsref{PropertiesOfZ}{PropertiesOfPhiTilde} constitute the main results of this subsection.

\subsection{Topological properties}
\label{TopologicalProperties}
We will next discuss the consequences of \equsref{PropertiesOfZ}{PropertiesOfPhiTilde} for the topology of the corresponding superconducting phase.
For this purpose, we first need to analyze its antiunitary symmetries. In order to go beyond a mean-field description, we have to discuss these symmetries on the level of Green's functions. 

By design, the Nambu Green's function defined in \equref{PathIntegralDef} satisfies the particle-hole symmetry (\ref{ChargeConjugation}). 
TRS is a much more interesting property of a superconductor in the sense that it can be spontaneously broken by the formation of the condensate. However, it is straightforward to check that \equref{TRSOfGreesFunc} is satisfied as a consequence of $Z_s(k)$ and  $\widetilde{\Phi}_s(k)$ being real valued and invariant under in $(s,k) \rightarrow (s_\K,-k)$ [\cf\equsref{PropertiesOfZ}{PropertiesOfPhiTilde}] together with $\epsilon_{\vec{k}s} = \epsilon_{-\vec{k}s_\K}$ resulting from the TRS of the high-temperature phase. Therefore, no spontaneous TRS breaking is possible in the weak-pairing limit if superconductivity is due to electron-phonon coupling. 

Consequently, the resulting system is invariant both under charge conjugation $\Xi$ with $\Xi^2=\mathbbm{1}$ as well as under time-reversal $\widetilde{\Theta}_{\vec{k}s}\mathcal{I}$ satisfying $(\widetilde{\Theta}_{\vec{k}s}\mathcal{I})^2 = -\mathbbm{1}$ and, thus, belongs to class DIII\cite{ALclasses}. In 1D and 2D, the superconductor is classified by a $\mathbbm{Z}_2$ and in 3D by a $\mathbbm{Z}$ topological invariant \cite{Bernevig}. To calculate these invariants, we will use the topological Hamiltonian approach\cite{TopHam1,TopHam2}: For a system with a finite gap, the topological properties of the many-body system described by the full Green's function $\mathcal{G}$ are calculated from the effective mean-field Green's function 
\begin{equation}
 \mathcal{G}^t_s(i\omega_n,\vec{k}) = \left(i\omega_n - h^t_{\vec{k}s} \right)^{-1} \label{DefinitionTopHamGF}
\end{equation} 
where the ``topological Hamiltonian'' is given by
\begin{equation}
  h^t_{\vec{k}s} := -\mathcal{G}^{-1}_s(i\omega = 0,\vec{k}). \label{DefOfTopHam}
\end{equation} 
For the calculation of the topological invariant, we have to go to zero temperature. In \equref{DefOfTopHam}, and similarly in the following expressions, $i\omega = 0$ has to be understood as the limit $T\rightarrow 0$ of the function evaluated at the Matsubara frequency $\omega_0=\pi/\beta$ (or equally well $\omega_{-1} =-\pi /\beta$). For this purpose, we assume that no additional topological phase transition occurs in the temperature range between the onset of superconductivity and $T=0$. In more mathematical terms, it means that, upon lowing $T$ to zero, the structure of the solution of the (nonlinear) Eliashberg equations does not change in a way that affects the topological invariant. Under this assumption, the topological properties of the superconducting phase can be inferred from the solution of the linear Eliashberg equations (\ref{LinearizedEliashbergEquations}).

Due to $\widetilde{\Phi}_s(0,\Omega) > 0$, the superconductor is fully gapped and the Green's function $\mathcal{G}_s(i\omega,\vec{k})$ must be an analytic function of $\omega$ in a finite domain containing the imaginary axis. Consequently, $\left.i\omega Z_s(i\omega,\vec{k})\right|_{i\omega=0} = 0$ as $i\omega Z_s(i\omega,\vec{k})$ is an odd function of $\omega$ [cf.~\equref{PropertiesOfZ}]. 
Therefore, the topological Hamiltonian becomes 
\begin{equation}
 h^t_{\vec{k}s} = \begin{pmatrix} \epsilon_{\vec{k}s} & \widetilde{\Phi}_s(i\omega=0,\vec{k})e^{-i\varphi_{\vec{k}}^s} \\ \widetilde{\Phi}_s(i\omega=0,\vec{k})e^{i\varphi_{\vec{k}}^s} & -\epsilon_{\vec{k}s} \end{pmatrix}, \label{ExpressionForTopHam}
\end{equation} 
which is manifestly Hermitian. Furthermore, it is readily checked to be particle-hole and time-reversal symmetric with $\Xi$ and $\widetilde{\Theta}_{\vec{k}s}$ as given in \equsref{ChargeConjugation}{TRSOfGreesFunc}.

The resulting topological properties are most easily inferred by reading the approximation of the general mean-field Hamiltonian in \equref{MeanFieldHamiltonian} to the weak-pairing description (\ref{BdGHamiltonian}) in reverse:
Comparison of $\mathcal{G}^t$ in \equref{DefinitionTopHamGF} and \equref{MeanFieldGreensFunction} shows that $h^t_{\vec{k}s}$ can be seen as the weak-pairing approximation of some mean-field Hamiltonian of the form of \equref{MeanFieldHamiltonian} with the property
\begin{equation}
 \widetilde{\Delta}_{s}(\vec{k}) \equiv \braket{\psi_{\vec{k}s}|\Delta_{\vec{k}} T^\dagger|\psi_{\vec{k}s}} = \widetilde{\Phi}_s(i\omega=0,\vec{k}). \label{OrderParameterAssociation}
\end{equation} 

In \refcite{ZhangTopInv}, it has been shown that the topological class-DIII invariant $\nu$ of a mean-field Hamiltonian of the form of \equref{MeanFieldHamiltonian} is, within the weak-pairing limit, determined by the sign of the order-parameter matrix elements $\widetilde{\Delta}_{s}(\vec{k})$ on the different Fermi surfaces of the system. More explicitly, in the 3D case, it holds
\begin{equation}
 \nu = \frac{1}{2} \sum_{s} \sign\left(\widetilde{\Delta}_{s}(\vec{k}_s)\right) C_{1s}, \label{3DTopInvariant}
\end{equation} 
where $\vec{k}_s$ is an arbitrary point on and $C_{1s}$ denotes the first Chern number of the Fermi surface $s$ (for the definition of $C_{1s}$ we refer to \equref{DefinitionOfFSChernNumber}). 
Due to \equsref{PropertiesOfPhiTilde}{OrderParameterAssociation}, we find
\begin{equation}
 \nu = \frac{1}{2} \sum_{s} C_{1s} = 0, \label{Trivial3D}
\end{equation} 
i.e., a topologically trivial superconductor. In the second equality of \equref{Trivial3D}, we have used that the total Chern number of all Fermi surfaces vanishes \cite{ZhangTopInv}. 

In lower dimensions, the expression (\ref{3DTopInvariant}) for the topological invariant assumes the form
\begin{equation}
 \nu =  \prod_{s} \left[\sign\left(\widetilde{\Delta}_{s}(\vec{k}_s)\right)\right]^{m_s} \label{12DTopInvariant}
\end{equation} 
as has been in shown in \refcite{ZhangTopInv} by means of dimensional reduction. In \equref{12DTopInvariant}, $m_s$ denotes the number of TRIM enclosed by Fermi surface $s$ in case of a 2D system, whereas $m_s=1$ for a 1D superconductor. Again, we find, in both dimensions, a trivial superconductor ($\nu = 1$) resulting from $\sign(\widetilde{\Delta}_{s}(\vec{k}_s))=1$.

Taken together, superconductivity in noncentrosymmetric systems that arises due to electron-phonon coupling alone can neither break TRS nor any point symmetry of the system and must necessarily be topologically trivial. This has been derived under very general assumptions: The inversion-symmetry breaking is assumed to be sufficiently strong for the weak-pairing approximation to be valid ($E_{\text{so}} \gtrsim  T_c$). The Eliashberg approach is controlled in the limit of adiabatic ionic motion ($m/M \ll 1$), in principle, allowing for arbitrarily strong interactions $\mathcal{V}$. Also the analysis of topological invariants is performed beyond the mean-field level. 
We emphasize that, despite looking deceptively like a mean-field description, the topological Hamiltonian approach we use is equivalent\cite{TopHam1,WangZhangTopSC} to the expressions for the topological invariants involving frequency integrals of the full Green's functions (see, e.g, \refcite{WangZhangGreens}). Thus, also interaction effects without static mean-field counterpart are captured. This is important as dynamical fluctuations can indeed change the topological properties of the system as has been demonstrated in \refcite{PoleExpansion}. 

Note that these conclusions are not altered when electronic renormalization effects of the phononic dispersion are taken into account since $\omega_{\vec{q}l}$ in \equref{PhononicHamiltonian} can already be regarded as the fully renormalized spectrum. In \secref{UnconventionalPairing} we will show, using exact relations derived from the spectral representation, that the same holds even if frequency-dependent corrections to the phonon propagator are considered. 

It is instructive to compare this result valid for noncentrosymmetric systems with the situation where inversion symmetry is preserved. In this case, all Fermi surfaces are doubly degenerate and the superconducting state can only be either singlet or triplet. As has been shown in \refcite{Brydon} using mean-field theory, singlet and triplet states will be degenerate if the electron-phonon coupling satisfies certain symmetries such that already an infinitesimal amount of residual Coulomb repulsion favors the triplet state that breaks inversion symmetry and has nontrivial topological properties\cite{FuBerg}. 
In our case, there are two main differences: Firstly, the absence of inversion symmetry generally mixes singlet and triplet components. Secondly, e.g., for a Fermi surface enclosing the $\Gamma$-point, breaking point symmetries necessarily implies the presence of nodes.  
   
In the remainder of the paper, we will discuss several generalizations of the considerations presented above including coupling to generic collective bosonic modes (\secref{UnconventionalPairing}), residual Coulomb interactions as well as disorder (\secref{TopSupPhon}) rendering topologically nontrivial properties possible in the weak-pairing limit.

\section{Unconventional superconductivity}
\label{UnconventionalPairing}
\begin{figure}[tb]
\begin{center}
\includegraphics[width=0.75\linewidth]{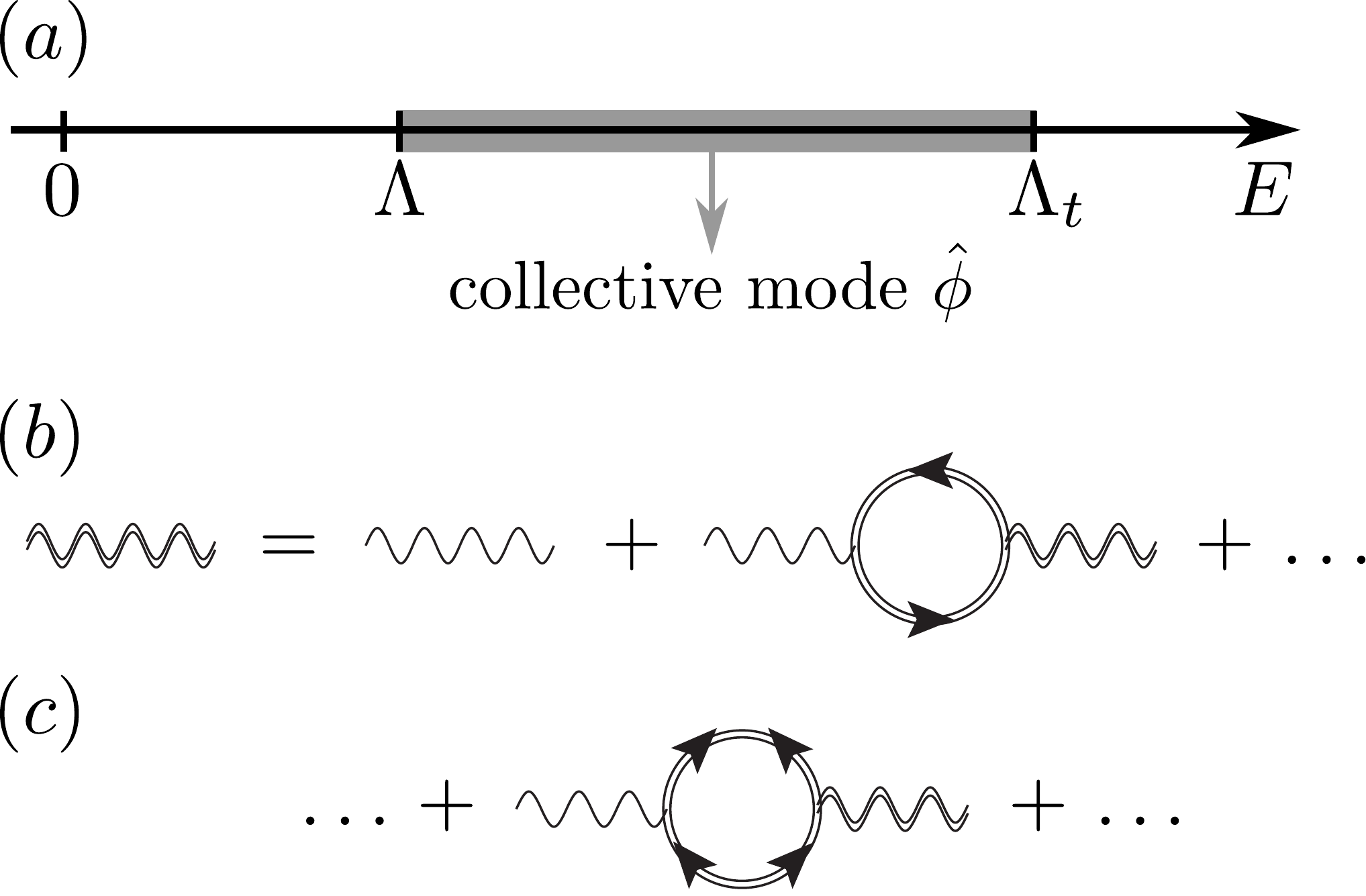}
\caption{In (a), the energy scales of the effective low-energy approach we use to describe unconventional pairing are illustrated. Part (b) and (c) show self-energy corrections of the bosonic propagator due to normal conducting electrons and the superconducting order parameter, respectively. The latter type of corrections are at least of quadratic order in $\Phi$.}
\label{EnergyScales}
\end{center}
\end{figure}
In this section, we will extend the analysis to unconventional superconductors, i.e., systems where superconductivity is not based on electron-phonon interaction but arises from a purely electronic mechanism. Eventually, it is the Coulomb interaction which, strongly renormalized depending on the microscopic details of the system, gives rise to the superconducting instability. 
Here we will fully neglect the electron-phonon interaction and treat the interacting electron problem in the following low-energy approach \cite{JoergChubukovRev}: We are not interested in the behavior of the system at high energies, e.g., of the order of the bandwidth $\Lambda_t$, but only focus on the physics for energies smaller than some cutoff $\Lambda < \Lambda_t$. 
As shown schematically in \figref{EnergyScales}(a), it is assumed that processes at energies between $\Lambda$ and $\Lambda_t$ drives the system close to some instability that we describe by the collective real ($\hat{\phi}_{\vec{q}j}^\dagger = \hat{\phi}_{-\vec{q}j}^\pdagger$) bosonic mode $\hat{\phi}_{\vec{q}j}$, $j=1,2,\dots , N_B$. For simplicity, we will first assume that the associated order parameter is either even ($t=+$) or odd ($t=-$) under time-reversal, which means mathematically
\begin{equation}
 \hat{\Theta} \hat{\phi}_{\vec{q}j} \hat{\Theta}^\dagger = t \, \hat{\phi}_{-\vec{q}j}. \label{PhiTRS}
\end{equation} 
The proximity to, e.g., (real) charge-density or spin-density wave order correspond to time-reversal even (TRE), $t=+1$, or time-reversal odd (TRO), $t=-1$, fluctuations, respectively.

Furthermore, we assume that the interaction processes at energies larger than $\Lambda$ neither destroy the Fermi-liquid behavior of the fermions nor lead to TRS breaking. Consequently, the noninteracting part of the fermionic Hamiltonian is still of the form of \equref{NonInteractingHam} satisfying \equref{HTRS}. The fermions are coupled to the bosons via
\begin{equation}
 \hat{H}_{\text{int}} = \sum_{\vec{k},\vec{q}} \hat{c}^\dagger_{\vec{k}+\vec{q}\alpha} \lambda^{(j)}_{\alpha\beta} \hat{c}^\pdagger_{\vec{k}\beta} \, \hat{\phi}_{\vec{q}j} \label{BFInteraction}
\end{equation} 
and all other residual electron-electron interactions will be neglected since, by assumption, the channel described by the collective mode $\hat{\phi}_{\vec{q}j}$ is dominant. In \equref{BFInteraction}, the matrices $\{ \lambda^{(j)} \}$ have to be Hermitian and satisfy
\begin{equation}
 \Theta \lambda^{(j)} \Theta^\dagger = t\, \lambda^{(j)} \label{TRSlambda}
\end{equation} 
resulting from $\hat{\phi}$ being real and \equref{PhiTRS}, respectively. In case of a model with only a single orbital where $\alpha$ just refers to the spin of the electrons, one could have $\{ \lambda^{(j)} \}=\{\sigma_0\}$, $N_B =1$, in case of $t=+$ and $\{ \lambda^{(j)} \}=\{\sigma_1,\sigma_2,\sigma_3\}$, $N_B=3$, for $t=-$. Here $\sigma_j$, $j=0,1,2,3$, denote the Pauli matrices in spin space. For simplicity of the presentation of the results, the discussion of more complex fermion-boson couplings will be postponed to \secref{Generalizations}.

The dynamics of the bosons will be described by the action
\begin{equation}
 S_{\text{col}} = \frac{1}{2} \int_q \phi_{qj} \left(\chi_0^{-1}(i\Omega_n,\vec{q})\right)_{jj'} \phi_{-qj'},
\end{equation} 
where $\phi$ is the field variable corresponding to the operator $\hat{\phi}$ and $\chi_0(i\Omega_n,\vec{q})$ is the bare susceptibility with respect to the order parameter of the competing particle-hole instability the system is close to. The full susceptibility $\chi(i\Omega_n,\vec{q})$, renormalized by particle-hole fluctuations as shown in \figref{EnergyScales}(b), is more important since it is experimentally accessible, e.g., via neutron scattering or NMR relaxation rate\cite{JoergChubukovRev}, and because it will enter the superconducting self-consistency equations discussed in \secref{SuperconductingInstab}.

As shown in \appref{SpectralRepSymOrderParameterSusc}, $\chi$ has to satisfy (the same holds for $\chi_0$) the exact relations
\begin{subequations}
\begin{align}
  \chi(i\Omega_n,\vec{q}) &= \chi^T(-i\Omega_n,-\vec{q}), \label{chiPropI} \\
  \chi(i\Omega_n,\vec{q}) &= \chi^\dagger(-i\Omega_n,\vec{q}), \label{chiPropII} \\
  \chi(i\Omega_n,\vec{q}) &= \chi(-i\Omega_n,\vec{q}). \label{chiPropIII}
\end{align}\label{chiProp}\end{subequations}
The first identity is just a consequence of $\chi$ being a correlator of twice the same operator $\hat{\phi}$ evaluated at $\vec{q}$ and $-\vec{q}$, whereas the second line is based on Hermiticity, $\hat{\phi}_{\vec{q}j}^\dagger = \hat{\phi}_{-\vec{q}j}^\pdagger$. Finally, the third relation follows from TRS of the system.

Being Hermitian, $\chi(i\Omega_n,\vec{q})$ has real eigenvalues all of which have to be positive as required by stability: By assumption, the competing instability will not occur and, hence, the bosons have to have a finite mass. 

In the following, we will proceed in a manner very similar to \secref{ElectronPhononCoupling}: Writing the entire model in the field integral representation and integrating out the bosons leads to an effective electron-electron interaction of the form of \equref{InteractionInEigenbasis} with
\begin{align}
 \begin{split}
&V^{s_1s_2}_{s_3s_4}(k_1,k_2,q) \\ & \,\, = -\frac{1}{2} \vec{\Lambda}^T_{s_2s_3}(\vec{k}_2-\vec{q},\vec{k}_2) \chi(q) \vec{\Lambda}_{s_1s_4}(\vec{k}_1+\vec{q},\vec{k}_1), \label{InteractionKernel} \end{split}
\end{align} 
where we have introduced the $N_B$-component vector of matrix elements
\begin{equation}
 \vec{\Lambda}_{ss'}(\vec{k},\vec{k}') = \psi_{\vec{k}s}^\dagger \vec{\lambda} \psi_{\vec{k}'s'}^{\phantom{*}}
\end{equation} 
in analogy to $G$ in \equref{DefinitionOfG}. Note that we use $\chi$ instead of the bare $\chi_0$ such that the interaction vertex in \equref{InteractionKernel} is already fully renormalized by particle-hole fluctuations. Due to \equsref{TRSOfESs}{TRSlambda}, TRS of the system implies
\begin{equation}
 \vec{\Lambda}_{ss'}(\vec{k},\vec{k}') = t \, e^{i(\varphi_{\vec{k}'}^{s'}-\varphi_{\vec{k}}^s)} \left(\vec{\Lambda}_{s^{\phantom{.}}_\K s'_\K}(-\vec{k},-\vec{k}')\right)^* \label{TRSOfLambda},
\end{equation} 
which constitutes the obvious generalization of \equref{TRSOfG} including not only TRE (such as phonons) but also TRO fluctuations.

We have seen in \secref{EliashbergTheory} that, in the weak-pairing approximation, the superconducting properties are fully determined by the Cooper and the forward scattering channel shown in \figref{DiagramsOfEliashberg}(a) and (b). As before, we still find that these two interaction channels are determined by the same matrix elements,
\begin{subequations}\begin{align}
 V^{s's'_\K}_{s_\K s} (k,-k,k'-k) &= t \, e^{i(\varphi_{\vec{k}}^s-\varphi_{\vec{k}'}^{s'})} \mathcal{V}_{s's}(k';k), \label{UnconvMatrixElementsC} \\ 
\mathcal{F}_{s's}(k',k) &= \mathcal{V}_{s's}(k';k) \label{SymmetryBetweenFCScattering}
\end{align}\label{UnconvMatrixElements}\end{subequations} 
with
\begin{equation}
 \mathcal{V}_{s's}(k';k) = -\frac{1}{2} \vec{\Lambda}^\dagger_{s's}(\vec{k}',\vec{k}) \chi(k'-k) \vec{\Lambda}^\pdagger_{s's}(\vec{k}',\vec{k}). \label{VUnconventional}
\end{equation} 
To show this, TRS (\ref{TRSOfLambda}) and Hermiticity, $\vec{\lambda}^\dagger = \vec{\lambda}$, have been taken advantage of. 

Recalling that stability forces $\chi(q)$ to be positive definite, we conclude that $\mathcal{V}_{s's} < 0$.
We see that the forward scattering amplitude $\mathcal{F}$ is, exactly as in case of phonons, negative for all states on the Fermi surfaces, whereas the global sign of the Cooper channel is reversed in case of TRO fluctuations as compared to phonons (or TRE electronic fluctuations). Note that, in general, this only holds for the renormalized electron-electron interaction since only $\chi$ and not the bare $\chi_0$ has to be positive definite.

\subsection{Superconducting instability}
\label{SuperconductingInstab}
Let us next analyze the consequences for the possible superconducting phases. As before, we apply Eliashberg theory that is frequently used for studying superconductivity caused by collective bosonic modes other than phonons (see, e.g., \refcite{ChubukovJoergEliashberg} and references therein). As in this case, $m/M \ll 1$ will not hold in general, one expects this approach only to be applicable in the weak-coupling regime. However, in the limit of large numbers of fermion flavors, neglecting vertex corrections is also justified in the strong-coupling case\cite{ChubukovJoergEliashberg}. While there are complications with this large-$N$ theory in 2D\cite{SSLee}, some efforts have been made to develop controlled approaches for this case as well\cite{MrossSenthil}.

Using Hermiticity of $\vec{\lambda}$, \equsref{TRSOfLambda}{chiPropIII}, it is straightforward to check that the three properties (\ref{VProp}) of the vertex function $\mathcal{V}$ are still satisfied. Consequently, the linearized Eliashberg equations are again of the form of \equref{LinearizedEliashbergEquations} with $\mathcal{V}_{s's}(k';k)$ now given by \equref{VUnconventional} and an additional prefactor of $t$ on the right-hand side of the gap equation (\ref{EliashbergGap}), i.e., $v$ is replaced by $tv$. Note that the renormalized propagator $\chi$ is taken into account which, diagrammatically, corresponds to replacing the bare bosonic line in \figref{DiagramsOfEliashberg}(c) by the full line (see \figref{EnergyScales}(b)). We emphasize that, for the linearized Eliashberg equations, there are no anomalous propagators entering the full bosonic line: Any term in the bosonic self-energy involving the anomalous self-energy $\Phi$, such as the one shown in \figref{EnergyScales}(c), is at least of quadratic 
order in $\Phi$ and, hence, does not contribute. Therefore, we can safely use the TRS constraint (\ref{chiPropIII}) near the transition without a priori knowledge about the time-reversal properties of the superconducting condensate.

Repeating the arguments presented in \secref{EliashbergTheory}, we directly conclude that \equref{PropertiesOfZ} is still valid.
The kernel $tv$ of the gap equation is symmetric [\cf\equref{VPropI}], real and, hence, diagonalizable.
As shown in \appref{UnconvLargeT}, the leading superconducting instability is again determined by its largest eigenvalue.

To begin with \textit{TRE fluctuations}, $t=+$, the kernel has, exactly as $\mathcal{V}$ in \equref{VUnconventional}, only positive components, such that the Perron-Frobenius theorem can be applied. It follows that the resulting superconducting order parameter satisfies $\delta > 0$ and, thus, preserves TRS and has no sign changes, neither on a given Fermi surface nor between different Fermi surfaces. It must transform under the trivial representation of the point group. Again \equref{PropertiesOfPhiTilde} is satisfied and, according to our analysis of \secref{TopologicalProperties}, the associated state is topologically trivial -- exactly as in the case of electron-phonon coupling. 
 
For \textit{TRO fluctuations}, we have $t=-$ such that $\delta$ now belongs to the eigenspace of $v$ with the smallest eigenvalue. This has two crucial consequences. Firstly, we cannot generically exclude spontaneous TRS breaking since it is no longer guaranteed that this eigenspace is one-dimensional. Although all eigenvectors of the real matrix $v$ can always be chosen to be real valued, the superconducting order parameter can be a complex superposition of the degenerate eigenvectors which makes TRS breaking possible. Note that, apart from accidental degeneracies which we will neglect here, these degeneracies can be enforced by symmetry if the point group of the system allows for multidmensional or complex irreducible representations\cite{DesignPrinciples,*DesignPrinciplesUnpub,Lax}.  
Secondly, the eigenvectors with minimal eigenvalue can have many sign changes which, depending on the form of the Fermi surfaces, can break any point symmetry of the system and lead to nodal points. 

To proceed, we will assume that the resulting superconducting state preserves TRS and, thus, belongs to class DIII. This is not very restrictive since it has been shown\cite{DesignPrinciples,*DesignPrinciplesUnpub} that, for 2D systems, spontaneous TRS breaking can only occur in the weak-pairing limit if there is a threefold rotation symmetry perpendicular to the plane.
Furthermore, as we want to discuss topological properties of superconductors, we will focus on fully gapped systems, where the sign changes take place \textit{between} different Fermi surfaces. Being fully gapped, we can apply the weak-pairing expressions in \equsref{3DTopInvariant}{12DTopInvariant} for the DIII invariant to the topological Hamiltonian (\ref{ExpressionForTopHam}).

To derive necessary conditions for the emergence of nontrivial topological invariants, we will next discuss an approximate symmetry that is expected to be applicable to many noncentrosymmetric systems. It will give rise to an asymptotic symmetry of the gap equation (\ref{EliashbergGap}) and, in turn, constrain the possible superconducting order parameters and associated topological properties.

\subsection{Asymptotic symmetry and necessary conditions for topological superconductivity}
\label{ApproximateSymmetry}
\begin{figure}[tb]
\begin{center}
\includegraphics[width=0.7\linewidth]{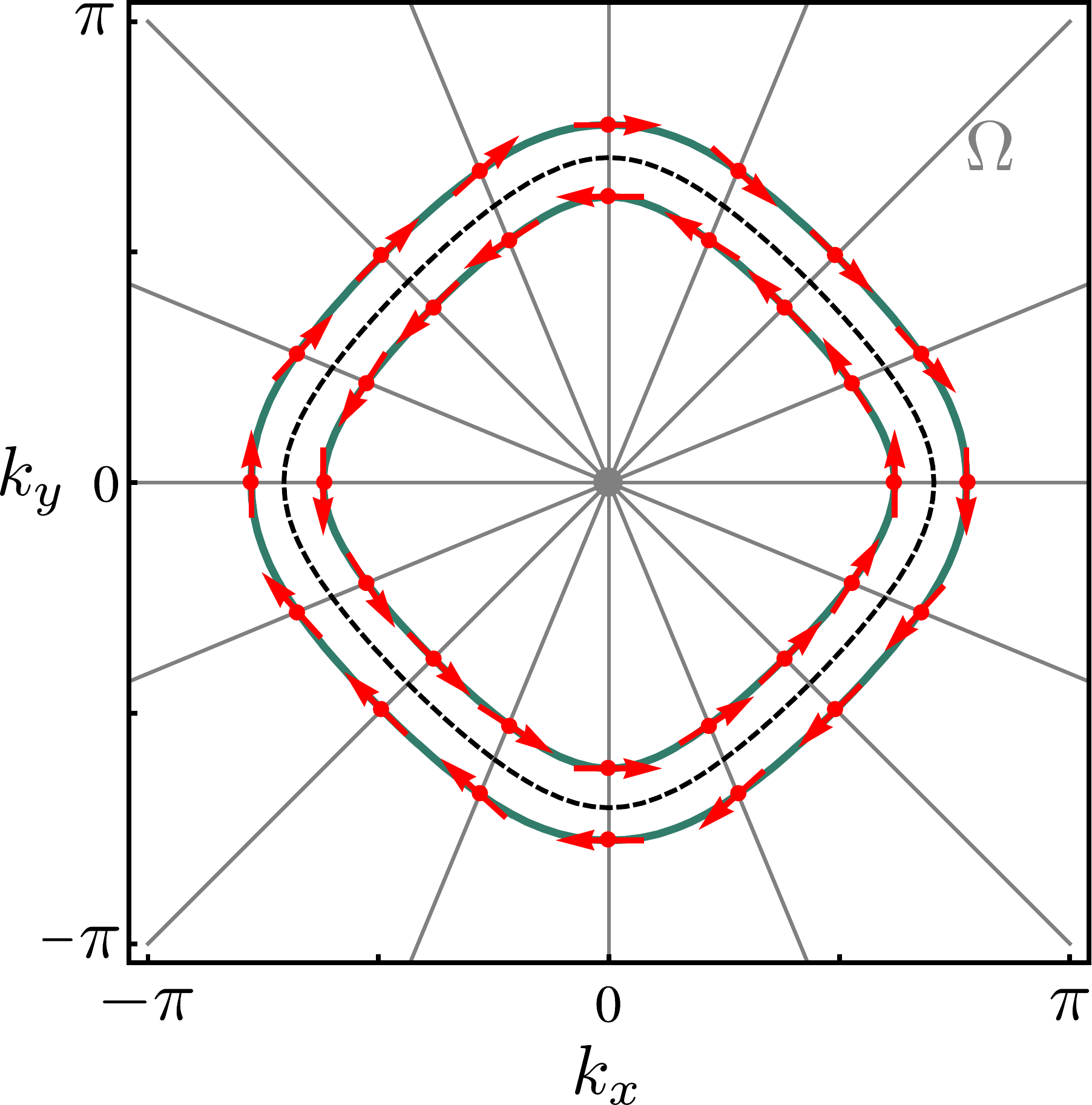}
\caption{(Color online) Fermi surfaces (solid green lines) and pseudospin orientation (red arrows) of the Rashba model defined in the main text using $\alpha=0.25t$, $\mu=-0.4 t$. The black dashed line is the doubly degenerate Fermi surface for $\alpha=0$.  In agreement with \equref{ApproximateSymmetryEigenstates}, the pseudospin orientation is approximately antiparallel on the two Fermi surfaces for states with the same polar angle $\Omega$.}
\label{WeakSpinSplitting}
\end{center}
\end{figure}

To deduce this approximate symmetry which relates the wavefunctions at different spin-orbit split Fermi surfaces and discuss the limit where it becomes exact, let us split the quadratic Hamiltonian (\ref{NonInteractingHam}) of the fermions according to
\begin{equation}
 h_{\vec{k}}^\pdagger= h^{\text{S}}_{\vec{k}} + h^{\text{A}}_{\vec{k}} \label{DecompH}
\end{equation} 
with a term $h^{\text{S}}$ that is symmetric and a term $h^{\text{A}}$ that is antisymmetric under inversion. We first diagonalize the centrosymmetric part of the Hamiltonian. The corresponding eigenvalues $\epsilon^{\text{S}}_{\vec{k}j}$, $j=1,2,...,N$, must be doubly degenerate due to the combination of inversion and TRS. Note that $h^{\text{S}}$ in general also includes spin-orbit coupling, which entangles the spin and orbital degrees of freedom of the electrons. Nonetheless, as is easily seen by construction, one can still introduce a $\vec{k}$-space local pseudospin basis $\{\ket{\vec{k},\sigma}\}$ which has the same transformation properties under TRS and inversion as the physical spin. Denoting the Pauli matrices in this basis by $s_i$, $i=0,1,2,3$, we have
\begin{equation}
 \braket{j,\sigma|h^{\text{A}}_{\vec{k}}|j',\sigma'} = \delta_{j,j'} \, \vec{g}^{(j)}_{\vec{k}} \cdot \vec{s}_{\sigma \sigma'}, \label{FormOfhA}
\end{equation} 
where $\vec{g}_{\vec{k}} = -\vec{g}_{-\vec{k}}$ and no term $\propto s_0$ can be present as dictated by TRS and $h^{\text{A}}$ being odd under inversion. In \equref{FormOfhA}, we have neglected all matrix elements between different $j$ which is justified as long as the energetic separation between the different bands $\epsilon^{\text{S}}_{\vec{k}j}$ is much larger than $|\vec{g}|$. A finite $\vec{g}$ can only arise if inversion symmetry is broken, e.g., at the surface of a system, at an interface between two different materials or in the bulk of a noncentrosymmetric crystal. It will lift the degeneracy of the bands $\epsilon^{\text{S}}_{\vec{k}j}$ as is illustrated by the 2D example shown in \figref{WeakSpinSplitting}, where $N=1$, $\epsilon^{\text{S}}_{\vec{k}} = -t(\cos k_x + \cos k_y)-\mu$ and the standard Rashba spin-orbit coupling, $\vec{g}_{\vec{k}} = \alpha(-\sin k_y,\sin k_x,0)^T$, have been assumed: The doubly degenerate Fermi surface (dashed black line) associated with $\epsilon^{\text{S}}$ is 
split 
into two (solid green lines).

Due to the decomposition of $h_{\vec{k}}$ in \equref{DecompH}, its eigenstates $\psi_{\vec{k}s}$ satisfying 
\begin{equation}
   \vec{g}^{(j)}_{\vec{k}} \hspace{-0.3em} \cdot \vec{s} \, \, \psi_{\vec{k}s} = \nu  \left|\vec{g}^{(j)}_{\vec{k}}\right| \psi_{\vec{k}s}  ,\quad  \nu=\pm, \label{EigenValueEqu}
\end{equation} 
and eigenvalues $\epsilon_{\vec{k}s} = \epsilon^{\text{S}}_{\vec{k}j} +  \nu |\vec{g}^{(j)}_{\vec{k}}|$ can now be labeled by the composite index $s=(j,\nu)$. If $\vec{g}^{(j)}_{\vec{k}}$ varies slowly on the separation $|\vec{g}|/v_F$ ($v_F$ denotes the Fermi velocity) of the Fermi surfaces $s=(j,\nu)$ and $s_\R\equiv(j,-\nu)$, we can approximate $\vec{g}^{(j)}_{\vec{k}}\approx \vec{g}^{(j)}(\Omega)$ and, hence,  $\psi_{\vec{k}s} \approx \psi_{\Omega s}$ in \equref{EigenValueEqu}.
As $\Theta \vec{s} \Theta^\dagger = - \vec{s}$, we obtain the asymptotic symmetry
\begin{equation}
  \psi_{\Omega s} \sim e^{i\gamma^s_\Omega} \Theta \psi_{\Omega s_\R} \label{ApproximateSymmetryEigenstates}
\end{equation} 
that becomes exact in the limit $\vec{g}^{(j)} \rightarrow 0$. Here $\gamma^s_\Omega$ are phase factors that depend on the choice of the eigenstates. 
Note that this relation is structurally similar to that based on TRS in \equref{TRSOfESs}: Both are $\vec{k}$-nonlocal antiunitary symmetries. TRS conntects a state at $\vec{k}$ and its Kramers partner at $-\vec{k}$, whereas \equref{ApproximateSymmetryEigenstates} relates wavefunctions of necessarily different Fermi surfaces -- the state $(s,\Omega)$ and its ``Rashba partner'' $(s_\R,\Omega)$.

Physically, \equref{ApproximateSymmetryEigenstates} means that, for given $\Omega$, the pseudospin orientation of the wavefunctions on the ``Rashba pair'' of Fermi surfaces $\{s,s_\R\}$ is antiparallel. As can be seen in \figref{WeakSpinSplitting}, where $\Omega$ is chosen to be the polar angle of $\vec{k}$, \equref{ApproximateSymmetryEigenstates} represents a very good approximation even for the moderately large value of the spin-orbit coupling used in the plot. 
Naturally, the validity of \equref{ApproximateSymmetryEigenstates} for discussing superconducting properties crucially depends on the bandstructure of the system. Typically, one expects $\vec{g}_{\vec{k}}$ to vary on momentum scales of order of the size the Brillouin zone, i.e., \equref{ApproximateSymmetryEigenstates} to be valid for $E_{\text{so}} \ll \Lambda_t$.

In the following, we will assume that \equref{ApproximateSymmetryEigenstates} holds and analyze its consequences. Firstly, taking advantage of the aforementioned similarity to TRS, we obtain
\begin{equation}
 \vec{\Lambda}_{ss'}(\Omega,\Omega') = t \, e^{i(\gamma_{\Omega'}^{s'}-\gamma_{\Omega}^s)} \left(\vec{\Lambda}_{s^{\phantom{.}}_\R s'_\R}(\Omega,\Omega')\right)^*. \label{RashbaRelation}
\end{equation} 
As a second step, this, together with \equref{chiProp}, allows for rewriting the central interaction matrix element (\ref{VUnconventional}) as follows
\begin{align}
 & \mathcal{V}_{s's}(i\omega_{n'},\Omega';i\omega_n,\Omega)  \\ & = -\frac{1}{2} \vec{\Lambda}^\dagger_{s'_\R s^{\phantom{.}}_\R}(\Omega',\Omega) \chi(i\omega_{n'}-i\omega_n,\vec{k}-\vec{k}') \vec{\Lambda}_{s'_\R s^{\phantom{.}}_\R}(\Omega',\Omega), \nonumber
\end{align} 
already using the approximation and notation introduced in \equref{SeparationApprox}.
The right-hand side of this equation only equals $\mathcal{V}_{s'_\R s^{\phantom{.}}_\R}(i\omega_{n'},\Omega';i\omega_n,\Omega)$ and leads to the symmetry
\begin{equation}
  \mathcal{V}_{s's}(i\omega_{n'},\Omega';i\omega_n,\Omega)  = \mathcal{V}_{s'_\R s^{\phantom{.}}_\R}(i\omega_{n'},\Omega';i\omega_n,\Omega) \label{AsympPropertyOfV}
\end{equation} 
under two assumptions: Firstly, similar to $\vec{g}$, the susceptibility $\chi(i\Omega_n,\vec{q})$ must be slowly varying in $\vec{q}$ on the scale $|\vec{g}|/v_F$. This is a very natural assumption as the description in terms of a collective mode will only be sensible if there are long-wavelength fluctuations. 
Secondly, and much more importantly, $\chi(i\Omega_n,\vec{q})$ must be an even function of momentum $\vec{q}$. We see from \equref{chiProp} that TRS alone does not determine the behavior under $\vec{q} \rightarrow -\vec{q}$ such that further information about the system is required. For this purpose, let us assume that there is a symmetry relating the fermionic momenta $\vec{k}$ and $-\vec{k}$, i.e., the full Hamiltonian of the system commutes with the unitary operator $\hat{\mathcal{S}}$ defined via
\begin{equation}
 \hat{\mathcal{S}} \hat{c}^\dagger_{\vec{k}\alpha} \hat{\mathcal{S}}^\dagger = \hat{c}^\dagger_{-\vec{k}\beta} S_{\beta\alpha}, \quad S^\dagger S = \mathbbm{1}. \label{SSym}
\end{equation} 
In a 2D system this symmetry can be realized as a two-fold rotation $C_2^z$ perpendicular to the plane of the system. Since $C_2^z$ commutes with all other symmetry operations, the irreducible representations of the point group must be either even or odd under this operation (see \appref{TwoFoldRotation}) and the same holds for the order parameter of the competing particle-hole instability. This means that
\begin{equation}
 \hat{\mathcal{S}} \hat{\phi}_{\vec{q}j} \hat{\mathcal{S}}^\dagger = \pm \hat{\phi}_{-\vec{q}j} \label{PhiIrredRep}
\end{equation} 
leading to the required relation
\begin{equation}
 \chi(i\Omega_n,\vec{q}) = \chi(i\Omega_n,-\vec{q}) \label{EvenInq}
\end{equation} 
as shown in \appref{SpectralRepSymOrderParameterSusc}. One might expect that, even in the absence of a two-fold rotation symmetry or in 1D and 3D, \equref{EvenInq} still constitutes a valid approximation for $E_{\text{so}} \ll \Lambda_t$: Although the superconducting instability, arising from infrared singularities, is essentially influenced by the splitting $|\vec{g}|$ of the Fermi surfaces due to the broken inversion symmetry, the susceptibility $\chi$ might not. As the tendency of the system towards the competing instability mainly results from processes at energies comparable to $\Lambda_t$ [\cf\figref{EnergyScales}(a)], the inversion-symmetry-breaking terms are expected to be negligible in the limit $E_{\text{so}} \ll \Lambda_t$ for the calculation of $\chi$. In that sense, \equref{SSym} is realized as an approximate inversion symmetry again yielding \equref{EvenInq}.

Let us now deduce the implications of the resulting asymptotic property (\ref{AsympPropertyOfV}) of the interaction matrix. Due to $\rho_{s} \sim \rho_{s_\R}$ in the limit considered, we have $Z_s \sim Z_{s_\R}$ as can be seen directly from \equref{EliashbergQPResidue}. Therefore, the kernel $v$ of the gap equation (\ref{EliashbergGap}) has the same symmetry as $\mathcal{V}$ in \equref{AsympPropertyOfV}. We conclude that the anomalous self-energy of any resulting superconducting state must be of the form 
\begin{equation}
 \widetilde{\Phi}_s(i\omega_n,\Omega) = p \, \widetilde{\Phi}_{s_\R}(i\omega_n,\Omega) \label{RashbaParity}
\end{equation} 
with either $p=+$ or $p=-$ for all $s$, $\omega_n$ and $\Omega$. This is a central result of this section. It highly constraints the possible order parameters and allows them to be grouped into two basic classes: The relative sign of the order parameter at Rashba partners can only be either positive ($p=+$) or negative ($p=-$) for \textit{all} Rashba pairs. In the following, the corresponding pairing states will be denoted by ``Rashba even'' and ``Rashba odd'', respectively.

This has also crucial consequences for the possible topological properties of the superconductor. To discuss this, we will, as already mentioned above, have to assume that the superconducting state is fully gapped and time-reversal symmetric. 
Note that \equref{RashbaParity} will also hold if these two additional assumptions are not satisfied.

Let us first focus on the 2D case where the topological invariant is determined by \equref{12DTopInvariant} with $\widetilde{\Delta}_s(\vec{k}) = \widetilde{\Phi}_s(0,\Omega_{\vec{k}})$. 
As long as the Rashba splitting $|\vec{g}|/v_F$ is much smaller than the size the Brillouin zone, we can assume $m_{(j,\nu)} = m_{(j,-\nu)} \equiv m_j$. It follows that any Rashba even state will be topologically trivial and, in case of Rashba odd pairing, the invariant is given by
\begin{equation}
 \nu = \prod_{j=1}^{N} (-1)^{m_j} = (-1)^{\sum_j m_j}. \label{2DSimplifiedInv}
\end{equation} 
The same also holds in 1D, where $m_j = 1$. Consequently, the total number of TRIM enclosed by Rashba pairs of Fermi surfaces must be necessarily odd for the interaction-induced superconductor to be topological.

To continue with 3D, we first note that the Fermi surface Chern numbers of Rashba partners must be equal in magnitude but opposite in sign, $C_{1(j,+)} = -C_{1(j,-)}\equiv C_{1j}$, which readily follows from \equref{ApproximateSymmetryEigenstates} as shown in \appref{ChernNumberRashba}. From \equref{3DTopInvariant} we then immediately see that any Rashba even state must again be trivial. For Rashba odd pairing, we get
\begin{equation}
 \nu = \sum_{j} C_{1j} \sign\left(\widetilde{\Phi}_{(j,+)}(0,\Omega)\right). \label{3DSimplifiedInv}
\end{equation} 
Note that the right-hand side does not depend on $\Omega$ as the sign of the order parameter of a fully gapped superconductor cannot change on a given Fermi surface.

\begin{figure}[tb]
\begin{center}
\includegraphics[width=\linewidth]{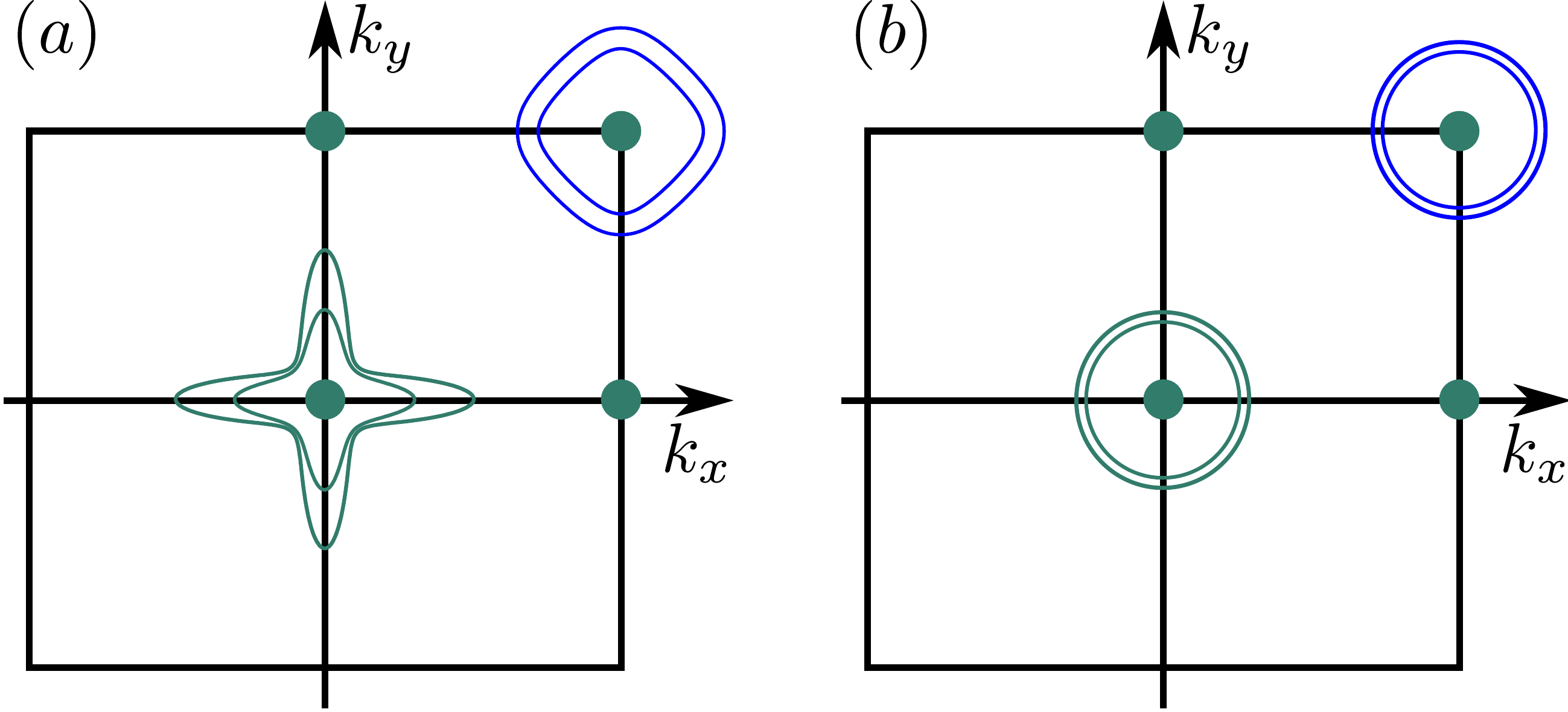}
\caption{(Color online) Illustration of nesting configurations in the simplest nontrivial case of $N=2$ Rashba pairs of Fermi surfaces. In (a), we expect $\chi$ to be peaked at momenta connecting states within the two Rashba pairs leading to Rashba odd pairing. The resulting state will nonetheless be topologically trivial as the number of TRIM enclosed by Rashba pairs is even. In case of (b), a Rashba even state is expected.}
\label{NestingConfigurations}
\end{center}
\end{figure}

Irrespective of the dimensionality of the system, we have seen that Rashba odd pairing is required to make topologically nontrivial superconductivity possible. 
We expect this to be realized when the strongest nesting occurs between Rashba partners such that $\chi$ is dominated by momenta connecting the Rashba-split Fermi surfaces. An example for $N=2$ is shown in \figref{NestingConfigurations}(a) where a Rashba odd state is expected. On the other hand, if different Rashba pairs are most strongly nested, as in case of the Fermi surfaces of \figref{NestingConfigurations}(b), a topologically trivial Rashba even state will arise. 

More specifically, we can conclude that for just a single Rashba pair, $N=1$, only Rashba odd pairing is possible. This simply follows from the fact that, in case of TRO fluctuations discussed here, the interaction is fully repulsive in the Cooper channel such that the superconducting state must have at least one sign change. Focusing on fully gapped superconductors, this sign change must occur between the two Fermi surfaces. Irrespective of the dimensionality of the system, the superconductor will be automatically topological if the number of TRIM enclosed by the Rashba pair is odd. 
In 1D and 2D, this is directly seen from \equref{2DSimplifiedInv}, whereas, for the 3D case, the relation\cite{ZhangTopInv} $(-1)^{C_{1j}}=(-1)^{m_j}$ implying $C_{1j}\neq 0$ for odd $m_j$ has to be taken into account in \equref{3DSimplifiedInv}.

\subsection{More general couplings}
\label{Generalizations}
Here we will generalize the previous analysis by considering more general forms of the fermion-boson coupling (\ref{BFInteraction}).

\subsubsection{Momentum-dependent complex order parameter}
To allow for the most general particle-hole order parameter we now investigate the coupling Hamiltonian
\begin{equation}
 \hat{H}_{\text{int}} = \sum_{\vec{k},\vec{q}} \hat{c}^\dagger_{\vec{k}+\vec{q}\alpha} m^{(j)}_{\alpha\beta}(\vec{k}+\vec{q},\vec{k}) \hat{c}^\pdagger_{\vec{k}\beta} \, \hat{\varphi}_{\vec{q}j} + \text{H.c.}, \label{GeneralizedCoupling}
\end{equation} 
where $\hat{\varphi}_{\vec{q}j}$ are $N_B'$-component complex bosons ($\hat{\varphi}^\dagger_{\vec{q}j} \neq \hat{\varphi}_{-\vec{q}j}$) and $m^{(j)}(\vec{k}+\vec{q},\vec{k})$ are potentially momentum-dependent, generally non-Hermitian matrices. The momentum dependence is essential, e.g., when discussing current fluctuations, where $m(\vec{k}+\vec{q},\vec{k}) \propto  (\vec{k} + \vec{q}/2) \sigma_0$ with $\sigma_j$ being Pauli matrices in spin space. The generalization to non-Hermitian order parameters is relevant, e.g., in case of imaginary spin- ($m^{(j)} = i \sigma_j$) or charge-density waves ($m = i \sigma_0$).

By decomposing both $\hat{\varphi}_{\vec{q}j}$ and the fermion bilinear into their Hermitian and Antihermitian parts, one can reduce \equref{GeneralizedCoupling} to the coupling to real bosons $\hat{\phi}_{\vec{q}j}$ with $N_B = 2N_B'$ components:
\begin{equation}
 \hat{H}_{\text{int}} = \sum_{\vec{k},\vec{q}} \hat{c}^\dagger_{\vec{k}+\vec{q}\alpha} M^{(j)}_{\alpha\beta}(\vec{k}+\vec{q},\vec{k}) \hat{c}^\pdagger_{\vec{k}\beta} \, \hat{\phi}_{\vec{q}j}, \label{GeneralizedCoupling2}
\end{equation} 
where $(M^{(j)}(\vec{k},\vec{k}'))^\dagger = M^{(j)}(\vec{k}',\vec{k})$. Let us for the moment again focus on either TRE or TRO fluctuations forcing $T\vec{M}^*(-\vec{k},-\vec{k}') T^\dagger = t\, \vec{M}(\vec{k},\vec{k}')$. Below we will comment on the situation of having both components at the same time. 

Repeating the analysis presented above, one readily finds that $Z$ must still satisfy all three properties in \equref{PropertiesOfZ}. In case of TRE fluctuations, spontaneous TRS breaking cannot occur with the resulting superconducting state being necessarily topologically trivial.

To derive the property (\ref{RashbaParity}) which is central for our analysis of superconductivity induced by TRO fluctuations, \equref{RashbaRelation} with $\vec{\Lambda}_{ss'}(\vec{k},\vec{k}') = \psi_{\vec{k}s}^\dagger \vec{M}(\vec{k},\vec{k}') \psi_{\vec{k}'s'}^\pdagger$ must hold. Due to the additional momentum dependence of the order parameter, this is only the case (with $t$ replaced by $s\, t$ in \equref{RashbaRelation}) if $\vec{M}(\vec{k},\vec{k}')$ changes little on the separation $|\vec{g}|/v_F$ of Rashba partners and if 
\begin{equation}
 \vec{M}(\vec{k},\vec{k}') = s\, \vec{M}(-\vec{k},-\vec{k}'), \quad s = \pm 1. \label{EvenFunctionOfK}
\end{equation} 
Note that \equref{EvenFunctionOfK} is satisfied by all examples discussed above. However, it can be violated when, e.g., current fluctuations and spin-density wave fluctuations are simultaneously relevant.

Again assuming the presence of the unitary symmetry introduced in \equsref{SSym}{PhiIrredRep}, all constraints on the topological properties discussed in \secref{ApproximateSymmetry} also hold for momentum dependent, complex order parameters with coupling (\ref{GeneralizedCoupling}) as long as \equref{EvenFunctionOfK} is satisfied.

\subsubsection{Frequency-dependent fermion-boson vertex}
So far we have assumed that the fermion-boson interaction can be described by a Hamiltonian in the low-energy theory. If this interaction obtains significant frequency-dependent renormalization corrections resulting from processes at energies between $\Lambda_t$ and $\Lambda$ [see \figref{EnergyScales}(a)], an action description,
\begin{equation}
 S_{\text{int}} = \int_{k,q} \bar{c}_{k+q\alpha} \Gamma^{(j)}_{\alpha\beta}(k+q;k) c_{k\beta} \, \phi_{qj}, \label{VertexFunction}
\end{equation} 
is required. Here $\Gamma^{(j)}$ is the generally momentum- and frequency-dependent vertex function. Similar to our treatment of $\chi$, the vertex will not be explicitly specified in the following. We will only take into account the exact relations resulting from TRS and Hermiticity (see \appref{SpectralRepFermionBosonVertex}). To begin with the former symmetry, it holds
\begin{equation}
 \Gamma_{\alpha\beta}^{(j)}(k;k') = t \, T_{\alpha \alpha'} \left[\Gamma_{\alpha'\beta'}^{(j)}(-k;-k')\right]^* T^\dagger_{\beta' \beta}, \label{VertexTRS}
\end{equation} 
which reduces to \equref{TRSlambda} for the coupling (\ref{BFInteraction}). The full effective electron-electron vertex is the same as in \equref{InteractionKernel} with $\vec{\Lambda}_{s,s'}(\vec{k},\vec{k}')$ replaced by the renormalized $\vec{\Lambda}^{\Gamma}_{ss'}(k;k') = \psi_{\vec{k}s}^\dagger \vec{\Gamma}(k;k') \psi_{\vec{k}'s'}^\pdagger$
which now becomes frequency dependent. Due to the constraint (\ref{VertexTRS}), $\vec{\Lambda}^{\Gamma}$ satisfies the analogue of \equref{TRSOfLambda} such that we find the same structure as in \equref{UnconvMatrixElementsC} with $\mathcal{V}$ being positive: As before, the interaction in the Cooper channel is either fully attractive or fully repulsive depending on whether the system is coupled to TRE or TRO fluctuations. 

To analyze the forward-scattering channel defined in \figref{DiagramsOfEliashberg}(b), we need to take into account the Hermiticity relation
\begin{equation}
 \Gamma_{\alpha\beta}^{(j)}(i\omega_n,\vec{k};i\omega_{n'},\vec{k}') = \left[\Gamma_{\beta\alpha}^{(j)}(-i\omega_{n'},\vec{k}';-i\omega_n,\vec{k}) \right]^*, \label{VertexHermiticity}
\end{equation} 
which reduces to $\vec{\lambda}^\dagger = \vec{\lambda}$ for the coupling (\ref{BFInteraction}). One can show that, despite the sign change of the frequencies on the right-hand side of \equref{VertexHermiticity}, the resulting quasiparticle weight $Z$ still satisfies \equref{PropertiesOfZ}. Furthermore, we find that, again, no TRS breaking is possible and $\delta > 0$ for TRE fluctuations.

To discuss the case of $t=-$, we have to take into account the implications for the vertex function resulting from the asymptotic symmetry introduced in \secref{ApproximateSymmetry}. In  \appref{SpectralRepFermionBosonVertex} it is shown that, as long as \equref{EvenFunctionOfK} is satisfied for the bare fermion-boson vertex, this imposes the constraint
\begin{align}
 \begin{split}
&\vec{\Lambda}^{\Gamma}_{ss'}(i\omega_n,\Omega;i\omega_{n'},\Omega') \\ &\quad = s\, t \, e^{i(\gamma_{\Omega'}^{s'}-\gamma_{\Omega}^s)} \left[ \vec{\Lambda}^{\Gamma}_{s_\R^{\phantom{.}} s'_\R}(-i\omega_n,\Omega;-i\omega_{n'},\Omega') \right]^* \end{split} \label{ApproxSymConstr}
\end{align} 
on the fully renormalized vertex function, which constitutes the obvious generalization of relation (\ref{RashbaRelation}).

Using the constraints resulting from the invariance of the system under $\hat{\mathcal{S}}$ defined in \equsref{SSym}{PhiIrredRep} on the bosonic propagator, \equref{EvenInq}, as well as on the vertex function,
\begin{equation}
 \vec{\Lambda}^{\Gamma}_{ss'}(i\omega_n,\Omega;i\omega_{n'},\Omega') = \pm \vec{\Lambda}^{\Gamma}_{s^{\phantom{.}}_\K s'_\K}(i\omega_n,\Omega_\K^{\phantom{.}};i\omega_{n'},\Omega'_\K), \label{SSymOfVertexFunc}
\end{equation} 
we recover the Rashba symmetry (\ref{AsympPropertyOfV}) of the Cooper channel. Similarly, one can show that $Z_s = Z_{s_\R}$ still holds. Consequently, the possible superconducting states can again be classified into Rashba even and Rashba odd according to \equref{RashbaParity}. 
Assuming a finite gap and a TRS-preserving order parameter, we find exactly the same conclusions concerning the topology of the superconducting state as before.

\subsubsection{General time-reversal properties}
Finally, let us discuss the situation when the dominant fluctuations are neither fully TRE nor TRO, i.e., if the bosons coupling to the fermions according to \equref{GeneralizedCoupling2} satisfy
\begin{equation}
 \hat{\Theta} \hat{\phi}_{\vec{q}j} \hat{\Theta}^\dagger = t_j \, \hat{\phi}_{-\vec{q}j}
\end{equation} 
with $t_j = +$ and $t_j = -$ for the TRE and TRO components of the fluctuations, respectively. 

From the analysis presented above, it is clear that the interaction cannot be generally repulsive or attractive in the Cooper channel. 
The constraint (\ref{chiPropIII}) now assumes the generalized form
\begin{equation}
 \chi_{jj'}(i\Omega_n,\vec{q}) = t_j t_{j'} \, \chi_{jj'}(-i\Omega_n,\vec{q}), \label{chiPropIIIMod}
\end{equation}
such that $\chi(q)$ is not Hermitian anymore. However, as long as we assume that all components $M^{(j)}$ in the bare coupling (\ref{GeneralizedCoupling2}) satisfy \equref{EvenFunctionOfK}, the properties (\ref{VertexTRS}) and (\ref{ApproxSymConstr}) with $t$ replaced by $t_j$ are still valid and it can be shown that \equref{AsympPropertyOfV} as well as $Z_s = Z_{s_\R}$ hold. 
Consequently, the possible superconducting order parameters must obey \equref{RashbaParity} leading to the same conclusions as discussed in \secref{ApproximateSymmetry} as far as fully gapped, time-reversal symmetric superconducting phases are concerned.

\section{Topological superconductivity from phonons}
\label{TopSupPhon}
In \secref{ElectronPhononCoupling}, we have shown that, in a clean system, electron-phonon coupling alone can never lead to topological superconductivity. 
Now let us ask how an electron-phonon interaction dominated superconductor can nonetheless be topologically nontrivial. 

Firstly, there might be some finite residual Coulomb interaction. Although typically small\cite{PseudoPotentials} due to renormalization group corrections at energies between the Fermi energy and the Debye frequency, it will induce sign changes between Fermi surfaces if the electron-phonon interaction has a favorable structure: Focusing for simplicity on a two-band model with Fermi surfaces, e.g., as shown in \figref{WeakSpinSplitting}, the topologically trivial superconducting state with $\widetilde{\Delta}_s$ having the same sign on both Fermi surfaces ($s^{++}$) will be (nearly) degenerate with the nontrivial $s^{+-}$ state (sign changes between the two Fermi surfaces) if the interband Cooper scattering is negligibly small. 
In this case, already a small residual repulsion can favor the topological $s^{+-}$ state. 
The analogous discussion in centrosymmetric superconductors can be found in \refcite{Brydon}, where it is shown that a topological odd-parity state can be induced by residual repulsions.

Secondly, one might ask whether disorder can induce a transition from an electron-phonon driven trivial superconductor to a topologically nontrivial state. From our discussion of particle-hole fluctuations in \secref{UnconventionalPairing}, we expect TRA, usually referred to as ``magnetic'', disorder to be a promising driving force for unconventional pairing. Indeed, recalling former studies of magnetic disorder in multiband superconductors \cite{MazinUnconv,PnictidesScatteringDolgov,PnictidesScattering,DisorderPaper}, we know that the $s^{+-}$ superconductor is only affected by intraband scattering while the $s^{++}$ state is prone to both intra- and interband processes. Consequently, if, as already discussed above, the interband Cooper scattering is sufficiently small, the critical temperatures of the $s^{+-}$ and $s^{++}$ state are nearly degenerate. Then, a small amount of magnetic disorder can lead to a transition from the trivial $s^{++}$ to the topological $s^{+-}$ state [see \figref{DisorderFigures}(
a)]. 
In the remainder of this section, we will substantiate this expectation by an explicit calculation and then discuss the impact of magnetic disorder on the resulting MBS.

\subsection{Disordered Ginzburg-Landau expansion}
\label{DisorderGLExp}
Let us first investigate the general multiband superconducting system described by the noninteracting Hamiltonian (\ref{NonInteractingHam}) and the electron-phonon induced interaction in \equref{MicroscopicInteraction}. As before, we assume that all bands are singly degenerate and that the weak-pairing approximation is applicable. Using the low-energy approach introduced in \secref{EffectiveElElInteraction}, the interaction in the Cooper channel reads
\begin{equation}
 \hat{H}_{\text{C}} = \sum_{\vec{k},\vec{k}'} e^{i(\varphi_{\vec{k}}^s-\varphi_{\vec{k}'}^{s'})} \mathcal{V}_{s's}(\vec{k}';\vec{k}) \, \hat{f}^\dagger_{\vec{k}'s'} \hat{f}^\dagger_{-\vec{k}'s'_\K} \hat{f}^{\phantom{\dagger}}_{-\vec{k}s^{\phantom{.}}_\K} \hat{f}^{\phantom{\dagger}}_{\vec{k}s} , \label{CooperInteraction}
\end{equation} 
where $\hat{f}$ and $\hat{f}^\dagger$ are the operator analogues of the Grassmann variables $f$ and $\bar{f}$ introduced in \equref{TrafoInEigenbasis}. Due to \equref{InteractionMatrixElement}, the interaction is fully attractive, $\mathcal{V}<0$, making no sign change of the order parameter possible. Following the conventional BCS approach, we have neglected the frequency dependence of the interaction kernel $\mathcal{V}$ in \equref{CooperInteraction} in order to make a Hamiltonian description possible.

We introduce disorder as perturbations of the form
\begin{equation}
 \hat{H}_{\text{dis}} = \int_{\vec{x},\vec{x}'} \hat{c}_{\alpha}^\dagger(\vec{x}) W_{\alpha\alpha'}(\vec{x},\vec{x}') \hat{c}_{\alpha'}^\pdagger(\vec{x}') \label{DisorderPotential}
\end{equation} 
with $\hat{c}^\dagger(\vec{x})$ and $\hat{c}(\vec{x})$ being the Fourier transform of the microscopic creation and annihilation operators $\hat{c}_{\vec{k}}^\dagger$ and $\hat{c}_{\vec{k}}$.
Assuming that the system is self-averaging, we treat $W$ as a Gaussian distributed real ($W^\dagger = W$) random field. Restricting the analysis to spatially local configurations, $\vec{x}=\vec{x}'$ in \equref{DisorderPotential}, with $\delta$-correlated and homogeneous statistics, it holds
 \begin{align}
\begin{split}
&\braket{W_{\alpha_1\alpha_1'}(\vec{x}_1,\vec{x}_1')W_{\alpha_2\alpha_2'}(\vec{x}_2,\vec{x}_2')}_\text{dis} \\
&\,\, = \delta(\vec{x}_1-\vec{x}_1')\delta(\vec{x}_2-\vec{x}_2')\delta(\vec{x}_1-\vec{x}_2) \Gamma_{\alpha_1 \alpha_1',\alpha_2\alpha_2'},  \label{DefinitionCorrelator} \end{split}
\end{align}
where $\braket{\dots}_\text{dis}$ represents the disorder average. Averaging over $W$ produces an effective four-fermion interaction within replica theory \cite{ReplicaAppr} with bare vertex given by $\Gamma$. The correlator $\Gamma$ can always be expressed in terms of Hermitian basis matrices $\{w_\mu \}$,
\begin{equation}
 \Gamma_{\alpha_1 \alpha_1',\alpha_2\alpha_2'} = \sum_{\mu,\mu'} C_{\mu \mu'} (w_{\mu})_{\alpha_1\alpha_1'} (w_{\mu'})_{\alpha_2\alpha_2'}, \label{GammaExpansionRed}
\end{equation} 
with $C$ being real and symmetric \cite{DisorderPaper},
\begin{equation}
 C = C^* = C^T. \label{CProperties}
\end{equation} 
Although a generic disorder realization $W$ will break all spatial symmetries, the symmetries of the clean system must be restored on average. This means that the correlator $\Gamma$ must be fully invariant under all symmetry operations $g$ of the point group of the system or, formally, that 
\begin{align}
\begin{split}
 &\Gamma_{\alpha_1 \alpha_1',\alpha_2\alpha_2'} = \left(\mathcal{R}_\psi(g)\right)_{\alpha_1\widetilde{\alpha}_1} \left(\mathcal{R}_\psi(g)\right)_{\alpha_2\widetilde{\alpha}_2} \\ & \qquad \quad \times \Gamma_{\widetilde{\alpha}_1 \widetilde{\alpha}_1',\widetilde{\alpha}_2\widetilde{\alpha}_2'} \left(\mathcal{R}_\psi^\dagger(g)\right)_{\widetilde{\alpha}_2'\alpha_2'} \left(\mathcal{R}_\psi^\dagger(g)\right)_{\widetilde{\alpha}_1'\alpha_1'}, \label{PointSymmetriesConstr}\end{split}
\end{align} 
where  $\mathcal{R}_\psi(g)$ denotes the wavefunction representation of the operation $g$.

Coming back to spatial symmetries below, let us first only focus on the TRS properties of the disorder configurations: TRS (``nonmagnetic'') and TRA (``magnetic'') disorder is mathematically equivalent to restricting the expansion (\ref{GammaExpansionRed}) to matrices satisfying
\begin{equation}
 \Theta w_{\mu} \Theta^\dagger = t_\gamma w_{\mu} \label{MagneticNonMagnetic}
\end{equation} 
with $t_\gamma = +$ and $t_\gamma = -$, respectively.

\begin{figure*}[tb]
\begin{center}
\includegraphics[width=\linewidth]{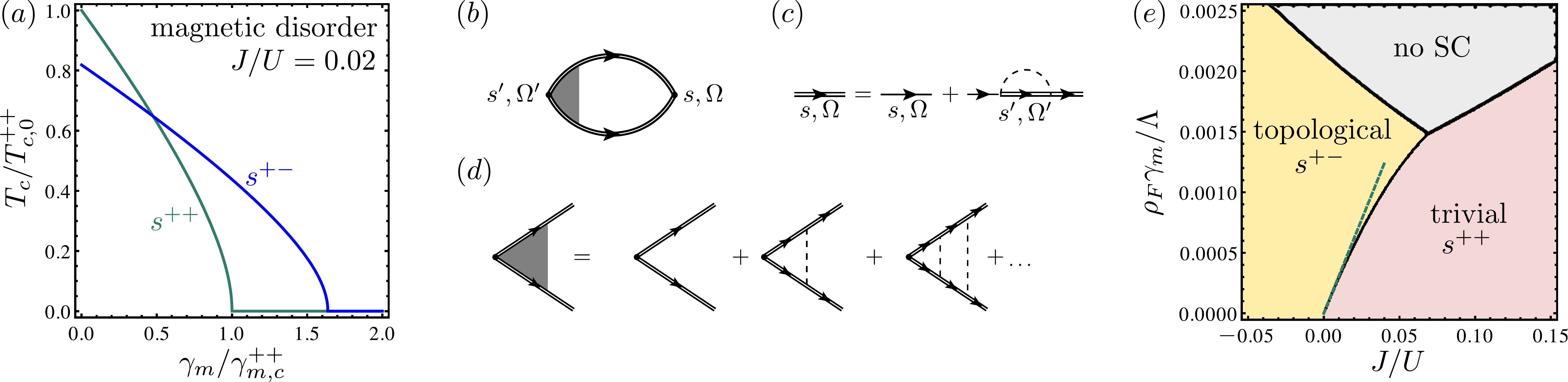}
\caption{(Color online) In (a), the transition temperatures of the $s^{++}$ (green) and $s^{+-}$ (blue) superconductors are shown as a function of the magnetic scattering strength $\gamma_m$ assuming that the intraband ($U$) is much larger than the interband Cooper scattering ($J$). Here $\gamma_{m,c}^{++}$ and $T_{c,0}^{++}$ are the critical scattering rate of the $s^{++}$ superconductor and its transition temperature in the absence of disorder, respectively. Part (b) shows the exact representation of the kernel of the quadratic Ginzburg-Landau expansion (\ref{GLExp}) in terms of the full Greens function (double line) and renormalized vertex (gray triangle). Focusing on the limit $k_Fl \gg 1$, the former and the latter only contain the noncrossing diagrams shown in (c) and (d), respectively. Here the impurity line (dashed) only enters in the combinations defined in \equsref{SDefinition}{VDefinition}. The full phase diagram together with the predictions of the asymptotic expression (\ref{CriticalScatteringRate}) 
for the critical scattering rate to enter the topological state (green line) are shown in (e). Here, ``no SC'' denotes the suppression of both superconducting states. In (a) and (e), we have used $\rho_F U = -0.4$.}
\label{DisorderFigures}
\end{center}
\end{figure*}

To solve for the dominant superconducting state in the presence of disorder, the interaction (\ref{CooperInteraction}) will be treated within mean-field approximation. We introduce the order parameter
\begin{equation}
 \widetilde{\Delta}_s(\vec{k}) = \sum_{\vec{k}',s'} \braket{ \hat{f}_{-\vec{k}'s'_\K} \hat{f}_{\vec{k}'s'}} e^{i\varphi_{\vec{k}'}^{s'}} \mathcal{V}_{ss'}(\vec{k};\vec{k}')
\end{equation} 
such that the resulting mean-field Hamiltonian is of form of \equref{BdGHamiltonian} directly revealing the connection to the topological properties of the associated superconducting state (see \secref{TopologicalProperties}). The transition temperatures of the competing superconducting states are obtained by calculating the disorder-averaged free energy $\braket{\mathcal{F}}_\text{dis}$ as a function of the order parameter. Assuming, exactly as in \equref{SeparationApprox}, that both $\widetilde{\Delta}_s(\vec{k})$ as well as the interaction matrix elements only depend on $s$ and $\Omega$, one finds
\begin{equation}
 \braket{\mathcal{F}}_\text{dis} \sim \sum_{s,s'}\int_s \diff \Omega \int_{s'} \diff \Omega' \, \widetilde{\Delta}^*_s(\Omega) \widetilde{D}_{\Omega s,\Omega' s'}(T) \widetilde{\Delta}_{s'}(\Omega') \label{GLExp}
\end{equation} 
as $\widetilde{\Delta}\rightarrow 0$, where 
\begin{equation}
 \widetilde{D}^{\phantom{-1}}_{\Omega s,\Omega' s'}(T) =  D_{\Omega s,\Omega' s'}^{\phantom{-1}}(T) - \mathcal{V}^{-1}_{\Omega s,\Omega' s'}.  \label{KernelOfGLExp}
\end{equation} 
Here, $\mathcal{V}^{-1}$ denotes the inverse of the interaction kernel $\mathcal{V}$ and the disorder averaged particle-particle bubble $D_{\Omega s,\Omega' s'}$ is represented diagrammatically in \figref{DisorderFigures}(b) in terms of the full Green's function (double line) and the dressed vertex (gray triangle). Focusing on weak disorder where the mean-free path $l$ is much larger than the inverse Fermi momentum $1/k_F$, all diagrams with crossed impurity lines, which are suppressed by a factor $(k_Fl)^{-1}$, can be neglected. The self-energy and vertex correction are thus simply given by the ``rainbow diagrams'' and ``Cooperon ladder'' as shown in \figref{DisorderFigures}(c) and (d), respectively.

In analogy to the Eliashberg approach (\cf \figref{DiagramsOfEliashberg}), the impurity line, which is just given by the transformation of the correlator (\ref{GammaExpansionRed}) into the eigenbasis of the normal state Hamiltonian, only enters in the form of two distinct index combinations: The self-energy is determined by forward scattering,  
\begin{align}
\begin{split}
\mathcal{S}^F_{\Omega s,\Omega' s'}  &:= \vcenter{\hbox{\includegraphics[height=3.5em]{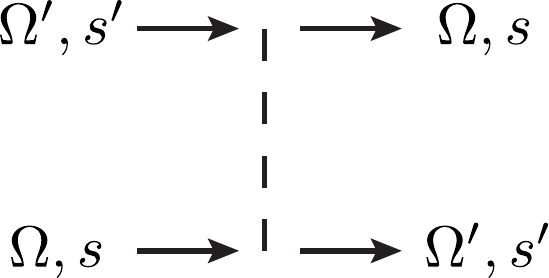}}} \label{SDefinition} \\
&= \sum_{\mu,\mu'}  \left(\psi^\dagger_{\Omega' s'} w_{\mu} \psi^{\phantom{\dagger}}_{\Omega s}\right)^* C_{\mu \mu'} \, \psi^\dagger_{\Omega' s'} w_{\mu'} \psi^{\phantom{\dagger}}_{\Omega s},
\end{split}
\end{align} 
which is real valued due to \equref{CProperties}.
The vertex-correction is determined by the Cooper scattering,
\begin{align}
\begin{split}
 \mathcal{S}^C_{\Omega s,\Omega' s'} &:= \vcenter{\hbox{\includegraphics[height=3.6em]{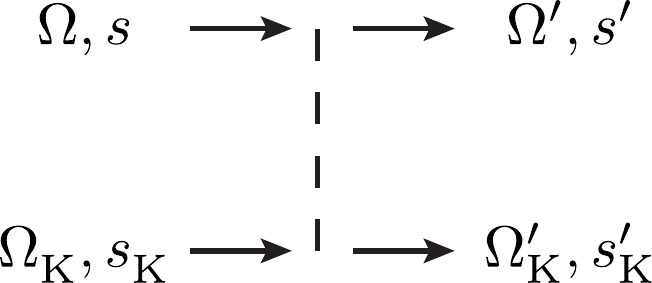}}} \label{VDefinition} \\
&= t_\gamma e^{i(\varphi_\Omega^s-\varphi_{\Omega'}^{s'})}\mathcal{S}^F_{\Omega s,\Omega' s'}.
\end{split}
\end{align} 
In the second line, \equsref{TRSOfESs}{MagneticNonMagnetic} have been taken into account. The relation (\ref{VDefinition}) between the Cooper and forward disorder scattering is the replica analogue of the relation for the electron-electron interaction in \equref{UnconvMatrixElements}. 

As shown in \appref{DisorderCalc}, summing up the diagrams in \figref{DisorderFigures}(c) and (d) yields the general result
\begin{equation}
 D = -T \sum_{\omega_n} \left( \mathcal{C}(\omega_n) - t_\gamma \, \mathcal{S}^S \right)^{-1} \label{GeneralExpreD}
\end{equation} 
with the symmetrized scattering vertex 
\begin{equation}
 \mathcal{S}^S_{\Omega s,\Omega' s'} = \mathcal{S}^F_{\Omega s,\Omega' s'} + \mathcal{S}^F_{\Omega s,\Omega'_\K s'_\K} \label{SymmetrizedSF}
\end{equation} 
and the diagonal matrix
\begin{align}
 \begin{split}
&\mathcal{C}_{\Omega s,\Omega' s'}(i\omega_n) \\ &= \frac{\delta_{s,s'}\delta_{\Omega,\Omega'}}{\rho_s(\Omega)} \left(\frac{|\omega_n|}{\pi} + \sum_{\tilde{s}} \int_{\tilde{s}}\diff\widetilde{\Omega} \, \rho_{\tilde{s}}(\widetilde{\Omega}) \, \mathcal{S}^S_{\Omega s,\widetilde{\Omega} \tilde{s}} \right) . \label{DiagonalMatrixDef}  \end{split}
\end{align} 
Note that the inverse in \equref{GeneralExpreD} refers to both $s$- and $\Omega$-space and that $\mathcal{S}^F_{\Omega s,\Omega' s'} = \mathcal{S}^F_{\Omega_\K^{\phantom{.}} s_\K^{\phantom{.}}, \Omega'_\K s'_\K}$ such that $\mathcal{S}^S$ in \equref{SymmetrizedSF} is symmetrized in both indices, $\mathcal{S}^S_{\Omega s,\Omega' s'} = \mathcal{S}^S_{\Omega_\K s_\K, \Omega' s'} = \mathcal{S}^S_{\Omega s, \Omega'_\K s'_\K}$.

Using the symmetry constraint (\ref{PointSymmetriesConstr}) on the disorder correlator, it is straightforward to show that 
\begin{equation}
 \mathcal{S}^S_{\Omega s,\Omega' s'} = \mathcal{S}^S_{\mathcal{R}_v(g)(\Omega s),\mathcal{R}_v(g)(\Omega' s')}, \label{SymConstrOfS}
\end{equation} 
where $\mathcal{R}_v(g)$ denotes the representation of the symmetry operation $g$ on the multi-index $(\Omega s)$. The symmetry of the spectrum implies that also the density of states $\rho_s(\Omega)$ is invariant under all symmetry operations $g$ of the point group. Consequently, $\mathcal{C}_{\Omega s,\Omega' s'}$ satisfies the same constraint as $\mathcal{S}^S$ in \equref{SymConstrOfS}. The point symmetries of the interaction imply that the very same holds for $\mathcal{V}_{\Omega s,\Omega' s'}$ and, hence, for the kernel (\ref{KernelOfGLExp}) of the disordered Ginzburg-Landau expansion. This shows that the resulting superconducting order parameter must again transform under one of the irreducible representations of the point group of the clean system although the symmetries are only preserved on average.

This allows to generalize the necessary condition of \refcite{DesignPrinciples,*DesignPrinciplesUnpub} for spontaneous TRS-breaking to disordered systems: We know from \refcite{DesignPrinciples,*DesignPrinciplesUnpub} that multidimensional or complex representations can be excluded for a 2D system in the weak-pairing limit for any point group without a threefold rotation symmetry perpendicular to the plane of the system. In that case, the resulting superconducting state has to be nondegenerate already on the quadratic level (\ref{GLExp}) of the Ginzburg-Landau expansion (neglecting accidental degeneracies).
Since $\mathcal{S}^F$ is real and symmetric, the same holds for $D$ and, hence, for the kernel $\widetilde{D}$. Due to the absence of degeneracy, this means that the superconducting order parameter $\widetilde{\Delta}_s(\Omega)$ can always be chosen to be real thus preserving TRS. 
This means that, irrespective of whether we consider magnetic or nonmagnetic disorder, the resulting superconducting state must be necessarily time-reversal symmetric in 2D if there is no threefold rotation symmetry in the clean system.

\subsection{Disorder induced topology}
\label{DisorderIndTop}
To show that disorder can drive an electron-phonon superconductor, that must be necessarily trivial in the clean limit, into a topological DIII state, let us focus for concreteness, e.g., on 2D systems with $C_{2v}$ point group. From the arguments presented above it is already clear that the resulting superconducting state must be time-reversal symmetric due to the absence of a threefold rotation symmetry.

Assuming that there are no additional orbital degrees of freedom, the normal state Hamiltonian can be written as
\begin{equation}
 h_{\vec{k}} = \epsilon_{\vec{k}} \sigma_0 + \vec{g}_{\vec{k}} \cdot \vec{\sigma}, \label{NonInterHam}
\end{equation}  
where $\sigma_j$ are Pauli matrices in spin space. 

The most general disorder correlator $\Gamma$ in case of nonmagnetic ($t_\gamma = +$) disorder reads in the microscopic basis as
\begin{equation}
 \Gamma_{\alpha_1 \alpha_1',\alpha_2\alpha_2'} = \gamma_0 (\sigma_0)_{\alpha_1\alpha_1'} (\sigma_0)_{\alpha_2\alpha_2'}, \label{NonMagneticDisorder}
\end{equation} 
whereas, in case of magnetic ($t_\gamma = -$) impurities, we have
\begin{align}
\begin{split}
& \Gamma_{\alpha_1 \alpha_1',\alpha_2\alpha_2'} =  \gamma_\parallel^{(1)}  (\sigma_1)_{\alpha_1\alpha_1'} (\sigma_1)_{\alpha_2\alpha_2'}     \\ & \quad + \gamma_\parallel^{(2)}(\sigma_2)_{\alpha_1\alpha_1'} (\sigma_2)_{\alpha_2\alpha_2'} + \gamma_\perp (\sigma_3)_{\alpha_1\alpha_1'} (\sigma_3)_{\alpha_2\alpha_2'}.  \label{MagneticDisorder}\end{split}
\end{align} 
The terms proportional to $\gamma^{(1,2)}_\parallel$ and $\gamma_\perp$ describe spin-magnetic impurities which are aligned in the plane and perpendicular to the plane of the 2D system, respectively. It is straightforward to show, without further assumptions about the structure of the spin-orbit vector $\vec{g}_{\vec{k}}$, that $\mathcal{S}^S_{\Omega s,\Omega' s'} = \gamma_0$ in case of $\Gamma$ given by \equref{NonMagneticDisorder} and $\mathcal{S}^S_{\Omega s,\Omega' s'} = \gamma^{(1)}_\parallel + \gamma^{(2)}_\parallel + \gamma_\perp =: \gamma_m$ for the correlator in \equref{MagneticDisorder}.

Due to the simple form of the scattering matrix, it is possible to perform the inversion in \equref{GeneralExpreD} analytically (see \appref{DisorderCalc} for details). To obtain a minimal phase diagram that captures the relevant physics, let us assume that $\rho_s(\Omega) \approx \text{const.}$ and that the interaction matrix elements can be parameterized as
\begin{equation}
 \mathcal{V}_{\Omega s,\Omega' s'} \approx \begin{pmatrix} U & J \\ J & U \end{pmatrix}_{s,s'}, \label{SimpleInteraction}
\end{equation} 
i.e., there is an intraband Cooper scattering ($U$) that is the same for both bands and an interband Cooper interaction ($J$) both of which are constant on the two Fermi surfaces.

By diagonalizing the resulting $\widetilde{D}$ in the free-energy expansion (\ref{GLExp}), one can deduce the transition temperatures and, hence, the dominant instability of the system as a function of the interaction parameters $U$, $J$ as well as of the disorder strength parameterized by $\gamma_0$ and $\gamma_m$ for nonmagnetic and magnetic disorder, respectively.

In case of nonmagnetic disorder, the critical temperature of the $s^{++}$ state which, due to the parameters used in the present calculation, has $\widetilde{\Delta}_1(\Omega) = \widetilde{\Delta}_2(\Omega) = \text{const.}$, is not affected by disorder. This is just a manifestation of the well-known Anderson theorem\cite{AT,ATAG1,ATAG2}. As a consequence of the sign change between the Fermi surfaces, the transition temperature of the competing $s^{+-}$ superconductor (here with $\widetilde{\Delta}_1(\Omega) = -\widetilde{\Delta}_2(\Omega) = \text{const.}$) is reduced by disorder. As $s^{++}$ dominates in the clean limit in case of electron-phonon pairing, no transition to the topological $s^{+-}$ state can be induced by nonmagnetic disorder.

This is different in case of magnetic impurities: \figref{DisorderFigures}(a) shows the transition temperatures of the $s^{++}$ and $s^{+-}$ superconductors as a function of the total amount of magnetic disorder $\gamma_m$. Despite being dominant in the clean limit, the $s^{++}$ state is more fragile against magnetic impurities since both inter- and intraband scattering act as pair breaking, while the $s^{+-}$ superconductor is only prone to the latter type of scattering events. This makes possible a finite range of impurity concentrations where the topological $s^{+-}$ state is stabilized. For larger $\gamma_m$ also the $s^{+-}$ condensate is destroyed by disorder and no superconducting instability occurs at all. 

The full phase diagram that shows the dependence on the ratio $J/U$ of the interaction parameters for a fixed (negative) value of $U$ can be found in \figref{DisorderFigures}(e). Here, $\rho_F := \sum_s \int\diff\Omega \rho_s(\Omega)$ denotes the total density of states of the system and $\Lambda$ is, as before, the energetic cutoff of the electron-phonon interaction. For completeness, we have also included positive values of $J$ where $s^{+-}$ is already dominant in the clean limit. Let us emphasize that $J>0$ cannot be realized by pure electron-phonon coupling as shown in \secref{EffectiveElElInteraction}.

As $s^{++}$ and $s^{+-}$ are degenerate for $J=0$, the critical scattering rate $\rho_F \gamma_m^*$ for stabilizing a topological phase must go to zero as $J \rightarrow 0$. For small $J/U$, it varies linearly with $J$ according to [\cf green line in \figref{DisorderFigures}(e)]
\begin{equation}
 \rho_F \gamma_m^* \sim \frac{16}{\pi^2} \, \frac{J}{U} \frac{1}{\rho_F U} T_{c,0}^{++}, \label{CriticalScatteringRate}
\end{equation} 
where $T_{c,0}^{++}$ is the critical temperature of the $s^{++}$ state in the clean limit.

If $J$ is sufficiently strong, the critical temperatures of both $s^{++}$ and $s^{+-}$ will go to zero as a function of the magnetic scattering strength $\gamma_m$ before a transition into the $s^{+-}$ phase can occur. This gives rise to a critical ratio $(J/U)_c$ for the impurity-induced topological transition. One finds (for $U<0$)
\begin{align}
\begin{split}
 \left(\frac{J}{U}\right)_c &= \frac{1}{\rho_F |U| \ln 2}\left(\sqrt{(\rho_F U \ln 2)^2+4}-2\right) \\
&\sim \frac{\ln 2}{4} \rho_F |U|, \quad \rho_F |U| \ll 1, \label{JoUCrit}
\end{split}
\end{align} 
showing that it scales linearly with $U$ in the weak-coupling limit. Our mean-field approach predicts $(J/U)_c$ to approach $1$ in the strong-coupling limit $\rho_F |U| \gg 1$.

At this point, two remarks are in order: 
Firstly, let us contrast our results with \refcite{MagneticDoping}, where magnetic scattering induces nontrivial topology with respect to a symmetry class with broken TRS: For the model considered in \refcite{MagneticDoping}, a Zeeman field is required\cite{Sau} to stabilize a topological superconductor. Our analysis shows that magnetic disorder can also drive the transition into a topological superconducting state that preserves TRS (class DIII). This means that while, locally, TRS is broken due to the presence of impurities it is restored globally in the sense that the resulting superconducting order parameter is time-reversal symmetric and that the impurities do not give rise to a net magnetic moment.

Secondly, we note that the disorder-induced formation of a topological phase has not been obtained in \refcite{DisorderPaper} since the analysis of \refcite{DisorderPaper} has been performed in the limit $J/U \rightarrow \infty$.

\begin{figure*}[tb]
\begin{center}
\includegraphics[width=\linewidth]{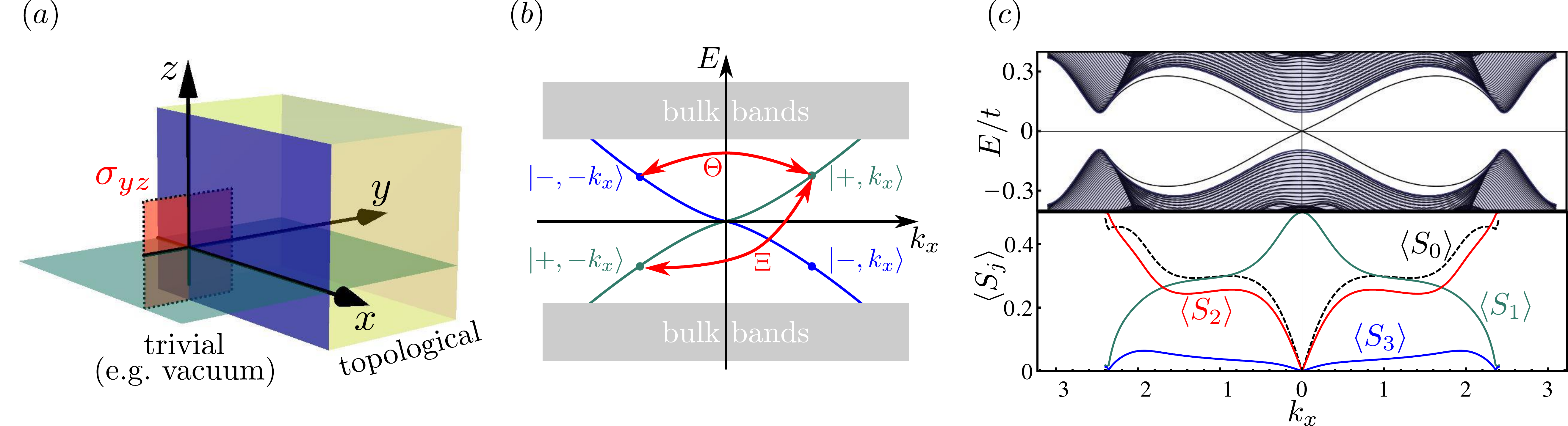}
\caption{(Color online) Point-symmetry protection of MBS against magnetic impurties. In (a), the geometry of a boundary (blue plane) at $y=0$ of two topologically distinct phases of a 2D system ($xy$-plane) is shown. If both phases have point group $C_{2v}$, there will be a residual reflection symmetry (red plane). A schematic of the spectrum of the system with edge state dispersion shown in blue and green is presented in (b). Part (c) shows the spectrum (upper panel) of the model defined in the main text using open along the $y$- and periodic boundary conditions along the $x$-direction. Due to the protection resulting from TRS and the reflection symmetry at the boundary, the matrix elements (lower panel) of charge impurities ($\braket{S_0}$) and magnetic impurities polarized along $y$ and $z$ ($\braket{S_2}$ and $\braket{S_3}$) vanish at $k_x=0$. Here we use the same parameters as in \figref{WeakSpinSplitting}, $\mu = -0.4t$, $\alpha = 0.25t$, for the normal state Hamiltonian and choose $\Delta_t = 0.
3t$, $\Delta_s = 0.1t$ for the superconducting order parameter.}
\label{ProtectionOfBoundStates}
\end{center}
\end{figure*}

\subsection{Protection of bound states}
\label{Protection}
One major consequence of the topologically nontrivial DIII bulk invariant is the existence of gapless counter-propagating Kramers partners of MBS at the interface of the superconductor to a topologically trivial phase such as the vacuum\cite{Bernevig}. The presence of these gapless modes is guaranteed by TRS. However, the magnetic impurities required to stabilize the bulk topology break TRS and might hence gap out the boundary states making them unobservable in experiments. 

Let us first notice that, at least theoretically, there exists a parameter range where the disorder-induced gap in the surface spectrum is irrelevant. As can be seen in \equref{CriticalScatteringRate}, the magnetic scattering rate required to induce a nontrivial bulk topology can be arbitrarily small as compared to the critical temperature and, hence, as compared to the gap of the superconductor at zero temperature. In this limit, the impact of the magnetic impurities on the Majorana modes can be neglected.

Secondly, unitary symmetries can protect the Kramers pair of MBS even if TRS is broken. E.g., in case of the 2D system with point group $C_{2v}$, the protection results from the residual reflection symmetry perpendicular to an interface along one of the crystallographic axes. To show this, let us assume that the system is located in the $xy$-plane with a boundary to a trivial phase at $y=0$ as illustrated in \figref{ProtectionOfBoundStates}(a). The presence of two distinct phases breaks all symmetries of the point group except for the invariance under reflection $\sigma_{yz}$ at the $yz$-plane.
Denoting the BdG Hamiltonian of the bulk system by $h^{\text{BdG}}_{\vec{k}}$, the spectrum and wavefunctions of the edge modes are determined by
\begin{equation}
 \widetilde{h}^{\text{BdG}}_{k_x} \ket{\pm,k_x} = E_{\pm} (k_x) \ket{\pm,k_x},
\end{equation} 
where $\widetilde{h}^{\text{BdG}}_{k_x}$ follows from $h^{\text{BdG}}_{\vec{k}}$ by replacing $k_y \rightarrow -i\partial_y$ and introducing some $y$-dependency to describe the boundary between the two topologically distinct phases. As illustrated in \figref{ProtectionOfBoundStates}(b), charge conjugation, $\Xi h^{\text{BdG}}_{\vec{k}} \Xi^{-1} = - h^{\text{BdG}}_{-\vec{k}}$ with antiunitary $\Xi$, and TRS of the BdG Hamiltonian lead to the constraints $E_\pm (k_x) = -E_\pm (-k_x)$ and $E_+ (k_x) = E_- (-k_x)$ on the edge state spectrum, respectively. 
Furthermore, charge-conjugation symmetry implies for the wavefunctions $\Xi \ket{\pm,k_x} = e^{i\alpha^{\pm}_{k_x}} \ket{\pm,-k_x}$ with some phases $\alpha^{\pm}_{k_x}$ and, in particular, $\Xi \ket{\pm,0} = e^{i\alpha^{\pm}_{0}} \ket{\pm,0}$ resulting from continuity in $k_x$. Denoting the spin operators in Nambu space by $S_j$, $j=1,2,3$, and noting that $\Xi S_j \Xi^{-1} = - S_j$, we find
\begin{equation}
 \braket{\mu,0|S_j|\mu',0} = -e^{i(\alpha_0^{\mu} - \alpha_0^{\mu'})}\braket{\mu',0|S_j|\mu,0} \label{ChargeConju}
\end{equation} 
with $\mu,\mu' = \pm$. From this, it already follows that the diagonal ($\mu=\mu'$) matrix elements of all spin operators $S_j$ must vanish. To restrict the off-diagonal components ($\mu\neq\mu'$), the mirror symmetry has be taken into account. Under $\sigma_{yz}$, it holds $k_x \rightarrow -k_x$ and $R_{\sigma_{yz}} S_j R_{\sigma_{yz}}^{-1} = p_j S_j$ with $p_1=1$ and $p_2=p_3=-1$ where $R_{\sigma_{yz}} = \exp(-i\pi S_1)$ is the representation of $\sigma_{yz}$ in Nambu space. It follows
\begin{equation}
 \braket{+,0|S_j|-,0} = -p_je^{-2i\alpha_0}\braket{-,0|S_j|+,0}, \label{MirrorProp}
\end{equation} 
where we have used $R_{\sigma_{yz}} \ket{+,k_x} = e^{i\alpha_{k_x}}\ket{-,-k_x}$, $\alpha_{k_x} \in \mathbbm{R}$, in the limit $k_x \rightarrow 0$.
The additional minus sign in \equref{MirrorProp} comes from $R_{\sigma_{yz}}^2 = -\mathbbm{1}$ which must hold for spin-$1/2$ fermions. Noting that $e^{2i\alpha_0} = e^{i(\alpha_0^{-} - \alpha_0^{+})}$, which follows from $[\Xi,R_{\sigma_{yz}}]=0$, the combination of \equsref{ChargeConju}{MirrorProp} implies that 
\begin{equation}
 \braket{\mu,0|S_j|\mu',0} = 0
\end{equation} 
for those component with $p_j = -1$, i.e., for $j=2,3$. This means that only impurities with finite spin polarization perpendicular to the mirror plane can open up a gap in the surface spectrum. This result is consistent with the numerical investigation of surface disorder in a model with $C_{4v}$ symmetry in \refcite{SchnyderDisorder}.

To further illustrate the protection of the MBS resulting from the symmetries of the system, let us investigate the standard Rashba single-band model defined by $\epsilon_{\vec{k}} = -t(\cos k_x + \cos k_y)-\mu$ and $\vec{g}_{\vec{k}} = \alpha (-\sin k_y,\sin k_x,0)^T$ in \equref{NonInterHam} and Fermi surfaces as shown in \figref{WeakSpinSplitting}. A natural Brillouin-zone regularization of the weak-pairing description of the $s^{+-}$ superconductor is given by the mean-field pairing term in \equref{MeanFieldHamiltonian} with pure triplet component $\Delta_{\vec{k}} = \Delta_t \vec{g}_{\vec{k}}\cdot \vec{\sigma} i \sigma_y/\alpha$. In the upper panel of \figref{ProtectionOfBoundStates}(c), we show the spectrum of the system with periodic boundary conditions along the $x$- and open boundary condition along the $y$-axis ($N_y = 100$ sites), where, for the sake of generality, also a small singlet component $\Delta_t i\sigma_y$ has been added. The edge state dispersions (doubly degenerate corresponding to the 
two 
edges of the system) crossing the Fermi level are clearly visible. In the lower panel of \figref{ProtectionOfBoundStates}(c), the maximum of the impurity matrix elements with respect to the subgap states at given $k_x$ is shown for both nonmagnetic, $\braket{S_0}$, as well as magnetic, $\braket{S_j}$, $j=1,2,3$, scatterers localized at one of the boundaries. We see that $\braket{S_0}$ vanishes for $k_x \rightarrow 0$ which is just a manifestation of the protection of the edge states resulting from TRS\cite{Bernevig}. Furthermore, also $\braket{S_2}$ and $\braket{S_3}$ vanish, in accordance with our general symmetry discussion above, whereas $\braket{S_1}$ assumes finite values at $k_x = 0$.

Consequently, if the impurities are, e.g., mainly polarized perpendicular to the plane of 2D system, $\gamma^{(1,2)}_\parallel = 0$ in \equref{MagneticDisorder}, a transition to a topological DIII superconductor can be induced by varying $\gamma_\perp$ without gapping the resulting boundary modes as long as the edges are along one of the crystallographic axes. Naturally, the same protection mechanism applies for all point groups $C_{nv}$, $n=1,2,3,4,6$, as long as the boundary is oriented perpendicular to one of the mirror planes of the bulk system.

\section{Application to materials}
\label{ApplicationToMaterials}
To illustrate the general results obtained above let us now discuss the implications for two physical systems that have attracted recent attention.

The first example is given by the LaAlO$_3$/SrTiO$_3$ heterostructures. Although conducting behavior has been observed\cite{Ohtomo,HerranzDEL,AnnadDEL} for all three different orientations of the interface, superconductivity has so far only been reported for the [001]\cite{Reyren} and [110]\cite{110SC} heterostructures.
The combination of the small transition temperature\cite{Reyren,110SC} of superconductivity and strong spin-orbit splitting\cite{CavigliaSOC,ShalomSOC,VenkatesanSOC} of the Fermi surfaces safely allows for applying the weak-pairing approximation in these systems. 

Let us first focus on the [001]-oriented interface, where superconductivity is associated\cite{Joshua2012,TrisconeConfExp} with the chemical potential entering the bands derived from the Ti $3d_{xz}$ and $3d_{yz}$ orbitals. If it is close to the bottom of these bands, there are only two singly-degenerate Fermi surfaces. In \refcite{ScheurerSchmalian}, it has been shown that, in this regime, a microscopically repulsive interaction will drive the system close to a spin-density wave instability with a competing superconducting instability that is topologically nontrivial. In case of electron-phonon coupling being dominant, the superconducting state is trivial. All of this is consistent with the analysis presented above which, on top of that, generalizes the absence of topological nontrivial structures in case of phonons beyond the weak-coupling limit considered in \refcite{ScheurerSchmalian} and shows that the time-reversal properties of the competing spin-density wave ($t=-1)$ are key to induce a nontrivial 
superconductor. 
Furthermore, when the energetically higher Rashba pair of bands\cite{ScheurerSchmalian} is populated, which can be induced via gate tuning\cite{CavigliaDome,CavigliaSOC,ShalomSOC}, the system becomes topologically trivial: Naively, two pairs of counter-propagating Majorana modes are expected at the boundary which, however, can be gapped out\cite{Bernevig} by surface perturbations that break neither particle-hole nor TRS.  
Finally, we have seen that, even if superconductivity is driven by electron-phonon coupling, magnetic disorder can nonetheless drive the system towards a topological phase. Note that magnetic scattering is expected to be particularly important for LaAlO$_3$/SrTiO$_3$ heterostructures where oxygen vacancies, which are initially nonmagnetic impurities, are believed\cite{DFToxygen1} to lead to local magnetic moments on the Ti sites.

Without taking into account further microscopic details, our results also allow for the following conclusions about the [110] interface. If the chemical potential is gate tuned\cite{VenkatesanSOC} to the lowest Rashba pair of bands (again arising from the $3d_{xz}$/$3d_{yz}$ orbitals\cite{ARPES110}), we obtain the same correspondence between the mechanism and the topology of superconductivity as in the [001] interface: Phonons alone lead to a trivial superconductor, whereas TRA fluctuations induce a topological state which will become trivial when populating the second lowest Rashba pair of bands. 

The second system we will discuss is single-layer FeSe on [001] SrTiO$_3$ which shows a superconducting transition\cite{FeSeObs} at temperatures significantly above $50\,\textrm{K}$.
Although the presence of the substrate manifestly breaks inversion symmetry, it is less obvious in this system whether the weak-pairing approximation can be applied due to the larger transition temperature. If it is appropriate for deducing the superconducting order parameter, we can conclude that, irrespective of the unknown pairing mechanism, the condensate will be topologically trivial. This follows from the experimental observation\cite{Liu2012} that there are two Rashba pairs of Fermi surfaces around the $M$-point leading to $\sum_j m_j =2$ in \equref{2DSimplifiedInv}. Note that this does not contradict the recent proposal\cite{TopFeSe} of topological superconductivity in this system since the analysis of \refcite{TopFeSe} has been performed in the opposite limit where inversion-symmetry breaking can be fully neglected for describing superconductivity.

\section{Conclusion}
To summarize, we have considered the relation between the mechanism driving the superconducting instability and the topological and time-reversal properties thereof in noncentrosymmetric systems. Our results are general in the sense that they only depend on very few properties of the system such as the time-reversal behavior of the bosonic fluctuations inducing superconductivity, the symmetries of the crystal and the topology of the Fermi surfaces with respect to the TRIM.
Throughout the paper, we have been using the weak-pairing approximation, formally defined in \equref{ParameterizationOfG}, which is justified as long as $E_{\text{so}} \gtrsim  T_c$. 
The superconducting properties are derived using Eliashberg theory\cite{Eliashberg}.

Firstly, we have seen that spontaneous TRS breaking is not possible for superconductivity resulting from electron-phonon coupling or from TRE particle-hole fluctuations, whereas it cannot be excluded when TRO particle-hole (e.g., spin-density wave) fluctuations are relevant. In the latter case, only the general necessary condition\cite{DesignPrinciples,*DesignPrinciplesUnpub}, the presence of a threefold rotation symmetry in the normal phase, for spontaneous TRS-breaking in 2D noncentrosymmetric systems can be used to gain information about the TRS properties of the condensate without detailed microscopic information.
We have shown that the criterion of \refcite{DesignPrinciples,*DesignPrinciplesUnpub} also holds in the presence of both nonmagnetic as well as magnetic (weak) disorder that conserves the symmetries of the high-temperature phase only on average.

Secondly, it has been proven that superconductivity arising from pure electron-phonon coupling will be fully gapped, neither break any point symmetry of the system nor be topologically nontrivial. This can be used to gain information about the pairing mechanism of a superconductor: The observation of topologically nontrivial properties, e.g., MBS at the edge of the sample, indicates that the mechanism cannot be purely conventional, i.e., other interaction channels or significant disorder scattering has to be taken into account for understanding superconductivity.
We have shown that exactly the same conclusions hold if superconductivity is driven by TRE particle-hole fluctuations. 

In case of TRO fluctuations, the superconducting order parameter naturally has sign changes making topologically nontrivial states possible. From the asymptotic symmetry (\ref{ApproximateSymmetryEigenstates}) valid for $E_{\text{so}} \ll \Lambda_t$, it follows that all resulting order parameters can be grouped into Rashba even ($p=+$) and Rashba odd ($p=-$) as defined in \equref{RashbaParity}. Focusing on TRS-preserving states, we have shown that only the latter can be topologically nontrivial which leads to constraints on the corresponding topological invariants. E.g., it implies that 1D and 2D noncentrosymmetric superconductors can only be topological if the number of TRIM enclosed by Rashba pairs of Fermi surfaces is odd. This necessary condition for topological superconductivity is readily accessible experimentally as the structure of the Fermi surfaces can be directly measured in photoemission experiments. Note that one does not have to 
resolve the spin-orbit splitting as the criterion only refers to Rashba \textit{pairs} of Fermi surfaces. E.g., for single-layer FeSe, photoemission data\cite{Liu2012} indicates that there are two Rashba pairs enclosing the $M$-point. If the weak-pairing approximation can be applied in this system, the condensate must be topologically trivial. 

Finally, we have seen that magnetic disorder can induce a transition from a topologically trivial superconductor, e.g., resulting from pure electron-phonon coupling, into a topological TRS-preserving phase. Focusing on 2D systems, it has been demonstrated that the resulting Kramers pair of MBS is protected against magnetic out-of-plane impurities if the associated edge is oriented perpendicular to one of the mirror planes of the bulk point group.       

Our analysis shows that all noncentrosymmetric systems, such as the oxide heterostructures, that show a strong tendency towards magnetism\cite{SpinSplitting,DFToxygen1} and have the necessary Fermi surface topology\cite{ScheurerSchmalian,ARPES110} are promising candidates for the realization of topologically nontrivial superconductivity.

\acknowledgments
The author is particularly grateful to J.~Schmalian for fruitful discussions. In addition, MS also thanks M.~Hoyer, B.~Jeevanesan, N.~Kainaris and D.~Mendler for discussions.
The author acknowledges partial support from the Deutsche Forschungsgemeinschaft (DFG) through the Priority Program SPP 1458 ``Hochtemperatur-Supraleitung in Eisenpniktiden'' (project-no.~SCHM 1031/5-1).

\appendix
\section{Exact relations following from the spectral representation}
\label{SpectralRepSym}
In this appendix, properties of the bosonic propagator, the Nambu Green's function, and the fermion-boson vertex function, which are consequences of certain unitary or antiunitary symmetries, are derived. These relations are most easily seen from the spectral representation of the corresponding $n$-point functions.

\subsection{Identities for the order-parameter susceptibility}
\label{SpectralRepSymOrderParameterSusc}
We begin with the bosonic propagator
\begin{equation}
 \chi_{jj'}(i\Omega_n,\vec{q}) := \frac{1}{2}\int_0^\beta \diff \tau e^{i\Omega_n \tau } \braket{T_\tau \hat{\phi}_{\vec{q} j'}(\tau) \hat{\phi}_{-\vec{q} j}(0)} \label{DefinitionOfBosonicProp}
\end{equation} 
as the discussion is most transparent in this case. In \equref{DefinitionOfBosonicProp}, $T_\tau$ denotes the time-ordering operator.
The spectral representation reads
\begin{align}
 \chi_{jj'}(i\Omega_n,\vec{q})  =\frac{1}{2} \sum_{n_1,n_2}\frac{\braket{n_1|\hat{\phi}_{\vec{q} j'}|n_2}\braket{n_2|\hat{\phi}_{-\vec{q} j}|n_1}}{i\Omega_n - (E_{n_2}-E_{n_1})}I_{n_1 n_2}^+, \label{SpectralRepOfChi}
\end{align} 
where $\{\ket{n}\}$ is a basis of exact eigenstates of the full many-body Hamiltonian with respective energies $E_n$ and
\begin{equation}
 I_{n_1 n_2}^\zeta := \frac{1}{Z}\left( e^{-\beta E_{n_2}}- \zeta\, e^{-\beta E_{n_1}} \right)
\end{equation} 
has been introduced with $Z$ denoting the partition function.
Upon relabeling $n_1 \leftrightarrow n_2$ in \equref{SpectralRepOfChi}, one readily finds that $\chi_{jj'}(q) = \chi_{j'j}(-q)$ which is already the first property in \equref{chiProp}.

Using Hermiticity, $\hat{\phi}_{\vec{q} j}^\dagger = \hat{\phi}_{-\vec{q} j}$, we can rewrite the spectral representation (\ref{SpectralRepOfChi}) as
\begin{align}
 \chi_{jj'}(i\Omega_n,\vec{q})  =\frac{1}{2} \sum_{n_1,n_2}\frac{\braket{n_1|\hat{\phi}_{\vec{q} j'}|n_2}\left(\braket{n_1|\hat{\phi}_{\vec{q} j}|n_2}\right)^*}{i\Omega_n - (E_{n_2}-E_{n_1})}I_{n_1 n_2}^+, \label{BosonSpecAlt}
\end{align} 
from which $\chi_{jj'}(i\Omega_n,\vec{q}) = \chi^*_{j'j}(-i\Omega_n,\vec{q})$, i.e., the second property (\ref{chiPropII}), can be read off.

To derive the constraint following from TRS, we rearrange the summation in \equref{BosonSpecAlt} by replacing $\ket{n_{1,2}} \rightarrow \hat{\Theta} \ket{n_{1,2}}$. Noting that $\ket{n}$ and $\hat{\Theta} \ket{n}$ have the same energy together with
\begin{align}
 \braket{\hat{\Theta} n_1|\hat{\phi}_{\vec{q} j}|\hat{\Theta} n_2} &= \braket{ n_1|\hat{\Theta}^\dagger \hat{\phi}_{\vec{q} j} \hat{\Theta}| n_2}^* \\
&= t \braket{ n_1| \hat{\phi}_{-\vec{q} j} | n_2}^*,
\end{align} 
where we used \equref{PhiTRS} in the second line, yields $\chi(i\Omega_n,\vec{q}) = \chi^T(i\Omega_n,-\vec{q})$. Applying \equref{chiPropI}, we arrive at the relation (\ref{chiPropIII}) stated in the main text.

Finally, the proof of \equref{EvenInq} proceeds very similarly to the discussion of TRS above: We rearrange the sums in the spectral representation (\ref{BosonSpecAlt}) such that $\ket{n_{1,2}}$ is replaced by $\hat{S} \ket{n_{1,2}}$, take advantage of the fact that the energies of $\ket{n}$ and $\hat{S} \ket{n}$ are identical and then write
\begin{equation}
 \braket{\hat{S} n_1|\hat{\phi}_{\vec{q} j}|\hat{S} n_2} = \pm \braket{ n_1|\hat{\phi}_{-\vec{q} j}| n_2}
\end{equation} 
where \equref{PhiIrredRep} has been applied. This directly leads to \equref{EvenInq}.

\subsection{Identities for the Nambu Green's function}
\label{TRSOfGF}
Let us begin with the derivation of the TRS constraint (\ref{TRSOfGreesFunc}) of the Nambu Green's function. For this purpose, it is convenient to first work in the microscopic basis and define
\begin{align}
  \mathcal{G}^m_{\alpha\beta}&(i\omega_n,\vec{k}) := -\int_0^\beta \diff \tau e^{i\omega_n \tau } \\ & \times \begin{pmatrix} \braket{T_\tau \hat{c}_{\vec{k}\alpha}^{\phantom{\dagger}}(\tau)\hat{c}^\dagger_{\vec{k}\beta}(0)} & \braket{T_\tau \hat{c}_{\vec{k}\alpha}^{\phantom{\dagger}}(\tau)\hat{c}^{\phantom{\dagger}}_{-\vec{k}\beta}(0)} \\ \braket{T_\tau \hat{c}_{-\vec{k}\alpha}^{\dagger}(\tau)\hat{c}^{\dagger}_{\vec{k}\beta}(0)} & \braket{T_\tau \hat{c}_{-\vec{k}\alpha}^{\dagger}(\tau)\hat{c}^{\phantom{\dagger}}_{-\vec{k}\beta}(0)} \end{pmatrix}. \nonumber
\end{align} 
Consider, e.g., the upper left component with spectral representation
\begin{align}
 \left(\mathcal{G}^m_{\alpha\beta}(k)\right)_{11}  = \sum_{n_1,n_2}\frac{\braket{n_1|\hat{c}_{\vec{k}\alpha}^{\phantom{\dagger}}|n_2}\braket{n_2|\hat{c}^\dagger_{\vec{k}\beta}|n_1}}{i\omega_n - (E_{n_2}-E_{n_1})}I_{n_1 n_2}^-. \label{SpectralRepOfGmic}
\end{align}
Exactly as in case of the bosons, we rewrite the summation and then use
\begin{equation}
 \braket{\hat{\Theta} n_1|\hat{c}_{\vec{k}\alpha}|\hat{\Theta} n_2} = T_{\alpha\alpha'} \braket{ n_1|\hat{c}_{-\vec{k}\alpha'}| n_2}^*
\end{equation} 
based on \equref{TRSOfOperators}. This yields 
\begin{equation}
  T_{\alpha\alpha'}^\pdagger \left(\mathcal{G}^m_{\alpha'\beta'}(-k)\right)^*_{11} T^\dagger_{\beta'\beta} = \left(\mathcal{G}^m_{\alpha\beta}(k)\right)_{11}. \label{TRSOneCompFermion}
\end{equation} 
Collecting the resulting behavior of all four components, one finds
\begin{equation}
 \mathcal{T}_{\alpha\alpha'}^\pdagger \left(\mathcal{G}^m_{\alpha'\beta'}(-k)\right)^* \mathcal{T}^\dagger_{\beta'\beta} = \mathcal{G}^m_{\alpha\beta}(k), \, \, \mathcal{T}_{\alpha\beta} = \begin{pmatrix} T_{\alpha\beta} &   \\   & -T_{\alpha\beta}^\dagger \end{pmatrix} \label{TRSMicrosBasis}
\end{equation} 
in accordance with the relations derived in \refcite{WangZhangTopSC}. 

To arrive at the constraint (\ref{TRSOfGreesFunc}), we have to transform \equref{TRSMicrosBasis} into the eigenbasis of the normal state Hamiltonian. Using \equref{TrafoInEigenbasis} in, e.g., the upper left component of the Nambu Green's function yields
\begin{equation}
 \left(\mathcal{G}^m_{\alpha\beta}(k)\right)_{11} = \left(\psi_{\vec{k} s}^{\phantom{*}}\right)_\alpha \left(\mathcal{G}_{ss'}(k)\right)_{11} \left(\psi_{\vec{k} s'}^*\right)_{\beta}
\end{equation} 
with $\mathcal{G}_{ss'}$ as defined in \equref{PathIntegralDef}. Inserting this into \equref{TRSMicrosBasis}, using the property (\ref{TRSOfESs}) of the wavefunctions and proceeding analogously for all four components, one finds that \equref{TRSMicrosBasis} is, within the weak-pairing approximation (\ref{ParameterizationOfG}), equivalent to
\begin{equation}
  e^{-i\varphi_{\vec{k}}^s\tau_3} \tau_3 \mathcal{G}^*_{s_\K}(-k) \tau_3 e^{i\varphi_{\vec{k}}^s\tau_3} = \mathcal{G}_s(k)
\end{equation}
as stated in the main text.

The derivation of the charge-conjugation symmetry (\ref{ChargeConjugation}) of the Green's function proceeds in two steps: Firstly, one can directly read off from the path integral definition (\ref{PathIntegralDef}) that
\begin{equation}
 \tau_1 \mathcal{G}_{ss'}(k) \tau_1 =  -\left(\mathcal{G}_{s'_\K s^{\phantom{.}}_\K}\right)^T(-k),
\end{equation} 
where $T$ only refers to particle-hole space. Secondly, applying the well-known relation (see, e.g., \refcite{WangZhangTopSC})
\begin{equation}
 \mathcal{G}(i\omega_n,\vec{k}) = \mathcal{G}^\dagger(-i\omega_n,\vec{k}), \label{HermiticityG}
\end{equation} 
which can also be shown from the spectral representation [using Hermiticity, $\braket{n|\hat{f}|n'}^* = \braket{n'|\hat{f}^\dagger|n}$], we find
\begin{equation}
 \tau_1 \mathcal{G}_{ss'}(i\omega_n,\vec{k}) \tau_1 =  -\mathcal{G}_{s^{\phantom{.}}_\K s'_\K }^*(i\omega_n,-\vec{k}).
\end{equation} 
In the weak-pairing approximation, this reduces to \equref{ChargeConjugation}.

\subsection{Identities for the fermion-boson vertex}
\label{SpectralRepFermionBosonVertex}
Let us now discuss exact relations of the renormalized fermion-boson vertex $\Gamma^{(j)}$ in \equref{VertexFunction}. To this end, we start by analyzing the associated three-point function
\begin{align}
 \begin{split}
C_{\alpha\beta}^{(j)}(k;k') = \int_0^\beta & \diff\tau\int_0^\beta\diff\tau' e^{i(\omega_{n'}\tau'-\omega_n\tau)} \\ & \times \braket{T_\tau \hat{c}_{\vec{k}\alpha}^\dagger(\tau) \hat{c}^{\phantom{\dagger}}_{\vec{k}'\beta}(\tau') \hat{\phi}_{\vec{k}-\vec{k}'j}(0)}. \end{split}
\end{align} 
with spectral representation
\begin{widetext}
\begin{align}
\begin{split}
 C_{\alpha\beta}^{(j)}(k;k') =  \sum_{n_1,n_2,n_3} \Bigl( & \braket{n_1|\hat{c}_{\vec{k}\alpha}^\dagger|n_2} \braket{n_2|\hat{c}^{\phantom{\dagger}}_{\vec{k}'\beta}|n_3} \braket{n_3|\hat{\phi}_{\vec{k}-\vec{k}'j}|n_1} \, I_{n_1 n_2 n_3}(\omega_n,\omega_{n'}) \\
 &-\braket{n_1|\hat{c}^{\phantom{\dagger}}_{\vec{k}'\beta}|n_2} \braket{n_2|\hat{c}_{\vec{k}\alpha}^\dagger|n_3} \braket{n_3|\hat{\phi}_{\vec{k}-\vec{k}'j}|n_1} \, I_{n_1 n_2 n_3}(-\omega_{n'},-\omega_n)  \Bigr).\end{split} \label{threepointSpec}
\end{align} 
Here we have introduced
\begin{equation}
 I_{n_1 n_2 n_3}(\omega,\omega') = \frac{1}{Z} \frac{e^{-\beta E_{n_3}} (\Delta_{21} + i\omega) + e^{-\beta E_{n_2}} (\Delta_{31} + i(\omega-\omega')) + e^{-\beta E_{n_1}} (\Delta_{32} - i\omega')}{(\Delta_{21} + i\omega) (\Delta_{31} + i(\omega-\omega'))(\Delta_{32} - i\omega')}
\end{equation} 
using the shortcut notation $\Delta_{ij} := E_{n_i} - E_{n_j}$.
\end{widetext}
Complex conjugation of \equref{threepointSpec}, relabeling $n_1 \leftrightarrow n_3$ and using that
\begin{align}
 I^*_{n_1 n_2 n_3}(\omega,\omega') &= I_{n_1 n_2 n_3}(-\omega,-\omega'), \\
 I_{n_3 n_2 n_1}(\omega,\omega') &= I_{n_1 n_2 n_3}(\omega',\omega),
\end{align}
one finds
\begin{equation}
 C_{\alpha\beta}^{(j)}(i\omega_n,\vec{k};i\omega_{n'},\vec{k}') = \left[C_{\beta\alpha}^{(j)}(-i\omega_{n'},\vec{k}';-i\omega_n,\vec{k}) \right]^*. \label{RelationOfC}
\end{equation}

The vertex function $\Gamma^{(j)}$ is related to $C^{(j)}$ according to
\begin{align}
 \begin{split}
C_{\beta\alpha}^{(j)}(k_2;k_1) =& \left(\mathcal{G}^m_{\alpha\alpha'}(k_1)\right)_{11} \Gamma^{(j')}_{\alpha'\beta'}(k_1;k_2) \\ & \times \left(\mathcal{G}^m_{\beta'\beta}(k_2)\right)_{11} \chi^{j'j}(k_1-k_2). \end{split}
\end{align} 
Inserting this in \equref{RelationOfC} and using $\chi^*(i\Omega,\vec{q}) = \chi(i\Omega,-\vec{q})$ [see \equref{chiProp}] as well as $\left(\mathcal{G}^m_{\alpha\beta}(i\omega_n,\vec{k})\right)_{11}^* = \left(\mathcal{G}^m_{\beta\alpha}(-i\omega_n,\vec{k})\right)_{11}$, which is readily shown from \equref{SpectralRepOfGmic} [and constitutes a special case of \equref{HermiticityG}], we arrive at the property (\ref{VertexHermiticity}) stated in the main text.

Using the same steps as in Appendix~\ref{SpectralRepSymOrderParameterSusc} and \ref{TRSOfGF}, one can analyze the consequences of TRS in the spectral representation (\ref{threepointSpec}) yielding
\begin{equation}
 C_{\alpha\beta}^{(j)}(k;k') = t \, T^\dagger_{\alpha \alpha'} \left[C_{\alpha'\beta'}^{(j)}(-k;-k')\right]^* T^\pdagger_{\beta' \beta}.
\end{equation} 
Taking into account the properties of the fermionic and bosonic propagator in \equsref{TRSOneCompFermion}{chiProp}, one finds the required identity (\ref{VertexTRS}).

Finally, we show that the asymptotic symmetry introduced in \secref{ApproximateSymmetry} leads to the constraint (\ref{ApproxSymConstr}) for the vertex function.
For this purpose, it is most convenient to work directly in the eigenbasis of the noninteracting fermionic Hamiltonian. Denoting the fermionic creation and annihilation operators in this basis by $\hat{f}^\dagger_{\Omega s}$ and $\hat{f}_{\Omega s}$ [\cf\equref{TrafoInEigenbasis} for their Grassmann analogues], we introduce the antiunitary Fock operator $\hat{R}$ via
\begin{equation}
 \hat{R} \hat{f}_{\Omega s} \hat{R}^\dagger = e^{i\gamma_\Omega^s} \hat{f}_{\Omega s_\R}, \quad \hat{R} \hat{\phi}_{\vec{q}j} \hat{R}^\dagger = s \, t\, \hat{\phi}_{\vec{q}j}.
\end{equation} 
In the asymptotic limit discussed in detail in \secref{ApproximateSymmetry}, the entire Hamiltonian commutes with $\hat{R}$: The quadratic fermionic Hamiltonian is invariant since the Fermi velocities of Rashba partners are asymptotically identical. The bare fermion-boson interaction (\ref{GeneralizedCoupling2}) commutes with $\hat{R}$ as long as \equref{RashbaRelation} with $\vec{\Lambda}_{ss'}(\vec{k},\vec{k}') = \psi_{\vec{k}s}^\dagger \vec{M}(\vec{k},\vec{k}') \psi_{\vec{k}'s'}^\pdagger$ as well as $t \rightarrow s\, t$ holds [which is guaranteed by assuming \equref{EvenFunctionOfK}] and the bosons  vary slowly on the scale $|\vec{g}|/v_F$. Any bosonic Hamiltonian quadratic in $\hat{\phi}$ must be invariant if, as required already before, the symmetry defined by \equsref{SSym}{PhiIrredRep} holds. This readily follows from the fact that $\hat{C} \hat{\phi}_{\vec{q}j} \hat{C}^\dagger = \pm s\, \hat{\phi}_{\vec{q}j}$ for the combined linear operator $\hat{C} = \hat{R}\hat{S}\hat{\Theta}$. Consequently, $\hat{C}$ is a symmetry of any bosonic Hamiltonian that is even in $\hat{\phi}$. If $\hat{C}$, $\hat{\Theta}$ and $\hat{S}$ are symmetries of the bosonic Hamiltonian, the same will hold for $\hat{R}$.

The analysis of the consequences of the invariance under $\hat{R}$ is completely analogous to $\hat{\Theta}$ as both are antiunitary symmetries of the many-body Hamiltonian:
Using the spectral representation of the normal component of the weak-pairing Green's function $\mathcal{G}_s$,
\begin{align}
 \left(\mathcal{G}_{s}(i\omega_n,\Omega)\right)_{11}  = \sum_{n_1,n_2}\frac{\braket{n_1|\hat{f}_{\Omega s}^{\phantom{\dagger}}|n_2}\braket{n_2|\hat{f}^\dagger_{\Omega s}|n_1}}{i\omega_n - (E_{n_2}-E_{n_1})}I_{n_1 n_2}^-,
\end{align}
we find
\begin{equation}
 \left(\mathcal{G}_{s}(i\omega_n,\Omega)\right)_{11} = \left(\mathcal{G}_{s_\R}(-i\omega_n,\Omega)\right)^*_{11}. \label{EigenbasisRSym}
\end{equation}
In the same way, the spectral representation of the three-point function $\widetilde{C}_{ss'}^{(j)}(k,k')$ in the eigenbasis of the normal state Hamiltonian can be used to proof the consequence
\begin{align}
 \begin{split}
&\widetilde{C}_{ss'}^{(j)}(i\omega_n,\Omega;i\omega_{n'},\Omega') \\ &\quad = s\, t \, e^{i(\gamma_{\Omega}^{s}-\gamma_{\Omega'}^{s'})} \left[ \widetilde{C}^{(j)}_{s_\R^{\phantom{.}} s'_\R}(-i\omega_n,\Omega;-i\omega_{n'},\Omega') \right]^* \end{split} \label{VertexRSym}
\end{align}  
of the symmetry under $\hat{R}$.
The combination of \equsref{EigenbasisRSym}{VertexRSym} then leads to the symmetry constraint (\ref{ApproxSymConstr}) of the main text.

By rearranging the sums in the spectral representations such that $\ket{n_j}$ is effectively replaced by $\hat{\mathcal{S}}\ket{n_j}$, one can proof \equref{SSymOfVertexFunc} straightforwardly.

\section{The leading superconducting instability}
\label{HighTemperature}
In this appendix, we show that the leading superconducting instability, i.e., the first nontrivial ($\delta \neq 0$) solution of \equref{EliashbergGap} when the temperature is decreased, is determined by the largest eigenvalue  (the so-called Perron root) of the positive matrix $v$ defined in \equref{DefinitionOfv}. 
As a first step, let us formally diagonalize $v$,
\begin{align}
\begin{split}
 \sum_{n'}\sum_{s'}\int_{s'}\diff\Omega'& v_{s,s'}(i\omega_n,\Omega;i\omega_{n'},\Omega') a^{j}_{s'}(i\omega_{n'},\Omega') \\ &= \lambda^{j}(\beta) a_{s}^{j}(i\omega_{n},\Omega). \label{EigenValueProblem}\end{split}
\end{align} 
Due to $v$ being symmetric and real, this is always possible, all eigenvalues $\lambda^{j}(\beta) \in \mathbbm{R}$ and the eigenvectors $\{a^{j}\}_j$ form an orthonormal basis such that \equref{EliashbergGap} assumes the simple form $\lambda^{j}(\beta) = 1$.
The set of degenerate eigenvalues that, upon lowering the temperature, first become $1$ determines the critical temperature and the order parameter $\delta$ must then be a superposition of the associated eigenvectors. So far, this is completely analogous to mean-field theory. However, due to the more indirect way the temperature enters in the Eliashberg equations, it is not clear whether the largest eigenvalue first becomes $1$. E.g., a finite subset of the eigenvalues could be larger than $1$ for all temperatures. In other words, we still have to show that all eigenvalues are smaller than $1$ in the limit of high temperatures $\beta \rightarrow 0$.

\subsection{Electron-phonon coupling}
\label{ElectronPhononLargeT}
Let us first focus on the effective electron-electron interaction resulting from electron-phonon coupling.
In the high temperature limit, the interaction matrix elements (\ref{InteractionMatrixElement}) become
\begin{equation}
 \mathcal{V}_{ss'}(k;k') \stackrel{\beta \rightarrow 0}{\longrightarrow} - \delta_{n,n'}\sum_l \frac{1}{\omega_{\vec{k}-\vec{k}'l}} \left|\mathcal{G}^{(l)}_{ss'}({\vec{k},\vec{k}'})\right|^2.
\end{equation} 
Using this in \equref{EliashbergQPResidue}, the quasiparticle residue becomes
\begin{equation}
Z_s(i\omega_{n},\Omega) \stackrel{\beta \rightarrow 0}{\longrightarrow} 1 + \frac{2}{|2n+1|} \sum_{s'}\int_{s'}\diff\Omega' \rho_{s'}(\Omega')f_{s,\Omega;s',\Omega'},
\end{equation} 
where, for notational convenience,  we have introduced
\begin{equation}
 f_{s,\Omega;s',\Omega'} = \sum_l \frac{1}{\omega_{\vec{k}-\vec{k}'l}} \left|\mathcal{G}^{(l)}_{ss'}({\vec{k},\vec{k}'})\right|^2.
\end{equation}
Here, $\vec{k}$ and $\vec{k}'$ denote the momenta associated with $(s,\Omega)$ and $(s',\Omega')$, respectively.
The interaction kernel then behaves asymptotically according to
\begin{equation}
 v_{s,s'}(i\omega_n,\Omega;i\omega_{n'},\Omega') \stackrel{\beta \rightarrow 0}{\longrightarrow} 2\delta_{n,n'}  \theta_{s,n,\Omega}\, f_{s,\Omega;s',\Omega'} \, \theta_{s',n',\Omega'} \label{HighTLimOfv}
\end{equation}
with
\begin{equation}
 \theta_{s,n,\Omega} = \frac{\sqrt{\rho_s(\Omega)}}{\sqrt{|2n+1| + 2 \sum_{\tilde{s}}\int_{\tilde{s}}\diff\tilde{\Omega} \, \rho_{\tilde{s}}(\tilde{\Omega})f_{s,\Omega;\tilde{s},\tilde{\Omega}} }}.
\end{equation} 
 
We thus see that, in the high temperature limit, the eigenvalue problem (\ref{EigenValueProblem}) decays into the different sectors characterized by a given Matsubara frequency. Since the right-hand side of \equref{HighTLimOfv} decays monotonically as a function of $|2n+1|$, we know from the Perron-Frobenius theorem\cite{PerronFrobTh} that the Perron root of $\lim_{\beta\rightarrow 0} v$ equals the Perron root of
\begin{equation}
 m_{s,\Omega;s',\Omega'} = \lim_{\beta\rightarrow 0} v_{s,s'}(i\omega_0,\Omega;i\omega_0,\Omega'),
\end{equation} 
which is a matrix only with respect to $s$ and $\Omega$.

We will show in the following that $m_{s,\Omega;s',\Omega'}$, as any matrix of the more general form
\begin{equation}
 M_{\mu\mu'} = \frac{\sqrt{\rho_\mu} f_{\mu\mu'} \sqrt{\rho_{\mu'}}}{\left(c+\sum_{\tilde{\mu}} f_{\mu\tilde{\mu}} \rho_{\tilde{\mu}}\right)^{1/2}\left(c+\sum_{\tilde{\mu}} f_{\mu'\tilde{\mu}} \rho_{\tilde{\mu}}\right)^{1/2}}
\end{equation} 
with $\rho_\mu, f_{\mu\mu'}, c \in\mathbbm{R}^+$, has a Perron root $r_M$ smaller than $1$ ($f$ does not have to symmetric for this to be true). For this purpose, let us rewrite $M = D_1D_2 f D_2 D_1$ where the diagonal matrices are defined according to
\begin{align}
 \left(D_1\right)_{\mu\mu'} &:=\delta_{\mu,\mu'} \frac{1}{\left(c+\sum_{\tilde{\mu}} f_{\mu\tilde{\mu}} \rho_{\tilde{\mu}}\right)^{1/2}}, \\ \left(D_2\right)_{\mu\mu'} &:=\delta_{\mu,\mu'}\sqrt{\rho_\mu}.
\end{align} 
Being similar, the two matrices $\widetilde{M} = D_1^2 f D_2^2$ and $M$ have the same spectrum and, in particular, the same Perron root. We thus know that
\begin{equation}
 r_M \leq \max_{\mu} \sum_{\mu'} \left|\widetilde{M}_{\mu\mu'}\right| = \max_{\mu} \frac{\sum_{\mu'} f_{\mu\mu'} \rho_{\mu'}}{c + \sum_{\tilde{\mu}} f_{\mu\tilde{\mu}} \rho_{\tilde{\mu}}} < 1.
\end{equation} 
Consequently, the eigenvalues of $m$ and, hence, of $\lim_{\beta\rightarrow 0} v$ are all smaller than $1$. Since all $\lambda^j(\beta)$ are continuous functions of $\beta$, the eigenvalue that first becomes $1$ must necessarily be the largest eigenvalue of $v$.

\subsection{Unconventional mechanism}
\label{UnconvLargeT}
The main question in case of unconventional pairing (see \secref{UnconventionalPairing}) concerns the high-temperature limit of the interaction matrix element $\mathcal{V}$ given by \equref{VUnconventional}. To analyze this limit, we first note that $\chi(i\Omega_n,\vec{q}) \rightarrow 0$ for $\Omega_n \rightarrow \infty$ is very reasonable to assume since the same must hold on the real axis in the limit of large energies. As before, the interaction becomes diagonal in Matsubara indices and the analysis is the same as in \secref{ElectronPhononLargeT} above.
The sole difference is that $v$ is replaced by $tv$ in the gap equation (\ref{EliashbergGap}) such that the order parameter $\delta$ of the leading instability belongs to the eigenspace of the positive matrix $v$ with largest or smallest eigenvalue depending on whether the driving fluctuations are TRE or TRO.

\section{Trivial representation}
\label{Representation}
Here, we show that the absence of sign changes of the order parameter, $\widetilde{\Phi}_s(i\omega_{n},\vec{k}) > 0$, implies that the condensate does not break any point symmetry of the normal state.

It is well-known that, for a second order phase transition, the order parameter must transform under one of the irreducible representations of the point group of the high-temperature phase. Denoting this irreducible representation and its dimension by $n_0$ and $d_{n_0}$, respectively, it holds for the order parameter in the weak-pairing description\cite{DesignPrinciples,*DesignPrinciplesUnpub}
\begin{equation}
 \widetilde{\Phi}_s(i\omega_{n},\vec{k}) = \sum_{\mu = 1}^{d_{n_0}} \eta_\mu \chi_{\vec{k} s}^{\mu n_0}(i\omega_n),
\end{equation} 
where $\chi^{\mu n_0}$ denote scalar basis functions transforming under $n_0$ with respect to $\vec{k}$ and $s$.
The grand orthogonality theorem\cite{Lax} of group theory implies that two sets of basis functions, $\{\ket{\chi^{\mu n}}\}$ and $\{\ket{\widetilde{\chi}^{\mu n}}\}$, satisfy $\braket{\widetilde{\chi}^{\mu' n'}|\chi^{\mu n}} \propto \delta_{n,n'}\delta_{\mu,\mu'}$. Let us take $\widetilde{\chi}_{\vec{k} s}(i\omega_n) = 1$, which transforms under the trivial representation, and assume that $n_0$ is a nontrivial representation. It then follows
\begin{equation}
 \sum_{s,\vec{k}} \widetilde{\Phi}_s(i\omega_{n},\vec{k}) = \sum_{\mu = 1}^{d_{n_0}} \eta_\mu \braket{\widetilde{\chi}(i\omega_n)|\chi^{\mu n_0}(i\omega_n)} = 0
\end{equation} 
conflicting with $\widetilde{\Phi}_s(i\omega_{n},\vec{k}) > 0$. The superconductor must thus transform under the trivial representation of the point group of the normal state.

\section{Consequences of a two-fold rotation}
\label{TwoFoldRotation}
In this appendix, it is shown that a two-fold rotation $C_2^{z}$ perpendicular to the plane of a 2D system forces all order parameters to be either even or odd under $C_2^{z}$ leading to \equref{PhiIrredRep}.  

As a first step, we have to show that $C_2^{z}$ commutes with all symmetry operations $h$ of the 2D point group. By construction, there cannot be any symmetry operation relating in-plane ($x$, $y$) and out-of-plane ($z$) coordinates such that the coordinate representation of any $h$ must have the form 
\begin{equation}
 M(h) = \begin{pmatrix} \, \, m(h) & \begin{matrix} 0 \\ 0 \end{matrix} \\ \begin{matrix} 0 & 0 \end{matrix} & c(h) \end{pmatrix}
\end{equation} 
in the basis $\{x,y,z\}$, where $m(h)$ is a real $2\times 2$ matrix and $c(h) \in \mathbbm{R}$. Obviously, $M(h)$ commutes with $M(C_2^z) = \text{diag}(-1,-1,1)$ and, hence, $[C_2^z,h] = 0$.

Therefore, it holds for any representation $\rho$
\begin{equation}
  [\rho(C_2^z),\rho(h)] = 0 \quad \forall h.
\end{equation} 
If $\rho$ is irreducible, Schur's lemma implies $\rho(C_2^z) = C \mathbbm{1}$ with $C \in \mathbbm{C}$. Due to $(C_2^z)^2 = E$, where $E$ is the identity operation, we have $C\in\{+1,-1\}$.

Assuming a second order phase transition, the competing order parameter must transform under one of the irreducible representations of the point group and, hence, can only be either even or odd under $C_2^z$.

\section{Chern numbers of Rashba partners}
\label{ChernNumberRashba}
Here we proof that the symmetry (\ref{ApproximateSymmetryEigenstates}) forces the Fermi surface Chern numbers defined by\cite{ZhangTopInv}
\begin{equation}
 C_{1s} :=\frac{i}{2\pi} \int_{s} \diff \omega^{jj'} \left(\partial_{k_{j'}} \psi^\dagger_{\vec{k}s} \partial_{k_j} \psi^\pdagger_{\vec{k}s}  - (j \leftrightarrow j') \right) \label{DefinitionOfFSChernNumber}
\end{equation} 
to satisfy $C_{1s} = -C_{1s_\R}$. In \equref{DefinitionOfFSChernNumber}, $\diff \omega^{jj'}$ are the surface element two forms of the Fermi surface $s$.

Using the notation $\psi_{\vec{k}s} \approx \psi_{\Omega s}$, we rewrite the integrand  according to
\begin{align}
 & \left(\partial_{k_{j'}} \psi^\dagger_{\Omega s}\right) \partial_{k_j} \psi^\pdagger_{\Omega s}  - (j \leftrightarrow j') \nonumber \\
 & =  \left(\partial_{k_{j'}} e^{-i\gamma^s_\Omega}  \psi^T_{\Omega s_\R} T^\dagger \right) \left( \partial_{k_j}e^{i\gamma^s_\Omega} T \psi^*_{\Omega s_\R} \right) - (j \leftrightarrow j') \nonumber \\
 & = -\left[ \left(\partial_{k_{j'}} \psi^\dagger_{\Omega s_\R}\right) \partial_{k_j} \psi^\pdagger_{\Omega s_\R}  - (j \leftrightarrow j') \right].
\end{align}
Here we have applied the symmetry (\ref{ApproximateSymmetryEigenstates}) in the second line and used $(\partial_{k_j} \psi^\dagger)\psi = -\psi^\dagger \partial_{k_j}\psi$ in the last line to show that all contributions involving derivatives of the phases $\gamma^s_\Omega$ vanish due to the antisymmetrization in $j$ and $j'$. 
Inserting this back into the integral of \equref{DefinitionOfFSChernNumber}, we obtain the required property.

\section{Derivation of the Ginzburg-Landau expansion and the transition temperatures}
\label{DisorderCalc}
In this appendix, we provide more details about how the results for disordered superconductors discussed in Secs.~\ref{DisorderGLExp} and \ref{DisorderIndTop} have been obtained.

Let us first derive the general form (\ref{GeneralExpreD}) of the disordered particle-particle diagram
\begin{align}\begin{split}
&D_{\Omega s,\Omega' s'} = -\frac{T^3}{2} \Biggl\langle \Biggl\langle \sum_{\omega_n,k_\perp} e^{i\varphi_\Omega^s}  f_{\Omega_\K -k_\perp-\omega_ns_\K} f_{\Omega k_\perp \omega_n s}  \\ &\quad \times\sum_{\omega_{n'},k_\perp'} e^{-i\varphi_{\Omega'}^{s'}} \bar{f}_{\Omega' k_\perp' \omega_{n'} s'} \bar{f}_{\Omega'_\K -k_\perp' -\omega_{n'} s_\K'}   \Biggr\rangle_0 \Biggr\rangle_{\text{dis}} \label{DefinitionParticleParticleBubble}\end{split}
\end{align}
that determines the Ginzburg-Landau expansion (\ref{GLExp}). In \equref{DefinitionParticleParticleBubble}, $k_\perp$ denote the momenta perpendicular to the Fermi surface and $\braket{\dots}_0$ represents the expectation value with respect to the normal state Hamiltonian (\ref{NonInteractingHam}) perturbed by a given disorder configuration (\ref{DisorderPotential}). 

As illustrated in \figref{DisorderFigures}(b), the calculation of $D$ proceeds in two steps: One first deduces the disorder-renormalized (normal) Green's function
\begin{equation}
 G_s(i\omega_n,\Omega,k_\perp) = \left[i\omega_n - \epsilon_s(\Omega,k_\perp)-\Sigma_s(i\omega_n,\Omega,k_\perp)\right]^{-1}.
\end{equation} 
The rainbow diagrams in \figref{DisorderFigures}(c) yield for the self-energy
\begin{equation}
 \Sigma_s(i\omega) = -i\pi\sign(\omega_n)\sum_{\tilde{s}} \int_{\tilde{s}}\diff\widetilde{\Omega} \, \rho_{\tilde{s}}(\widetilde{\Omega}) \, \mathcal{S}^S_{\Omega s,\widetilde{\Omega} \tilde{s}},
\end{equation} 
where we have used $\rho_s(\Omega) = \rho_{s_\K}(\Omega_\K)$ to symmetrize the scattering vertex.

Secondly, we have to take into account the vertex corrections and sum up the Cooperon ladder shown in \figref{DisorderFigures}(d). The central building block of the ladder is
\begin{equation}
 c_s(i\omega_n,\Omega) = \sum_{k_\perp} G_s(i\omega_n,\Omega,k_\perp) G_{s_\K}(-i\omega_n,\Omega_\K,-k_\perp), 
\end{equation} 
which describes the propagation of a Cooper pair $\{s,\Omega;s_\K,\Omega_\K \}$. One finds
\begin{equation}
 c_s(i\omega_n,\Omega) = \rho_s(\Omega)\left(\frac{|\omega_n|}{\pi} + \sum_{\tilde{s}} \int_{\tilde{s}}\diff\widetilde{\Omega} \, \rho_{\tilde{s}}(\widetilde{\Omega}) \, \mathcal{S}^S_{\Omega s,\widetilde{\Omega}\tilde{s}}\right)^{-1}.
\end{equation} 
Using the diagonal matrix $\mathcal{C} = \text{diag}(c^{-1})$ as explicitly defined in \equref{DiagonalMatrixDef}, one can write the particle-particle bubble as an infinite series
\begin{align}
 \begin{split}
& D_{\Omega s,\Omega' s'} = -T \sum_{\omega_n} \Bigl(\mathcal{C}^{-1}(i\omega_n) + t_\gamma \mathcal{C}^{-1}(i\omega_n) \mathcal{S}^S \mathcal{C}^{-1}(i\omega_n) \\ & \quad + t^2_\gamma \mathcal{C}^{-1}(i\omega_n) \mathcal{S}^S \mathcal{C}^{-1}(i\omega_n)\mathcal{S}^S \mathcal{C}^{-1}(i\omega_n) + \dots \Bigr)_{\Omega s,\Omega' s'}. \end{split}
\end{align}
Note that also the Cooper scattering (\ref{VDefinition}) enters in the symmetrized form (\ref{SymmetrizedSF}) resulting from the different Wick contractions. As required by gauge symmetry, the phases $\varphi_\Omega^s$ in \equsref{DefinitionParticleParticleBubble}{VDefinition} cancel. 
Summing up the geometric series one finds the compact form (\ref{GeneralExpreD}) stated in the main text.

In the simple case $\mathcal{S}^S_{\Omega s,\Omega' s'} = \gamma$ which is realized in the example discussed in the main text, one can invert $\mathcal{C} - t_\gamma \, \mathcal{S}^S$ analytically. Defining $\rho_F := \sum_s \int\diff\Omega \rho_s(\Omega)$, \equref{GeneralExpreD} becomes 
\begin{align}
 &D_{\Omega s,\Omega' s'} \\ &= -T \sum_{\omega_n}\frac{\rho_s(\Omega)}{\frac{|\omega_n|}{\pi} + \rho_F\gamma} \Biggl( \delta_{s,s'}\delta_{\Omega,\Omega'} + \frac{t_\gamma \rho_{s'}(\Omega')\gamma}{\frac{|\omega_n|}{\pi} + \rho_F \gamma(1-t_\gamma)} \Biggr) \nonumber
\end{align}
With the assumptions stated in the main text, $\rho_s(\Omega) \approx \text{const.}$ and \equref{SimpleInteraction}, we have $\widetilde{\Delta}_s(\Omega) =\widetilde{\Delta}_s$ and the kernel $\widetilde{D}$ of the free energy expansion (\ref{GLExp}) effectively becomes a $2 \times 2$ matrix. Its eigenvalues $\lambda^{++}$ and $\lambda^{+-}$ corresponding to the $s^{++}$ ($\widetilde{\Delta}_1=\widetilde{\Delta}_2$) and $s^{+-}$ ($\widetilde{\Delta}_1=-\widetilde{\Delta}_2$) state, the zeros of which determine the associated transition temperatures, read 
\begin{subequations}
\begin{align}
 \lambda^{++} &= -\frac{1}{U+J} -\frac{\rho}{2}\ln\left(\frac{2e^\gamma \Lambda}{\pi T}\right), \\
 \lambda^{+-} &= -\frac{1}{U-J} -\frac{\rho}{2}\left[\ln\left(\frac{\Lambda}{2\pi T}\right) - \psi\left(\frac{1}{2} + \frac{\gamma}{2T}\right)\right] \label{TcSpmNonMag}
\end{align}
\end{subequations}
for nonmagnetic $(t_\gamma = +1)$ disorder. Here $\psi$ denotes the digamma function. We see that the transition temperature $T_c^{++}$ of the $s^{++}$ state is unaffected by disorder as required by the Anderson theorem\cite{AT,ATAG1,ATAG2}. The critical temperature $T_c^{+-}$ of the $s^{+-}$ phase is reduced as can be seen in \equref{TcSpmNonMag} by noting that $\psi(x)$ is monotonically increasing for $x>0$. As long as $J<0$, it holds $T_c^{+-}< T_c^{++}$ irrespective of the disorder strength.

In case of magnetic ($t_\gamma = -1$) scattering, we have
\begin{subequations}
\begin{align}
 \lambda^{++} &= -\frac{1}{U+J} -\frac{\rho}{2}\left[\ln\left(\frac{\Lambda}{2 \pi T}\right) - \psi\left(\frac{1}{2} + \frac{2\gamma}{2T}\right)\right], \label{TcSppMag} \\	
 \lambda^{+-} &= -\frac{1}{U-J} -\frac{\rho}{2}\left[\ln\left(\frac{\Lambda}{2 \pi T}\right) - \psi\left(\frac{1}{2} + \frac{\gamma}{2T}\right)\right]. \label{TcSpmMag}
\end{align}\label{TcMag}\end{subequations}
In this case, both $T_c^{++}$ and $T_c^{+-}$ are reduced by disorder. However, $T_c^{++}$ is more strongly affected due to the additional factor of $2$ in front of $\gamma$ in \equref{TcSppMag}. Physically, $2\gamma$ has to be seen as the sum of intra- and interband scattering strengths which happen to be identical in the example considered, whereas $\gamma$ in \equref{TcSpmMag} is just the intraband contribution. 

\equsref{CriticalScatteringRate}{JoUCrit} are straightforwardly obtained by analyzing \equref{TcMag} in the associated asymptotic limits $J/U \rightarrow 0$ and $T	 \rightarrow 0$, respectively.
 
%

\end{document}